\documentclass[11pt]{article}
\usepackage{amsmath,amsfonts}
\usepackage{geometry}
\usepackage{caption}
\usepackage{cite}
\usepackage{setspace}
\usepackage{url}
\usepackage{comment}
\usepackage{hyperref}
\newcommand{\be}{\begin{equation}}
\newcommand{\ee}{\end{equation}}

\newcommand{\ON}[1]{\mathrm{O}( #1 )}
\newcommand{\SU}[1]{\mathrm{SU}( #1 )}
\newcommand{\SL}[1]{\mathrm{SL}( #1 )}
\newcommand{\GL}[1]{\mathrm{GL}( #1 )}
\newcommand{\SO}[1]{\mathrm{SO}( #1 )}
\newcommand{\EG}[1]{\mathrm{E}_{#1(#1)}}
\newcommand{\ODD}{\mathrm{O}(D,D)}
\newcommand{\Odd}{\mathrm{O}(d,d)}
\newcommand{\UO}{\mathrm{U}(1)}
\newcommand{\Spin}[1]{\mathrm{Spin}(#1)}
\newcommand{\GH}{\mathrm{G}_{\mathrm{half}}}
\newcommand{\GR}{\mathrm{G}_{R}}

\newcommand{\USp}[1]{\mathrm{USp}(#1)}

\newcommand{\D}{\mathfrak{D}}
\newcommand{\tr}{\mathrm{tr}}

\newcommand{\hg}{\bag}
\newcommand{\hK}{\hat{K}}
\newcommand{\K}{K}
\newcommand{\J}{J}
\newcommand{\LG}{J}
\newcommand{\LO}{\omega}
\newcommand{\TL}{\LG}
\newcommand{\hJ}{\hat{J}}

\newcommand{\obf}[1]{\overline{\mathbf{#1}}}
\newcommand{\mbf}[1]{\mathbf{#1}}
\newcommand{\qbf}[4]{\left(\mbf{#1},\mbf{#2},\mbf{#3},\mbf{#4}\right)}
\newcommand{\dbf}[2]{\left(\mbf{#1},\mbf{#2}\right)}
\newcommand{\tbf}[3]{\left(\mbf{#1},\mbf{#2},\mbf{#3}\right)}
\newcommand{\gL}{\mathcal{L}}

\newcommand{\gM}{\mathcal{M}}
\newcommand{\gH}{\mathcal{H}}
\newcommand{\tc}{{\cal W}}

\newcommand{\cRY}{{\cal R}^Y}
\newcommand{\cSY}{{\cal S}^Y}

\newcommand{\bR}{\bar{R}}

\newcommand{\vol}{\mathrm{vol}}

\newcommand{\bae}{\bar{e}}
\newcommand{\bag}{\bar{g}}
\numberwithin{equation}{section}

\newcommand\Tstrut{\rule{0pt}{3ex}}         % = `top' strut
\newcommand\Bstrut{\rule[-1.3ex]{0pt}{0pt}}   % = `bottom' strut

\begin{document}

\begin{titlepage}
\vfill

\begin{flushright}
LMU-ASC 39/17
\end{flushright}

\vfill

\begin{center}
   \baselineskip=16pt
   	{\LARGE \bf Half-maximal supersymmetry from \\\vskip.2cm exceptional field theory}
   	\vskip 2cm
 	{\large \bf Emanuel Malek}
   	\vskip .6cm
   	{\it Arnold Sommerfeld Center for Theoretical Physics, Department f\"ur Physik, \\ Ludwig-Maximilians-Universit\"at M\"unchen, Theresienstra{\ss}e 37, 80333 M\"unchen, Germany}
   	\vskip .6cm
   	{{E.Malek@lmu.de}} \\
   	\vskip 2cm
\end{center}
\vfill

\begin{abstract}
We study $D \geq 4$-dimensional half-maximal flux backgrounds using exceptional field theory. We define the relevant generalised structures and also find the integrability conditions which give warped half-maximal Minkowski$_D$ and AdS$_D$ vacua. We then show how to obtain consistent truncations of type II / 11-dimensional SUGRA which break half the supersymmetry. Such truncations can be defined on backgrounds admitting exceptional generalised $\SO{d-1-N}$ structures, where $d = 11 - D$, and $N$ is the number of vector multiplets obtained in the lower-dimensional theory. Our procedure yields the most general embedding tensors satisfying the linear constraint of half-maximal gauged SUGRA. We use this to prove that all $D\geq 4$ half-maximal warped AdS$_D$ and Minkowski$_D$ vacua of type II / 11-dimensional SUGRA admit a consistent truncation keeping only the gravitational supermultiplet.  We also show to obtain heterotic double field theory from exceptional field theory and comment on the M-theory / heterotic duality. In five dimensions, we find a new $\SO{5,N}$ double field theory with a $\left(6+N\right)$-dimensional extended space. Its section condition has one solution corresponding to 10-dimensional ${\cal N}=1$ supergravity and another yielding six-dimensional ${\cal N}=\left(2,0\right)$ SUGRA.
\end{abstract}

\vfill

\end{titlepage}

\tableofcontents

\newpage

\section{Introduction}
Exceptional field theory (EFT) \cite{Hull:2007zu,Berman:2010is,Hohm:2013vpa,Hohm:2013uia} is an $\EG{d}$-invariant extension of supergravity, which includes 11-dimensional and IIB SUGRA in a unified formalism based on an enlarged coordinate space . In particular, just like generalised geometry \cite{Gualtieri:2003dx,Hitchin:2004ut,Grana:2006hr,Coimbra:2011nw,Coimbra:2011ky,Coimbra:2012af}, and double field theory (DFT) \cite{Siegel:1993th,Siegel:1993xq,Hull:2009mi,Hull:2009sg,Hull:2009zb}, which is based on earlier work of \cite{Duff:1989tf,Duff:1990hn,Tseytlin:1990nb,Tseytlin:1990va}, exceptional field theory treats the metric and $p$-form gauge fields on an equal footing. This makes it a natural, and powerful, tool in the study of flux vacua of string theory.

For example, exceptional field theory, double field theory and generalised geometry have been successfully used to find consistent truncations of 10- and 11-dimensional supergravity. A truncation is consistent if all the solutions of the lower-dimensional theory are also solutions to the original, higher-dimensional theory. Finding such truncations is a notoriously hard problem because of the non-linearity of the equations of motion \cite{Duff:1984hn}.

The way this is tackled in exceptional field theory is to generalise the notion of a Scherk-Schwarz reduction on a group manifold \cite{Scherk:1979zr} to include fluxes. Such generalised Scherk-Schwarz Ans\"atze \cite{Aldazabal:2011nj,Geissbuhler:2011mx,Grana:2012rr,Dibitetto:2012rk,Aldazabal:2013sca,Berman:2012uy,Geissbuhler:2013uka,Berman:2013cli,Godazgar:2013dma,Godazgar:2013oba,Godazgar:2013pfa,Hohm:2014qga,Lee:2014mla,Lee:2015xga,Godazgar:2015qia,Cho:2015lha} can be used to describe consistent truncation on generalised parallelisable spaces \cite{Lee:2014mla}, which includes the $S^4$, $S^5$ and $S^7$ truncations of 11- and 10-dimensional supergravity \cite{Lee:2014mla,Hohm:2014qga,Baguet:2015sma}. In addition to finding new consistent truncations on spheres and hyperboloids \cite{Lee:2014mla,Hohm:2014qga,Cassani:2016ncu} and non-geometric backgrounds \cite{Lee:2015xga,Inverso:2016eet}, the set-up has been used to study the relationship between different consistent truncations \cite{Malek:2015hma,Malek:2015etj}, as well as to prove the consistency of Pauli reductions on group manifolds \cite{Baguet:2015iou}. It has also allowed uplifts of maximal and half-maximal dyonic gaugings \cite{Inverso:2016eet,Ciceri:2016hup}.\footnote{The closely-related approach of \cite{Ciceri:2014wya,Guarino:2015jca,Guarino:2015vca} has also been fruitful in finding consistent truncations.}

However, because they require globally well-defined generalised frame fields, these generalised Scherk-Schwarz truncations can only be performed consistently on backgrounds which preserve all supersymmetries. A much more interesting class of flux geometries is given by those breaking some of the supersymmetry. In \cite{Malek:2016bpu} it was shown how to describe half-maximal backgrounds in $\SL{5}$ exceptional field theory \cite{Berman:2010is} and how to use this to define consistent truncations yielding general half-maximal seven-dimensional gauged supergravities. These methods were also used in \cite{Malek:2016vsh} to show that exceptional field theory can be reduced to heterotic double field theory.

Here we extend these results by studying half-maximal supersymmetry in $D \geq 4$ dimensions within the framework of exceptional field theory. In particular, we will show how EFT can be used to study generic half-maximal flux geometries, which are naturally encoded in terms of exceptional generalised $G$ structures \cite{Coimbra:2014uxa}. We show how to describe the relevant ``$\GH$ structures'' using the exceptional field theory analogue of differential forms.

A key ingredient in this method is the so-called intrinsic torsion of the $\GH$ structure \cite{Coimbra:2014uxa}. This encodes to what extend the backgrounds break supersymmetry and violate the equations of motion. Mathematically, it is the obstruction to introducing a $\GH$-compatible torsion-free connection. We show how to find the intrinsic torsion by making use of the structures appearing in the EFT tensor hierarchy \cite{Cederwall:2013naa,Hohm:2013vpa,Hohm:2013uia,Hohm:2015xna,Wang:2015hca}. This allows us to define a suitable notion of (weak) integrability which leads to half-maximal Minkowski or AdS vacua.

Consistent half-maximal truncations are one of the main applications we consider in this paper. We show to define these on flux geometries which preserve only half the supersymmetries and demonstrate that this, in principle, leads to generic half-maximal gaugings. In particular, our method naturally leads to the most general gaugings satisfying the linear constraint of half-maximal gauged SUGRA \cite{Schon:2006kz,Bergshoeff:2007vb,Dibitetto:2015bia}. This includes those deformations which cannot be obtained via generalised Scherk-Schwarz reductions of double field theory but which are particularly interesting phenomenologically. For example, in $D = 4$ we naturally obtain half-maximal gauged SUGRA with $\SL{2}$ angles, which can otherwise only be obtained modifying the higher-dimensional double field theory as in \cite{Ciceri:2016hup}. The truncation Ansatz and the conditions which guarantee its consistency, are naturally encoded in the $\GH$ structure and its intrinsic torsion. The quadratic constraint is automatically satisfied if we satisfy the section condition.

One of the results that we find is that consistent truncations of type II and 11-dimensional supergravity only yield half-maximal gauged SUGRAs with a small number, $N \leq d-1$, of vector multiplets. Here $d = 11 - D$ is the rank of the exceptional group controlling the relevant EFT. Another is that we are able to prove the half-maximal version of the conjecture of \cite{Gauntlett:2007ma}: we show that for any half-maximal warped AdS$_D$ vacuum of type II or 11-dimensional SUGRA, with $D \geq 4$, there is a consistent truncation keeping only the gravitational supermultiplet. Our proof also extends to half-maximal warped Mink$_D$ vacua, thus proving a special case of the conjecture of \cite{Duff:1985jd}.

Finally, we use these tools to show how EFT can be reduced to heterotic double field theory. This requires half of the supersymmetry to be broken, which is why the method developed for consistent truncations is useful. In particular, heterotic DFT is obtained by modifying the consistent truncation Ansatz to allow the would-be lower-dimensional fields to still depend on the internal coordinates. This allows one to easily see which lower-dimensional theories can be obtained from truncations of both type II / M-theory and heterotic theory, thus making the duality between these manifest. We should also mention that the analogous method based on generalised Scherk-Schwarz reductions has recently been used to show that exceptional field theory contains both massive IIA \cite{Ciceri:2016dmd} and generalised IIB SUGRA \cite{Baguet:2016prz}.

In five dimensions, we find that the $\EG{6}$ EFT can be reduced to a new half-maximal $\SO{5,N}$ double field theory with a $6+N$ dimensional coordinate space. The section condition of this theory has two inequivalent solutions. While one of them allows dependence on five coordinates, thus corresponding to a reformulation of 10-dimensional heterotic or type I supergravity, the other only allows dependence on a single coordinate. We argue that this solution corresponds to a 5+1 split of six-dimensional ${\cal N}=\left(2,0\right)$ SUGRA. A similar phenomenon was found in ``double field theory at $\SL{2}$ angles'' \cite{Ciceri:2016hup}.

This paper is roughly organised into two parts. The first shows how to describe half-maximal supersymmetry in $D \geq 5$ dimensions, excluding chiral half-maximal supersymmetry in six dimensions, in exceptional field theory. In particular, in section \ref{s:GHalfDgeq5} we define the appropriate $\GH$ structures and their intrinsic torsion, and give the (weak) integrability conditions which imply that the flux geometry defines a half-maximal (AdS$_D$) Mink$_D$ vacuum. The details of this construction in the various dimensions can be found in appendices \ref{s:D=7} -- \ref{s:D=5}.

In section \ref{s:Reformulate} we show how to reformulate the EFT action in terms of the $\GH$ structure. This is the half-maximal analogue of the ``flux formulation'' of double and exceptional field theory \cite{Geissbuhler:2013uka,Berman:2013uda,Blair:2014zba}. We apply the technology developed to show how to define consistent truncations yielding half-maximal gauged SUGRA in section \ref{s:ConsTruncation}. This also allows us to prove the half-maximal case of the conjecture of \cite{Gauntlett:2007ma}: for every warped half-maximal AdS$_{D}$ vacuum of 10- or 11-dimensional SUGRA there is a consistent truncation keeping only the gravitational supermultiplet.

We end the first part of the paper by showing in section \ref{s:HetConsTruncation} how to obtain the heterotic double field theory as a consistent truncation of EFT. We also show in subsection \ref{s:ModDFT} that the $\EG{6}$ EFT contains a novel double field theory with $\SO{5,N}$ symmetry but with a $6+N$-dimensional enlarged space. This theory unifies half-maximal ten-dimensional and six-dimensional ${\cal N}=\left(2,0\right)$ SUGRA, similar to ``double field theory at $\SL{2}$ angles'' \cite{Ciceri:2016hup}.

The second part deals with the cases of half-maximal supersymmetry in four dimensions, chiral supersymmetry in six dimensions and half-maximal structures in double field theory. These are discussed in sections \ref{s:D=4}, \ref{s:D=6b} and \ref{s:HalfDFT}, respectively. For each of these cases, we introduce the appropriate $\GH$ structures and their intrinsic torsion, and discuss consistent truncations. We also prove the relevant cases of the conjecture of \cite{Gauntlett:2007ma}.

Those readers who are more familiar with double field theory than exceptional field theory might want to read section \ref{s:HalfDFT} first as a warm-up. This will hopefully allow them to grasp the main ideas of the half-maximal structure before reading the remaining sections which deal with exceptional field theory.

\section{Review of exceptional field theory} \label{s:Review}

Let us briefly introduce the key features of exceptional field theory that are relevant for our discussions. We refer the interested reader to the comprehensive reviews of this subject can be found in \cite{Aldazabal:2013sca,Berman:2013eva,Hohm:2013bwa}. Exceptional field theory is an extension of 10- and 11-dimensional supergravity that makes an $\EG{d}$ symmetry manifest. The relevant groups can be found in table \ref{t:GHalf}.

Just like in exceptional generalised geometry \cite{Coimbra:2011ky,Coimbra:2012af}, one starts by performing a Kaluza-Klein split of 11-dimensional SUGRA (or alternatively of type II SUGRA) into $D$ ``external'' dimensions and $d$ internal ones. Upon performing this split, the bosonic degrees of freedom can be organised into representations of $\EG{d}$. For example, the purely internal fields can be combined into a symmetric $\EG{d}$ matrix, $\gM$, parameterising the coset space
\begin{equation}
 \gM \in \frac{\EG{d}}{H_d} \,,
\end{equation}
where $H_d$ is the maximal compact subgroup of $\EG{d}$, see table \ref{t:GHalf}. The matrix $\gM$ is called the generalised metric and plays the analogous role to the metric in differential geometry.

The bosonic fields with fixed number, $i$, of external legs combine into the representations, $R_i$, of $\EG{d}$ which also appear in the tensor hierarchy of maximal gauged SUGRA \cite{deWit:2008ta,deWit:2008gc,Kleinschmidt:2011vu,Palmkvist:2011vz}. These representations can be found in table \ref{t:TensorHierarchy} and we will refer to these fields as $A_\mu$, $B_{\mu\nu}$, $C_{\mu\nu\rho}$, \ldots, where we suppress the EFT indices. For these gauge fields one can introduce field strengths, ${\cal F}_{\mu\nu}$, ${\cal H}_{\mu\nu\rho}$, \ldots, and we refer to \cite{Hohm:2013vpa,Hohm:2013uia,Hohm:2015xna,Abzalov:2015ega,Wang:2015hca,Musaev:2015ces,Berman:2015rcc,Bosque:2016fpi} for further details. Finally, the $d$ internal coordinates are viewed as part of an enlarged coordinate space $Y^M$ which forms the $R_1$ representation of $\EG{d}$.

So far, the discussion has essentially been fixed at a point in the internal space. A crucial point to extend this globally is that the diffeomorphisms and gauge transformations of 11-dimensional SUGRA act as local $\EG{d}$ transformation, which we call generalised diffeomorphisms. The generators of diffeomorphisms, a vector field, and gauge transformations, a set of $p$-form fields, combine into a generalised vector field, which transforms in the $R_1$ representation of $\EG{d}$. They generate the local symmetries via the generalised Lie derivative, which takes the form \cite{Coimbra:2011nw,Berman:2011cg,Coimbra:2011ky,Berman:2012vc}
\begin{equation}
 \gL_{\Lambda} V^M = \Lambda^N \partial_N V^M + \left(\mathbb{P}_{adj}\right)^{M}{}_N{}^P{}_Q V^N \partial_P \Lambda^Q  + \lambda V^M \partial_N \Lambda^N \,. \label{eq:GenLieDerivative}
\end{equation}
Here $V$ is a generalised vector field of weight $\lambda$, $M$ is an index of the $R_1$ representation of $\EG{d}$, $\partial_M$ denotes derivatives with respect to the $Y^M$ coordinates, and $\mathbb{P}_{adj}$ is the projector onto the adjoint of $\EG{d}$. For the remainder of the paper, it is useful to highlight that when $\lambda=\frac1{D-2}$, \eqref{eq:GenLieDerivative} can be rewritten as
\begin{equation}
 \gL_{\Lambda} V^M = \Lambda^N \partial_N V^M - V^N \partial_N \Lambda^M + Y^{MN}_{PQ} V^P \partial_N \Lambda^Q \,, \label{eq:GenLieDerivativeY}
\end{equation}
where $Y^{MN}_{PQ}$ is an $\EG{d}$ invariant \cite{Berman:2012vc}. Unless otherwise stated, we will from now on always take generalised vector fields to have weight $\frac{1}{D-2}$.

If the fields have arbitrary dependence on the $Y^M$ coordinates, the theory fails to be consistent. A minimal requirement that needs to be imposed is that the algebra of generalised diffeomorphisms closes. This requires us to impose the ``section condition''
\begin{equation}
 Y^{MN}_{PQ} \partial_M \otimes \partial_N = 0 \,, \label{eq:SectionCondition}
\end{equation}
where the derivatives are taken to act on any field of the exceptional field theory, including as double derivatives on the same field. Different solutions of the section condition lead to 11-dimensional or IIB SUGRA \cite{Hohm:2013pua,Blair:2013gqa,Hohm:2013vpa,Hohm:2013uia}, and in this sense it unifies these theories. One can also imagine spaces in which the solution to the section condition is not globally well-defined. These would correspond to non-geometric, or U-fold, backgrounds \cite{Hull:2004in,Hull:2006va,Hull:2009sg,Hellerman:2002ax}.

One can write a unique action for exceptional field theory that is invariant under generalised diffeomorphisms, up to the section condition \cite{Hohm:2013vpa,Hohm:2013uia,Wang:2015hca,Abzalov:2015ega,Musaev:2015ces,Berman:2015rcc}. Upon solving the section condition \eqref{eq:SectionCondition} this reduces to the action of 10- or 11-dimensional SUGRA. While here we have focused only on the bosonic part of exceptional field theory, it is also possible to include fermions, as was done explicitly for $\EG{7}$ and $\EG{6}$ in \cite{Godazgar:2014nqa,Musaev:2014lna}. It should be noted that fermions have been included in the appropriate exceptional generalised geometry \cite{Coimbra:2012af}. Finally, although not relevant to our considerations here, the $\EG{8}$ EFT has also been constructed in \cite{Hohm:2014fxa} and supersymmetrised in \cite{Baguet:2016jph}.

\section{Exceptional $\GH$ structures in $D \geq 5$ dimensions} \label{s:GHalfDgeq5}
We want to consider reductions of type II or 11-dimensional SUGRA on some ``internal manifold'', $M_d$, to a $D$-dimensional half-maximal SUGRA. In order to obtain a half-maximal theory in $D$ dimensions, $M_d$ must admit a half-maximal number of spinors. If we started in 11-dimensions and had no flux on $M_d$, these would have to be spinors of $\SO{d}$. \footnote{Note that if we are considering truncations of type II SUGRA, $M_d$ is actually a $\left(d-1\right)$-dimensional manifold.}

However, this it not so if we include flux. In this case, the flux terms in the supersymmetry variations, schematically of the type $F_{\mu_1\ldots\mu_4} \gamma^{\mu_1\ldots\mu_4} + \ldots$, generate a $H_d$ action on the spinors, see e.g. \cite{Coimbra:2012af,Godazgar:2014nqa}. This means that we should be working with spinors of $H_d \supset \SO{d}$, just like we work with bosonic $E_{d(d)}$ tensors rather than $\GL{d}$ tensors. Doing this also allows us to treat the type II theories and 11-d SUGRA on the same footing.

We now return to the condition of obtaining a $D$-dimensional half-maximal theory, without assumptions on the flux. In light of the above comments, we see that $M_d$ must admit a half-maximal set of $H_d$ spinors. This requirement is naturally phrased in the language of $G$ structures: $M_d$ must have an exceptional generalised $\GH = \SO{d-1}$ structure \cite{Coimbra:2014uxa}, since this is the stabiliser of a half-maximal set of spinors in $H_d$, see table \ref{t:GHalf}.\footnote{Throughout this paper we will ignore discrete group factors, for the sake of simplicity.} This means that the structure group of the exceptional generalised tangent bundle, which in general is $E_{d(d)}$, can be reduced to $\GH \subset E_{d(d)}$, in analogy to ordinary $G$ structures. One can already see that the language here is natural for discussing supersymmetry: the R-symmetry of the half-maximal supergravity is the maximal commutant of $\GH \subset H_d$,  and also listed in table \ref{t:GHalf}.

\vskip1em
\noindent\makebox[\textwidth]{
 \begin{minipage}{\textwidth}
  \begin{center}
  \begin{tabular}{|c|c|c|c|c|}
  \hline
  $D$ & $E_{d(d)}$ & $H_d$ & $\GH$ & $\GR$ \Tstrut\Bstrut \\ \hline
  7 & $\SL{5}$ & $\USp{4}$ & $\SU{2}$ & $\SU{2}$ \\
  6a & $\Spin{5,5}$ & $\USp{4}\times\USp{4}$ & $\SU{2}\times\SU{2}$ & $\SU{2}\times\SU{2}$ \\
  6b & $\Spin{5,5}$ & $\USp{4}\times\USp{4}$ & $\USp{4}$ & $\USp{4}$ \\
  5 & $\EG{6}$ & $\USp{8}$ & $\USp{4}$ & $\USp{4}$ \\
  4 & $\EG{7}$ & $\SU{8}$ & $\SU{4}$ & $\SU{4} \times \mathrm{U}(1)$ \\
   \hline
  \end{tabular}
  \vskip-0.5em
  \captionof{table}{\small{$\GH$ structures and R-symmetry groups in various dimensions. 6a and 6b refer to the non-chiral and chiral 6-dimensional half-maximal supergravities, respectively.}}  \label{t:GHalf}
  \end{center}
 \end{minipage}}
\vspace{1em}

If we want the reduction on $M_d$ to yield a half-maximal Minkowski$_D$ or AdS$_D$ vacuum, we also need to impose certain differential constraints on the exceptional generalised $\GH$ structure. These are known as integrability, or ``holonomy'' constraints, which we will discuss in sections \ref{s:Vacua} and \ref{s:4DVacua}. We will also show in section \ref{s:ConsTruncation}, that even if we want to obtain a consistent truncation rather than just vacua, we still need to impose a set of differential constraints on the $\GH$ structure, although these are weaker than the integrability constraints.

Both sets of constraints again naturally fit into the framework of generalised $G$ structures. They make use of the so-called generalised intrinsic torsion of the generalised $\GH$ structure, which we will define in subsection \ref{s:IntTorsion}.\footnote{In the case of $\SU{3}$ and $G_2$ structures, the components of the intrinsic torsion are also known as the torsion classes \cite{Joyce}.} From now onwards, we will often drop the adjectives ``exceptional'' and ``generalised'' to avoid clutter, with the understanding that all structures are defined on the exceptional generalised tangent bundle. 

The $\GH$ structures can conveniently be defined in terms of a set of nowhere-vanishing generalised tensor fields which is stabilised by the generalised structure group. One can think of these generalised tensor fields as being a generalisation of differential forms. To be more precise, there is a natural graded product between these differential forms, generalising the wedge product, and a nilpotent derivative. We will make extensive use of these in describing the $\GH$ structures and their intrinsic torsion in sections \ref{s:GHalfStr} and \ref{s:IntTorsion}.

\subsection{Generalised differential forms} \label{s:DiffForms}
As has already been observed in \cite{Cederwall:2013naa,Hohm:2015xna,Wang:2015hca} the sections of the exceptional vector bundles which appear in the tensor hierarchy of exceptional field theory can be thought of as differential forms. Here we will extend this analogy and make use of it to describe the $\GH$ structures. We label the relevant exceptional vector bundles $\bar{{\cal R}}_i$. Their fibres are the vector spaces $R_i$, listed in table \ref{t:TensorHierarchy}. Note that the base space is the full space, not just $M_d$, i.e. tensors in these bundles can depend on all, internal and external, coordinates.

\vspace{1em}
\noindent\makebox[\textwidth]{
 \begin{minipage}{\textwidth}
  \begin{center}
  \begin{tabular}{|c|c|c|c|c||c|}
  \hline
  $D$ & $R_1$ & $R_2$ & $R_{3}$ & $R_4$ & $R_c$ \Tstrut\Bstrut \\ \hline
  7 & $\mbf{10}$ & $\obf{5}$ & $\mbf{5}$ & $\obf{10}$ & $\emptyset$ \\
  6 & $\mbf{16}$ & $\mbf{10}$ & $\obf{16}$ & N/R & $\mbf{1}$ \\
  5 & $\mbf{27}$ & $\obf{27}$ & $\mbf{78} \oplus \mbf{1}$ & N/R & $\mbf{27}$ \\
  4 & $\mbf{56}$ & $\mbf{133}$ & $\mbf{912}$ & N/R & $\mbf{1539}$ \\
   \hline
  \end{tabular}
  \vskip-0.5em
  \captionof{table}{\small{Tensor hierarchy representations relevant here. The representations $R_i$ correspond to the fibres of the exceptional vector bundles $\bar{{\cal R}}_i$. The bundles which are marked N/R are not relevant for the purposes of this paper.}}\label{t:TensorHierarchy}
  \end{center}
 \end{minipage}}
\vspace{1em}

It is in fact more natural to consider weighted bundles, which we denote as ${\cal R}_i = \bar{{\cal R}}_i \otimes {\cal S}^{i}$, where ${\cal S}^{i}$ is the rank zero exceptional vector bundle isomorphic to a power of the determinant bundle $\det \left(T^*M\right)^{i/(D-2)}$. The sections of ${\cal S}^{i}$ are thus scalar densities of weight $\frac{i}{D-2}$ under the generalised Lie derivative.

We begin by setting up our conventions, by introducing a natural graded product between these tensors, which for obvious reasons we refer to as the wedge product, as well as some nilpotent derivatives \cite{Cederwall:2013naa,Hohm:2015xna,Wang:2015hca}. The wedge product maps, for $1 \leq i \leq D-4$ and $1 \leq j \leq D-3-i$,
\begin{equation}
 \wedge: R_i \otimes R_j \longrightarrow R_{i+j} \,, \label{eq:Wedge}
\end{equation}
and explicitly for $A \in R_i$ and $B \in R_j$,
\begin{equation}
 A \wedge B = \left( A \otimes B \right)\vert_{R_{i+j}} \,. \label{eq:WedgeVector}
\end{equation}
It can similarly be defined for the exceptional vector bundles, irrespective of their weight,
\begin{equation}
 \begin{split}
  \wedge: {\cal R}_i \otimes {\cal R}_j \longrightarrow {\cal R}_{i+j} \,, \\
  \wedge: \bar{{\cal R}}_i \otimes \bar{\cal R}_j \longrightarrow \bar{\cal R}_{i+j} \,.
 \end{split}
\end{equation}
In terms of the underlying geometry on $M_d$, sections of the ${\cal R}_i$ bundles consist of the formal sum of vector fields and differential forms. The wedge product between the ${\cal R}_i$ bundles is similar to the Clifford action of $\ODD$ generalised vectors on $\ODD$ spinors \cite{Gualtieri:2003dx,Coimbra:2011nw,Hohm:2011dv}, i.e. it consists of the contraction of the vectors with forms and wedge products of forms.

Let us also label the adjoint representation of $E_{d(d)}$ by $P$ and the corresponding exceptional adjoint bundle by ${\cal P}$. Note that for $0 < i \leq D-3 $, $R_{D-2-i} = R^*_{i}$, and
\begin{equation}
 {\cal R}_{D-2-i} = {\cal R}^*_{i} \otimes {\cal S}^{D-2} \,.
\end{equation}
Thus we can define the wedge products, for $i \leq D-3$,
\begin{equation}
 \begin{split}
  \wedge: R_i \otimes R_{D-2-i} &\longrightarrow \mbf{1} \,, \\
  \wedge_P: R_i \otimes R_{D-2-i} &\longrightarrow P \,, \label{eq:WedgeSP}
 \end{split}
\end{equation}
by projecting the tensor products onto the singlet representation, $\mbf{1}$, and $P$, respectively. Similarly we define for the exceptional vector bundles
\begin{equation}
 \begin{split}
  \wedge: {\cal R}_i \otimes {\cal R}_{D-2-i} &\longrightarrow {\cal S}^{D-2} \,, \\
  \wedge_P: {\cal R}_i \otimes {\cal R}_{D-2-i} &\longrightarrow {\cal P} \otimes {\cal S}^{D-2} \,. \label{eq:WedgeSPVector}
 \end{split}
\end{equation}

For $D \geq 5$, we will also make use of a generalisation of the wedge product which maps
\begin{equation}
 \begin{split}
  \bullet: R_{i} \otimes R_{j} &\longrightarrow R_{i+j+2-D} \,,
 \end{split}
\end{equation}
for $i + j > D-2$, and where we let $R_0 = \mathbf{1}$. Note that this does not fit into the usual tensor hierarchy discussion \cite{deWit:2008ta,deWit:2008gc,Kleinschmidt:2011vu,Palmkvist:2011vz} and this is why we do not denote it by $\wedge$. It is also convenient to define
\begin{equation}
 \bullet_P: R_i \otimes R_j \longrightarrow R^P_{i+j+2-D} \,,
\end{equation}
where
\begin{equation}
 \begin{split}
  R^P_0 &= P \,, \\
  R^P_i &= R_i \,, \,\, \textrm{for i } \neq 0 \,.
 \end{split}
\end{equation}
Thus $\bullet$ and $\bullet_P$ only differ when acting on $R_i \otimes R_{D-2-i}$. Given a scalar density, $\bullet$ and $\bullet_P$ can similarly be defined on the bundles ${\cal R}_i$.

Let us now define the nilpotent differential operator. For $V \in \Gamma\left({\cal R}_i\right)$, with $2 \leq i \leq D-3$, this is
\begin{equation}
 d: \Gamma\left({\cal R}_i\right) \longrightarrow \Gamma\left({{\cal R}_{i-1}}\right) \,, \qquad dV = \left( \partial \otimes V \right) \vert_{{\cal R}_{i-1}} \,,
\end{equation}
where $\partial$ denotes the internal derivatives, $\partial_M$. An explicit computation shows that for $D \geq 6$, 
\begin{equation}
 d^2 = 0 \,,
\end{equation}
i.e. the differential operator is nilpotent. Note that the derivative only maps generalised tensor to generalised tensors when defined on the weighted bundles ${\cal R}_i$. We will show how to define this derivative operator acting on certain sections of the weighted adjoint bundle in section \ref{s:D=4}.

We will make use of two further identities of \cite{Hohm:2015xna,Wang:2015hca}. The first is that the generalised Lie derivative acting on a section $B \in \Gamma\left({\cal R}_2\right)$ can be expressed using the wedge product and differential as
\begin{equation}
 \begin{split}
  \gL_{A} B &= A \wedge d B + d \left( A \wedge B \right) \,,
 \end{split}
\end{equation}
for any $A \in \Gamma\left({\cal R}_1\right)$. The second identity is that for $A_1, A_2 \in \Gamma\left({\cal R}_1\right)$,
\begin{equation}
 \gL_{A_1} A_2 + \gL_{A_2} A_1 = d \left( A_1 \wedge A_2 \right) \,.
\end{equation}
We will give explicit expressions for the wedge product and nilpotent derivative when discussing the specific cases in appendices \ref{s:D=7} -- \ref{s:D=5}.

\subsection{$\GH$ structures in $D \geq 5$ dimensions} \label{s:GHalfStr}
We can now define a $\GH$ structure in $D \geq 5$ in terms of the differential forms introduced above. We will here consider only the non-chiral six-dimensional half-maximal supergravities and discuss the chiral $D=6$ supergravity in section \ref{s:D=6b}, as well as the case $D=4$ in section \ref{s:D=4} since these cases follow a slightly different pattern to the remainder.

Notice first of all that $\GH \subset \SO{d-1,d-1} \subset E_{d(d)} \times \mathbb{R}^+$ and so we begin by first describing $\SO{d-1,d-1} \subset E_{d(d)} \times \mathbb{R}^+$ structures before further reducing the structure group to $\SO{d-1}$. A generalised $\SO{d-1,d-1} \subset E_{d(d)} \times \mathbb{R}^+$ structure is equivalent to having the following nowhere-vanishing fields:
\begin{itemize}
 \item a scalar density $\kappa$ of weight $\frac{1}{D-2}$, i.e. $\kappa \in \Gamma\left({\cal S}\right)$,
 \item a section $\K$ of the ${\cal R}_2$-bundle,
 \item and a section $\hK$ of the ${\cal R}_{D-4}$-bundle.
\end{itemize}
We will refer to these sections collectively as ${\cal \K} = \left( \K\,, \,\, \hK\,, \,\, \kappa \right)$. These sections must further satisfy the point-wise conditions
\begin{equation}
 \left(K \otimes K\right) \vert_{{\cal R}_c \otimes {\cal S}^{4}} = 0 \,, \qquad \left(\hK \otimes \hK\right)\vert_{{\cal R}^*_c \otimes {\cal S}^{2D-8}} = 0 \,, \label{eq:KPurity}
\end{equation}
and are subject to the compatibility condition
\begin{equation}
 \K \wedge \hK = \kappa^{D-2} \,. \label{eq:KCompatibility}
\end{equation}
We will call an $\SO{d-1,d-1} \subset E_{d(d)} \times \mathbb{R}^+$ structure a dilaton structure, because it corresponds to the dilaton scalar field in the half-maximal gauged SUGRA obtained after truncation. Equation \eqref{eq:KPurity} is analogous to the condition of having a decomposable differential form in ordinary geometry, and of having pure spinors in generalised geometry \cite{Gualtieri:2003dx}, and has also appeared in the discussion of the section condition of EFT \cite{Berman:2012vc}. As we will show in appendices \ref{s:D=7} to \ref{s:D=5} explicitly, these conditions imply that $\K$, $\hK$ and $\kappa$ are stabilised by a $\SO{d-1,d-1} \subset E_{d(d)} \times \mathbb{R}^+$ subgroup.

To further reduce the structure group to $\SO{d-1} \subset E_{d(d)} \times \mathbb{R}^+$ we introduce $d-1$ nowhere-vanishing generalised vector fields $\J_u \in \Gamma\left({\cal R}_1\right)$, where $u = 1, \ldots, d-1$ labels the $\SO{d-1}_R$ R-symmetry. These sections must satisfy the compatibility conditions
\begin{equation}
 \begin{split}
  \J_u \wedge \K &= 0 \,, \\
  \J_u \wedge \J_v &= \delta_{uv} \K \,. \label{eq:JCompatibility}
 \end{split}
\end{equation}
It is worthwhile noting that any $\SO{d-1}$ structures related by \emph{global} rescalings
\begin{equation}
 \begin{split}
  \kappa &\longrightarrow \kappa\, \lambda \,, \\
  \hK &\longrightarrow \hK\, \lambda^{D-2} \,, \\
  K &\longrightarrow K\, \sigma^2 \,, \\
  \J_u &\longrightarrow \J_u\, \sigma \,, \label{eq:GlobalRescaling}
 \end{split}
\end{equation}
are equivalent. In particular, the rescalings of $\kappa$ and $\hK$ correspond to a conformal transformation of the external $D$-dimensional metric
\begin{equation}
 g_{\mu\nu} \longrightarrow g_{\mu\nu} \, \lambda^2 \,.
\end{equation}

The generalised tensors $\J_u$, $\K$ and $\hK$ can be thought of as a flux- and higher-dimensional generalisation of almost hyper-complex structures on four-manifolds. For example, as we will show in \ref{s:K3Example}, in the case of 11-dimensional flux-less and intrinsic torsion-free (in the sense that we will define in section \ref{s:IntTorsion}) compactifications to $D=7$ they correspond to the complex and K\"ahler structure of K3.

The first equation in \eqref{eq:JCompatibility} implies that $\J_u$ transforms in the vector representation, $V_{d-1,d-1}$, of $\SO{d-1,d-1} \subset E_{d(d)} \times \mathbb{R}^+$. Further breaking $\SO{d-1,d-1} \longrightarrow \SO{d-1}$ the vector representation decomposes as $V_{d-1,d-1} \longrightarrow V_{S} \oplus V_{R}$, where $V_S$ denotes the vector representation of the $\SO{d-1}$ structure group, which we also denote $\SO{d-1}_S$ and $V_R$ denotes the vector representation of the $\SO{d-1}_R$ R-symmetry group. The $V_S$ and $V_R$ representations would appear in the second equation of \eqref{eq:JCompatibility} with opposing signs and thus this equation implies that $\J_u$ transforms as a vector of $\SO{d-1}_R$. Since there are $d-1$ nowhere vanishing $\J_u$'s, these in fact form a basis for $V_R$ at each point. This shows that the $\J_u$ are stabilised by $\SO{d-1}_S$ and thus, together with nowhere-vanishing $\K$, $\hK$ and $\kappa$ are equivalent to a reduced structure group $\SO{d-1}_S \subset E_{d(d)} \times \mathbb{R}^+$.

Note that given an $\SO{d-1,d-1}$ structure, one can always find $\J_u$ which are only well-defined up to $\SO{d-1}_R$ rotations. This corresponds to a reduced structure group $\SO{d-1} \times \SO{d-1} \subset \SO{d-1,d-1} \subset E_{d(d)} \times \mathbb{R}^+$. This is always possible as one can always reduce a structure group to its maximal compact subgroup. The $\SO{d-1}_R$ invariant set of $\J_u$'s corresponds to a $\SO{d-1,d-1}$ generalised metric. We will see this explicitly when defining consistent truncations \ref{s:ConsTruncation}. Here we instead want a $\SO{d-1}$ structure and thus the $\J_u$ here must be individually well-defined, not just up to $\SO{d-1}_R$ transformations.

Given the reduced structure group $\SO{d-1}_S$ we can introduce further invariant tensors which will be useful in what follows. Firstly, we can define $d-1$ sections of the ${\cal R}_{D-3}$ bundle
\begin{equation}
 \hJ_u = \J_u \wedge \K \,,
\end{equation}
which satisfy
\begin{equation}
 \hJ_u \wedge J_v = \delta_{uv} \kappa^{D-2} \,.
\end{equation}
Secondly, we can define the generators of the $\SO{d-1}_R$ symmetry as
\begin{equation}
 \LG_{uv} = \J_{[u} \wedge_P \hJ_{v]} \,. \label{eq:RGenerator}
\end{equation}
One can show that these generate the $\SO{d-1}_R$ algebra
\begin{equation}
 \left[ \LG_{uv}, \LG_{wx} \right] = 2 \kappa^{D-2} \left( \delta_{w[u} \LG_{v]x} - \delta_{x[u} \LG_{v]w} \right) \,,
\end{equation}
and act on $\J_u$ as
\begin{equation}
 \LG_{uv} \cdot \J_w = 2 \kappa^{D-2} \delta_{w[u} \J_{v]} \,,
\end{equation}
where $\cdot$ denotes the adjoint action of $E_{d(d)}$. In the case of 11-dimensional compactifications to $D=7$ without flux, the $\L_{uv}$ become the almost hypercomplex structures on the internal four-manifold.

It is easy to show that $\GH$ structures can also be constructed as spinor bilinears of the half-maximal set of spinors. This was shown for $D = 7$ in \cite{Malek:2016bpu}. The compatibility requirements \eqref{eq:KPurity}, \eqref{eq:KCompatibility} and \eqref{eq:JCompatibility} then follow from Fierz identities of the spinors. This shows that $\GH$ structures are equivalent to the manifold admitting a half-maximal set of spinors.

\subsection{Intrinsic torsion of exceptional $\GH$ structures} \label{s:IntTorsion}
We now consider the intrinsic torsion of the exceptional $\GH$ structures \cite{Coimbra:2014uxa} which, as we will see in section \ref{s:ConsTruncation}, is related to the embedding tensor of the corresponding half-maximal gauged SUGRAs. It is the flux-generalisation of the torsion classes known for $\SU{3}$ and $G2$ backgrounds \cite{Joyce}. Roughly speaking, the generalised intrinsic torsion measures the breaking of supersymmetry and violation of the equations of motion by the internal background. Mathematically, it is the obstruction to introducing a torsion-free $\GH$ compatible connection. As we will show in section \ref{s:ReformulatePot}, the scalar potential of the action can be rewritten in terms of the intrinsic torsion.

The intrinsic torsion is defined as follows. Consider a $\GH$ connection, i.e. a connection, $\nabla$, compatible with $\GH$ in the sense that
\begin{equation}
 \nabla \J_u = \nabla \K = \nabla \hK = \nabla \kappa = 0 \,.
\end{equation}
In general such a connection will have torsion, defined as the tensorial part of the connection. In fact, the torsion \cite{Coimbra:2011nw,Coimbra:2011ky,Hohm:2011si,Cederwall:2013naa} is given by
\begin{equation}
 \left( \gL^\nabla_{\Lambda} - \gL_{\Lambda} \right) V^M = \tau^M{}_{NP} \Lambda^N V^P \,, \label{eq:Torsion}
\end{equation}
for any generalised vector fields $\Lambda^M$ and $V^M$. Here $\gL^\nabla$ means the generalised Lie derivative with all derivatives replaced by the covariant derivatives $\nabla$ and $\tau$ is the torsion. As shown in \cite{Coimbra:2011ky,Cederwall:2013naa,Hohm:2013uia,Hohm:2013vpa}, $\tau$ only takes values in certain representations $\tau \subset W \subset R_1^* \otimes P$. Note that here we will abuse notation and not differentiate between vector spaces and the corresponding vector bundles, since this distinction is irrelevant for the following discussion.

Now consider two different $\GH$ structures. Their difference is a tensor valued in $K_{\GH} \equiv R_1^* \otimes \mathrm{adj}\left(\GH\right)$. One can define the torsion of this tensor as in \eqref{eq:Torsion}, and the associated map
\begin{equation}
 \tau: K_{\GH} \longrightarrow W \,.
\end{equation}
Clearly, $Im \tau \subset W$. Now the intrinsic torsion, $W_{int}$ is simply defined as the subset of $W$ that is independent of the choice of $\GH$ structure, i.e.
\begin{equation}
 W_{int} = W / Im \tau \,. \label{eq:WInt}
\end{equation}

The remainder of this section deals with finding explicit expression of the intrinsic torsion. To do this, we note from \eqref{eq:Torsion} that the intrinsic torsion is a generalised tensor involving one derivative, that by definition \eqref{eq:WInt} is independent of the $\GH$ connection. As a result, one should be able to define it without ever introducing a $\GH$ connection in the first place.

Before we move on, let us finish with two comments. Firstly, one can define an intrinsic torsion of any subgroup $G \subset E_{d(d)}$ following the above recipe. In particular, below we will begin by finding explicit expressions of the intrinsic torsion of the dilaton structure first. Recall that by this we mean the intrinsic torsion associated to the subgroup $\SO{d-1,d-1} \subset E_{d(d)} \times \mathbb{R}^+$.

Secondly, while there is an intrinsic torsion associated to any $G \subset E_{d(d)}$, in general one cannot rewrite the scalar potential of the action purely in terms of this intrinsic torsion. Typically, one requires that the $G$ structure is related to supersymmetry, i.e. implying a subset of well-defined spinors exists, for this to work. We will show in section \ref{s:ReformulatePot} that one can indeed rewrite the scalar potential purely in terms of the intrinsic torsion of $\GH$. This is crucial for the proof of consistency of the truncation in section \ref{s:ConsTruncation}.

\subsubsection{Intrinsic torsion of dilaton structure}
We begin by considering the intrinsic torsion of the dilaton structure. This has a universal piece, which is the same for all $D \geq 5$ and makes use of the exterior derivative $d: \Gamma\left({\cal R}_2\right) \longrightarrow \Gamma\left({\cal R}_1\right)$ and $\K$,
\begin{equation}
 d\K = \tc_{\K} \,. \label{eq:DilTorsion0}
\end{equation}
To see that $d\K$ gives components of the intrinsic torsion, consider replacing the partial derivative in $d$ by a covariant derivative $\nabla$. We label this new differential operator by $d^{\nabla}$. Because both $d\K$ and $d^{\nabla} \K$ are tensors, their difference is again a tensor and thus corresponds to a component of the torsion of $\nabla$, call this $- \tc_{\K}$. Thus,s
\begin{equation}
 d^\nabla\K = d\K - \tc_{\K} \,.
\end{equation}
For a $\SO{d-1,d-1}$ connection, $\nabla \K = 0$ and thus for such a connection,
\begin{equation}
 d\K = \tc_{\K} \,. \label{eq:dKTorsion}
\end{equation}
The left-hand side of \eqref{eq:dKTorsion} is clearly independent of the choice of $\SO{d-1,d-1}$ connection, and therefore $\tc_{\K}$ is an element of the intrinsic torsion of the dilaton structure.

We can now decompose $\tc_{\K}$ into its irreducible representations under $\SO{d-1,d-1}$. Decomposing $E_{d(d)} \longrightarrow \SO{d-1,d-1}$, one finds in $D=6,7$,
\begin{equation}
 R_1 \longrightarrow V_{d-1,d-1} \oplus \phi_{d-1,d-1} \,, \label{eq:R1Decomp}
\end{equation}
where $V_{d-1,d-1}$ is the vector and $\phi_{d-1,d-1}$ the spinor representation of $\SO{d-1,d-1}$ and we have ignored the weight factor of $R_1$. \footnote{In $D=6$, there are two different relevant spinor representations of $\SO{4,4}$ which we label as $\phi_{4,4}$ and $\tilde{\phi}_{4,4}$. It is the former that appears in \eqref{eq:R1Decomp}} In $D=5$ one instead has
\begin{equation}
 R_1 \longrightarrow V_{d-1,d-1} \oplus \phi_{d-1,d-1} \oplus \mbf{1} \,, \label{eq:R1DecompD=5}
\end{equation}
but as we will show in appendix \ref{s:D=5}, the singlet necessarily vanishes in \eqref{eq:DilTorsion0}. Thus, we can write
\begin{equation}
 \left(d \K\right)^M = \kappa^2 \tilde{T}^M + \kappa^{6-D} \left(\hK \bullet T_3\right)^M \,, \label{eq:DilTorsion1}
\end{equation}
where, $M$ is an index for the $R_1$ representation and $T_3 \in \Gamma\left(\bar{\cal R}_3 \otimes {\cal S}^{-1}\right)$. $\tilde{T}$ and $T_3$ are the two irreps of the intrinsic torsion and for later convenience we have chosen the factors of $\kappa$ so that the intrinsic torsion has weight minus one under the generalised Lie derivative.

Following \eqref{eq:R1Decomp} and \eqref{eq:R1DecompD=5}, $\tilde{T} \in V_{d-1,d-1}$ and $T_3 \in \phi_{d-1,d-1}$, and thus they satisfy
\begin{equation}
 \begin{split}
  \tilde{T} \wedge \K &= 0 \,, \\
  T_3 \bullet \K &= 0 \,, \label{eq:DilTorsionConstraints}
 \end{split}
\end{equation}
and for $D = 5$ also
\begin{equation}
 \left( T_3 \otimes \K \right)\vert_{\obf{351}} = 0 \,. \label{eq:5DDilTorsionConstraints}
\end{equation}
We will show this explicitly in appendices \ref{s:D=7} -- \ref{s:D=5}. Equation \eqref{eq:DilTorsion1} can be inverted to give expressions for $\tilde{T}$ and $T_3$ as
\begin{equation}
 \begin{split}
  T_3 &= -2 \kappa^{-4} d\K \wedge \K \,, \\
  \tilde{T} &= \kappa^{-2} d\K - 2 \kappa^{-D} \hK \bullet \left(d\K\wedge\K\right) \,.
 \end{split}
\end{equation}

There are further elements of the intrinsic torsion which we will discuss in detail in appendices \ref{s:D=7} to \ref{s:D=5}. In $D=6$, $7$ they are given by
\begin{equation}
 d\hK = \tc_{\hK} \,,
\end{equation}
which can be decomposed as
\begin{alignat}{2}
  d\hK &= \kappa \K P_1 + \kappa^3 P_2 \,, \qquad &&\textrm{for } D = 7 \,, \nonumber \\
  d\hK &= \kappa^2 \tilde{P} + \kappa^{6-D} \hK \bullet P_3 \,, \qquad &&\textrm{for } D = 6 \,, \label{eq:D67DilTorsionClasses}
\end{alignat}
where for $D = 7$, $P_1 \in \Gamma\left({\cal S}^{-1} \right)$ and $P_2 \in \Gamma\left(\bar{\cal R}_1 \otimes {\cal S}^{-1} \right)$ and satisfy
\begin{equation}
 \begin{split}
  P_2 \wedge \K &= 0 \,. \label{eq:7DDilTorsionConstraints}
 \end{split}
\end{equation}
For $D = 6$, $\tilde{P} \in \Gamma\left(\bar{\cal R}_1 \otimes {\cal S}^{-1} \right)$ and $P_3 \in \Gamma\left(\bar{\cal R}_3 \otimes {\cal S}^{-1} \right)$, and satisfy
\begin{equation}
 \begin{split}
  \tilde{P} \wedge \K &= 0 \,, \\
  P_3 \bullet \K &= 0 \,. \label{eq:6DDilTorsionConstraints}
 \end{split}
\end{equation}

In $D=5$, $\hK$ is a generalised vector and can be used as a generator of a generalised diffeomorphism. As we explain in appendix \ref{s:D=5}, in $D=5$ the remaining intrinsic torsion is given by
\begin{equation}
 \gL_{\hK} \K = \tc_{\hK} \,, \qquad \gL_{\hK} \kappa = \tc'_{\hK} \,,
\end{equation}
which can be decomposed as
\begin{equation}
 \begin{split}
  \gL_{\hK} \K &= \kappa \K P_1 + \kappa P_2 \bullet \K \,, \qquad \gL_{\hK} \kappa^{D-2} = \kappa^{D-1} P_1 \,,
  \label{eq:D5DilTorsionClasses}
 \end{split}
\end{equation}
with $P_1 \in \Gamma\left({\cal S}^{-1} \right)$ and $P_2 \in \Gamma\left({\cal P} \otimes {\cal S}^{-1} \right)$ and satisfying
\begin{equation}
 P_2 \bullet \hK = \left( P_2 \otimes \hK \right)\Big\rvert_{\mbf{351}} = 0 \,. \label{eq:5DDilTorsionConstraints2}
\end{equation}

Finally, we can define an integrable dilaton structure as one where all of its intrinsic torsion vanishes. Thus in $D=6, 7$ an integrable dilaton structure satisfies
\begin{equation}
 d\K = d\hK = 0 \,,
\end{equation}
while in $D=5$ it satisfies
\begin{equation}
 d\K = \gL_{\hK} \K = \gL_{\hK} \kappa = 0 \,.
\end{equation}
Similar to the ${\cal N}=2$ case \cite{Ashmore:2015joa}, and as we will discuss elsewhere, this is related to certain moment maps vanishing.

\subsubsection{Intrinsic torsion of the $\SO{d-1}$ structure}
We now consider the intrinsic torsion of the $\SO{d-1}$ structure. Thus we look for covariant derivatives of the $\SO{d-1}$ structure, $\left( \J_u\,,\,\, \K\,,\,\, \hK\,,\,\, \kappa \right)$ which do not involve a connection. The independent components of the intrinsic torsion are given by
\begin{equation}
 \begin{split}
  \gL_{\J_{[u}} \J_{v]} &= \tc_{\J\,uv} \,, \\
  \gL_{\J_u} \hK&= \tc_{C\,u} \,, \\
  d\K &= \tc_{\K} \,, \label{eq:SOdTorsion}
 \end{split}
\end{equation}
and
\begin{alignat}{2}
  d\hK &= \tc_{\hK} \,, \quad&& \textrm{for } D = 6\,,\, 7\,, \\
  \gL_{\hK} K &= \tc_{\hK} \,, \quad \gL_{\hK} \J_u = \tc_{\hK\,u} \quad &&\textrm{for } D = 5 \,. \label{eq:SOdTorsion2}
\end{alignat}
Here, the new intrinsic torsion arising from further reducing the structure group to $\SO{d-1} \subset SO{d-1,d-1}$ is given by $\tc_{\J\,uv}$ and $\tc_{C\,u}$ in equation \eqref{eq:SOdTorsion}.

We will now argue that one cannot build other generalised tensors involving one derivative of the $\SO{d-1}$ structure. Firstly, combinations such as $d\hJ_u$ are not independent because when $D=6,7$ one can write
\begin{equation}
 \gL_{\J_u} \hK = \J_u \wedge d \hK + d \left( \J_u \wedge \hK \right) \,,
\end{equation}
and using the definition $\hJ_u = \J_u \wedge \hK$ one finds
\begin{equation}
 d \hJ_u = \gL_{\J_u} \hK - \J_u \wedge d \hK = \tc_{C\,u} - \J_u \wedge \tc_{\hK} \,.
\end{equation}
On the other hand, when $D=5$
\begin{equation}
 \gL_{\J_u} \hK + \gL_{\hK} \J_u = d \left( \J_u \wedge \hK \right) = d \hJ_u \,,
\end{equation}
and hence $d\hJ_u$ is again determined by the other torsion classes listed above.

Other combinations one could have considered are $\gL_{\J_{(u}} \J_{v)}$, $\gL_{\J_u} \K$ and $\gL_{\J_u} \hJ_v$. But
\begin{equation}
 \begin{split}
  \gL_{\J_{(u}} \J_{v)} &= \frac12 d \left( \J_u \wedge \J_v \right) \\
  &= \frac12 \delta_{uv} d \K \\
  &= \frac1{d-1} \delta_{uv} \gL_{\J^w} \J_w \,,
 \end{split}
\end{equation}
is again not independent from the torsion classes listed above. Similarly,
\begin{equation}
 \begin{split}
  \gL_{\J_u} \K &= \J_u \wedge d\K + d \left( \J_u \wedge \K \right) \\
  &= \J_u \wedge d\K \\
  &= \J_u \wedge \tc_\K \,,
 \end{split}
\end{equation}
and
\begin{equation}
 \begin{split}
  \gL_{\J_u} \hJ_v &= \gL_{\J_u} \left( \J_v \wedge \hK \right) \\
  &= \gL_{\J_u} \J_v \wedge \hK + \J_v \wedge \gL_{\J_u} \hK \\
  &= \frac{1}{2}\delta_{uv} \tc_{\K} \wedge \hK + \tc_{\J\,uv} \wedge \hK + \J_v \wedge \tc_{C\,u} \,,
 \end{split}
\end{equation}
are also determined by the torsion classes in \eqref{eq:SOdTorsion} and \eqref{eq:SOdTorsion2}.

Let us now decompose the intrinsic torsion, $\tc$ into its irreducible representations under $\SO{d-1}_S \times \SO{d-1}_R$ at each point. We find
\begin{equation}
 \begin{split}
  d\K &= \kappa^2 T_1 + \kappa\, \J_u\, T_2{}^u + \kappa^{6-D} \hK \bullet T_3 \,, \\
  \gL_{\J_{[u}} \J_{v]} &= \kappa^2 R_{1\,uv} + \kappa\, R_{2\,uvw} \J^w + \kappa\, T_{2[u} \J_{v]} - \kappa^{8-2D} \hK \bullet \left( \LG_{uv} \cdot T_3 \right) \,, \\
  \gL_{\J_u} \hK &= \kappa^{D-3} S_{1\,u} + \kappa^{D-4} \J_u \wedge S_2 + \kappa \left( U_u - T_{2\,u} \right) \hK \,, \\
  \gL_{\J_u} \kappa^{D-2} &= \kappa^{D-1} U_u \,, \label{eq:JTorsionClasses}
  \end{split}
\end{equation}
where $\LG_{uv}$ are the $\SO{d-1}_R$-symmetry generators defined in \eqref{eq:RGenerator}. Note that when $D=5$, $S_2 \in {\cal P}$ and we let the wedge product $\J_u \wedge S_2$ be the adjoint action. The dimension-dependent intrinsic torsion in $D = 6, 7$ is
\begin{alignat}{2}
  d\hK &= \kappa \K P_1 + \kappa^3 P_2 \,, \qquad &&\textrm{for } D = 7 \,, \nonumber \\
  d\hK &= \kappa^2 P_1 + \kappa \J_u P_2{}^u + \hK \bullet P_3 \,, \qquad &&\textrm{for } D = 6 \,, \label{eq:D67TorsionClasses}
 \end{alignat}
In $D = 5$, $\hK \in \Gamma\left({\cal R}_1\right)$, and thus we can use it as a generator of generalised diffeomorphisms. Thus, the dimension-dependent intrinsic torsion is given by
\begin{equation}
 \begin{split}
  \gL_{\hK} \K &= \kappa \K P_1 + \kappa P_2 \bullet \K \,, \qquad \gL_{\hK} \kappa^3 = \kappa^4 P_1 \,, \\
  \gL_{\hK} \J_u &= \kappa^2 P_{3\,u} + \kappa P_{4\,uv} \J^{v} + \frac12 \kappa P_1 \J_u - \kappa^{-3} \left( \left( P_2 \bullet \K \right) \wedge_P \J_u \right) \cdot \hK \,,
  \label{eq:D5TorsionClasses}
 \end{split}
\end{equation}
The right-hand side of \eqref{eq:JTorsionClasses}, \eqref{eq:D67TorsionClasses} and \eqref{eq:D5TorsionClasses} follow from the compatibility conditions of the $\GH$ structure, equations \eqref{eq:KPurity}, \eqref{eq:KCompatibility} and \eqref{eq:JCompatibility}, as we will now explain.

The first equation of \eqref{eq:JTorsionClasses} is easily understood. Here we have just further decomposed $\tilde{T}$ and $T_3$ under $\SO{d-1,d-1}$. Recall that $\tilde{T} \in V_{d-1,d-1}$ which contains the irreducibles $V_{d-1,d-1} \longrightarrow V_S \oplus V_R$ under $\SO{d-1}_S \times \SO{d-1}_R$. These correspond to $T_1$ and $T_2{}^u$, respectively.

We will discuss the decomposition of $\gL_{\J_u} \hK = \tc_{C}$ in appendices \ref{s:D=7} to \ref{s:D=5}, as the details differ with dimension, $D$. Nonetheless, and somewhat miraculously, the answer is the same for all those dimensions and is as given in \eqref{eq:JTorsionClasses}. The only feature which is easily explained is that the irreducible in the $V_R$ representation is always given by $\left( U_u - T_u \right)$. This follows from
\begin{equation}
 \begin{split}
  \K \wedge \gL_{\J_u} \hK &= \gL_{\J_u} \kappa^{D-2} - \hK \wedge \gL_{\J_u} \K \\
  &= \gL_{\J_u} \kappa^{D-2} - \hK \wedge \left( \J_u \wedge d\K \right) \\
  &= \kappa^{D-1} \left( U_u - T_{2\,u} \right) \,.
 \end{split}
\end{equation}

We will prove the decomposition of $\gL_{\J_{[u}} \J_{v]} = \tc_{\J\,uv}$ given in \eqref{eq:JTorsionClasses} in appendix \ref{A:TorsionClassesDecomp}, since it is somewhat more lengthy. Finally, the decomposition given in \eqref{eq:D67TorsionClasses} and \eqref{eq:D5TorsionClasses} clearly depends on the dimension and we will give it in appendices \ref{s:D=7} -- \ref{s:D=5}.

In section \ref{s:ConsTruncation} we will show how this decomposition of the intrinsic torsion is related to the linear constraint of half-maximal gauged SUGRA. There we will also see that the torsion classes in equations \eqref{eq:D67TorsionClasses} and \eqref{eq:D5TorsionClasses} are related to the dimension-specific components of the embedding tensor of half-maximal gauged SUGRA, see e.g. \cite{Bergshoeff:2007vb}.

The torsion classes in \eqref{eq:JTorsionClasses} satisfy
\begin{equation}
 \begin{split}
  T_1 \wedge \J_u &= T_1 \wedge \hJ_{u} = T_1 \wedge \K = 0 \,, \\
  T_3 \wedge \hJ_u &= T_3 \bullet \K = 0 \,, \\
  R_{1\,uv} \wedge \J_w &= R_{1\,uv} \wedge \hJ_w = R_{1\,uv} \wedge \K= 0 \,, \\
  S_{1\,u} \wedge \K &= S_{1\,u} \bullet_P \hJ^u = 0 \,, \\
  S_2 \wedge \hK &= 0 \,. \label{eq:TorsionClassRepConstraint}
 \end{split}
\end{equation}
In $D=6$, we additionally have $S_{1\,u} \wedge \hK = 0$. The seven-dimensional torsion classes in \eqref{eq:D67TorsionClasses} satisfy equations \eqref{eq:7DDilTorsionConstraints}, while the six-dimensional ones satisfy
\begin{equation}
 \begin{split}
  P_1 \wedge \J_u &= P_1 \wedge \hJ^u = P_1 \wedge \K = 0 \,, \\
  P_{2\,u} \wedge \K &= 0 \,, \\
  P_3 \bullet \K &= 0 \,.
 \end{split}
\end{equation}
Those in \eqref{eq:D5TorsionClasses} satisfy
\begin{equation}
 \begin{split}
  P_2 \bullet \hK &= \left(P_2 \otimes \hK \right)\Big\vert_{\mbf{351}} = 0 \,, \\
  P_{3\,u} \wedge \J_v &= P_{3\,u} \wedge \hJ_v = P_{3\,u} \wedge \K = 0 \,.
 \end{split}
\end{equation}
These equation imply that the torsion classes transform in certain irreducible representations, as we make explicit in appendices \ref{s:D=7} to \ref{s:D=5}.

\subsection{Half-maximal Minkowski and AdS vacua} \label{s:Vacua}
We can also use our set-up to formulate the conditions for half-maximal warped Minkowski and AdS vacua. In this case, Lorentz / AdS symmetry requires the gauge fields of the EFT tensor hierarchy to vanish. Furthermore, the $\GH$ structure must be independent of the external space.

One can derive the conditions that the internal space must satisfy from the SUSY variations of the SUGRA, as was shown for Minkowski vacua in \cite{Coimbra:2014uxa,Coimbra:2016ydd}. There it was shown that these require the $\GH$ structure to have vanishing intrinsic torsion, i.e.
\begin{equation}
 \begin{split}
  \gL_{\J_u} \J_v &= \gL_{\J_u} \hK = \gL_{\J_u} \kappa = d\K = 0 \,,
 \end{split}
\end{equation}
and
\begin{alignat}{2}
  d\hK &= 0 \,, \qquad &&\textrm{for } D = 6\,,\, 7\,, \nonumber \\
  \gL_{\hK} K &= \gL_{\hK} \J_u = 0 \,, \qquad &&\textrm{for } D = 5 \,.
\end{alignat}
In this case we will say that we have an integrable $\GH$ structure. As we will show elsewhere, one can understand the above relations as the vanishing of certain moment maps. This is similar to the integrability conditions of ${\cal N}=2$ flux vacua \cite{Ashmore:2015joa}.

We can also weaken the integrability conditions to allow for AdS vacua, as also discussed in the ${\cal N}=1$ case in \cite{Coimbra:2015nha}. By comparing to the half-maximal gauged SUGRA conditions for AdS vacua \cite{Louis:2015mka,Karndumri:2016ruc,Louis:2015dca} that preserve all of the supersymmetries, we find that for AdS vacua we require
\begin{equation}
 \begin{split}
  d\K &= \gL_{\J_u} \hK = \gL_{\J_u} \kappa = 0 \,, \\
  \gL_{\J_u} \J_v &= \bR_{uvw} \J^w \,, \\
 \end{split}
\end{equation}
and
\begin{alignat}{2}
  d\hK &= -\frac14 \epsilon_{uvw} \bR^{uvw} \K \,, \qquad && \textrm{for } D = 7 \,, \nonumber \\
  d\hK &= -\frac{1}{18} \epsilon_{uvwx} \J^u \bR^{vwx} \,, \qquad &&\textrm{for } D = 6 \,, \\
  \gL_{\hK} \J_u &= - \frac1{3\sqrt{2}} \epsilon_{uvwxy} \J^v \bR^{xwy} \,, \qquad \gL_{\hK} \K = \gL_{\hK} \kappa = 0 \,, \qquad && \textrm{for } D = 5 \,. \nonumber \label{eq:WeakIntegrability}
\end{alignat}
We will call such $\GH$ structures ``weakly integrable''. Here $\bR_{uvw}$ encodes the cosmological constant in a way which breaks the R-symmetry to that of the corresponding lower-dimensional superconformal algebra, see table \ref{t:AdS}.

\vspace{1em}
\noindent\makebox[\textwidth]{
 \begin{minipage}{\textwidth}
  \begin{center}
  \begin{tabular}{|c|c|c|c|}
  \hline
  $D$ & $\SO{d-1}_R$ & Rep of $\bR_{uvw}$ & Unbroken R-symmetry \Tstrut\Bstrut \\ \hline
  7 & $\SU{2}$ & $\mbf{1}$ & $\SU{2}$ \\
  6 & $\SU{2} \times \SU{2}$ & $\left(\mbf{2},\mbf{2}\right)$ & $\SU{2}$ \\
  5 & $\USp{4}$ & $\mbf{10}$ & $\SU{2} \times \mathrm{U}(1)$  \\
   \hline
  \end{tabular}
 \vskip-0.5em
 \captionof{table}{\small{R-symmetry of half-maximal Minkowski vacua, the representation of $\bR_{uvw}$ and the unbroken R-symmetry, which is that of the AdS vacuum in $D \geq 5$ dimensions.}} \label{t:AdS}
  \end{center}
 \end{minipage}
}
\vspace{1em}

\subsection{Relation to ${\cal N}=2$ structures} \label{s:N2Embedding}
In \cite{Grana:2009im,Ashmore:2015joa,Ashmore:2016qvs,Ashmore:2016oug} general ${\cal N}=2$, i.e. quarter-maximal, flux Minkowski and AdS vacua in $D=4, 5, 6$ dimensions were studied using exceptional generalised geometry. Since every half-maximal background is also quarter-maximal, the $\GH$ structure we have described so far should contain within it an exceptional generalised ${\cal N}=2$ structure. Let us briefly indicate how this works in the cases $D = 5, 6$.

\subsubsection{$D=6$ ${\cal N}=2$ structures from the $\GH$ structure}

For $D=6$, the ${\cal N}=2$ structure is a $\USp{4} \times \SU{2} \subset \SO{5,5} \times \mathbb{R}^+$ structure which is defined bosonically by the existence of a $\SU{2}_R^{{\cal N}=2}$ triplet of adjoint tensors ${\cal J}_i$, with $i = 1, 2, 3$ in the adjoint of $\SU{2}_R^{{\cal N}=2}$, and a section of the ${\cal R}_2$ bundle, ${\cal Q}$. These two tensors must satisfy the compatibility conditions
\begin{equation}
 \begin{split}
  \left[ {\cal J}_i,\, {\cal J}_j \right] &= \kappa^2 \epsilon_{ijk} {\cal J}^k \,, \qquad {\cal J} \cdot {\cal Q} = 0 \,, \\
  \tr \left( {\cal J}_i {\cal J}_j \right) &= - \delta_{ij} \eta_{IJ} {\cal Q}^I {\cal Q}^J = - \delta_{ij} \kappa^{4} \,, \label{eq:D6N2Compat}
 \end{split}
\end{equation}
where $\cdot$ represents the adjoint action of $\SO{5,5}$, and the ${\cal J}_i$ generate a highest weight $\SU{2}$ subalgebra of $\SO{5,5}$.

These tensors are contained within our $\GH$ structure as follows. Firstly, the scalar density $\kappa$ is the same in both cases because we use the convention
\begin{equation}
 \kappa^{here} = \sqrt{\kappa^{there}} \,,
\end{equation}
where $\kappa^{there}$ refers to the scalar density in \cite{Ashmore:2015joa}. Next, recall that from the 4 $\J_u$'s one can define generators of the $\SO{4}^{{\cal N}=4}_R$ symmetry as
\begin{equation}
 \LG_{uv}{}^\alpha = \kappa^{-4} \left(t^\alpha\right)^{M}{}_N \J_{[u}{}^N \hJ_{v]M} \,.
\end{equation}
These satisfy
\begin{equation}
 \left[ \LG_{uv},\, \LG_{wx} \right] = \delta_{u[w} \LG_{x]v} - \delta_{v[w} \LG_{x]u} \,.
\end{equation}
One can identify the ${\cal J}_i$'s by breaking $\SO{4}_R^{{\cal N}=4} \longrightarrow \SU{2}_R^{{\cal N}=2}$ and picking the three generators corresponding to this $\SU{2}_R^{{\cal N}=2}$, and rescaling by $\kappa$ to obtain the correct weight. On the other hand, the ${\cal R}_2$ section ${\cal Q}$ is given by
\begin{equation}
 {\cal Q} = \frac1{\sqrt{2}} \left( \K + \hK \right) \,.
\end{equation}
This ensures that ${\cal Q}$ satisfies the compatibility conditions \eqref{eq:D6N2Compat}.

\subsubsection{$D=5$ ${\cal N}=2$ structures from the $\GH$ structure}
The case of $D=5$ is very similar to that of $D=6$, with a $\SU{2}_R^{{\cal N}=2}$ triplet of adjoint tensor ${\cal J}_i$ generating the $\SU{2}_R^{{\cal N}=2}$ algebra. In addition there is now a generalised vector field, i.e. a section of the ${\cal R}_1$ bundle, ${\cal K}$. The compatibility conditions are now
\begin{equation}
 \begin{split}
  \left[ {\cal J}_i,\, {\cal J}_j \right] &= \kappa^2 \epsilon_{ijk} {\cal J}^k \,, \qquad {\cal J} \cdot {\cal K} = 0 \,, \\
  \tr \left( {\cal J}_i {\cal J}_j \right) &= - \delta_{ij} d_{MNK} {\cal K}^M {\cal K}^N {\cal K}^K = - \delta_{ij} \kappa^{3} \,, \label{eq:D5N2Compat}
 \end{split}
\end{equation}
where $\cdot$ represents the adjoint action of $\EG{6}$ and the ${\cal J}_i$ generate a highest weight $\SU{2}$ subalgebra of $\EG{6}$. Here we have rescaled the scalar density $\kappa$ relative to that in \cite{Ashmore:2015joa} by
\begin{equation}
 \left(\kappa^{here}\right)^{3/2} = \kappa^{there} \,.
\end{equation}

The adjoint generators ${\cal J}_i$ are again given by rescaled version of a $\SU{2}_R^{{\cal N}=2}$ subalgebra of the generators $\LG_{uv} \in {\cal P}$. On the other hand, the generalised vector ${\cal K}^M$ is given by
\begin{equation}
 {\cal K}^M = \hK^M + v^u \J_u{}^M \,,
\end{equation}
where $v^u$ has to be a singlet under the $\SU{2}^{{\cal N}=2}_R \subset \SO{5}^{{\cal N}=4}_R$ subgroup and satisfy $v^u v_u = 1$. This ensures that ${\cal K}^M$ satisfies the compatibility conditions \eqref{eq:D5N2Compat}.

\subsection{Examples} \label{s:GHalfExamples}
Let us conclude this section by giving examples of $\GH$ structures. The easiest, non-trivial ones are M-theory backgrounds of the form $K3 \times T^n$. We will now show how these backgrounds are encoded in the $\GH$ structure.

\subsubsection{M-theory on $K3$} \label{s:K3Example}
We begin by considering M-theory on $K3$ with a seven-dimensional external space. This example has also been studied in \cite{Malek:2016bpu}. Recall that the generalised tensors $K, \hK, \J_u$ and $\kappa$ are combinations of certain vector fields, which we denote by $v$, and $p$-forms, which we denote by $\omega_{(p)}$. With vanishing fluxes, one finds
\begin{equation}
 \begin{split}
  \kappa^5 &= \sqrt{g} \,, \\
  \K &= \omega_{(4)} + \omega_{(1)} \,, \\
  \hK &= \omega_{(3)} + \omega_{(0)} \,, \\
  \J_u &= v_u + \omega_{(2)u} \,,
 \end{split}
\end{equation}
where $\sqrt{g}$ is the measure of the four-dimensional internal space, with $\vol_{(4)}$ its four-form.

The compatibility requirements for $K$ and $\hK$, equations \eqref{eq:KPurity}, \eqref{eq:KCompatibility}, become
\begin{equation}
 \omega_{(0)} \wedge \omega_{(4)} + \omega_{(1)} \wedge \omega_{(3)} = \vol_4 \,.
\end{equation}
For $K3$ this is solved by taking $\omega_{(4)} = vol_4$ and $\omega_{(0)} = 1$. With this choice, the compatibility requirements for $\J_u$, \eqref{eq:JCompatibility}, are
\begin{equation}
 \begin{split}
  \imath_{v_u} \omega_{(4)} &= 0 \,, \\
  \omega_{(2)u} \wedge \omega_{(2)v} &= \delta_{uv} \vol_{(4)} \,.
 \end{split}
\end{equation}
This is solved by taking the three two-forms to be the Hyperk\"ahler structure on $K3$, i.e. two of the three two-forms are given by the real and imaginary parts of the holomorphic 2-form and the third by the K\"ahler structure. Let us denote these three two-forms by $\Omega_u$, thus
\begin{equation}
 \omega_{(2)u} = \Omega_{u} \,.
\end{equation}

All in all, we have
\begin{equation}
 \begin{split}
  \kappa^4 &= \sqrt{g} \,, \\
  \K &= \vol_{(4)} \,, \\
  \hK &= 1 \,, \\
  \J_u &= \Omega_u \,.
 \end{split}
\end{equation}
With this choice, the intrinsic torsion becomes
\begin{equation}
 \begin{split}
  d\K &= d\hK = 0 \,, \\
  \gL_{\J_u} \J_v &= 0 \,, \\
  \gL_{\J_u} \hK &\propto d\Omega_u \,, \\
  \gL_{\J_u} \kappa^5 &= 0 \,.
 \end{split}
\end{equation}
Thus, we see that almost all components of the intrinsic torsion vanish automatically. The only one which is not automatically zero, $\gL_{\J_u} \hK$, vanishes because the K\"ahler and holomorphic 2-form are closed, i.e.
\begin{equation}
 \gL_{\J_u} \hK \propto d\Omega_u = 0 \,.
\end{equation}
We see that in this simple example, the geometric and exceptional $\SU{2}$ structures coincide.

\subsubsection{M-theory on $K3 \times S^1$} \label{s:K3S1Example}
We now consider M-theory on $K3 \times S^1$, or equivalently type IIA SUGRA on $K3$. The external space is thus six-dimensional and we obtain a non-chiral half-maximal SUGRA. With vanishing fluxes, the generalised tensor fields we must consider are given by
\begin{equation}
 \begin{split}
  \kappa^4 &= \sqrt{g} \,, \\
  \K &= \omega_{(1)} + \omega_{(4)} \,, \\
  \hK &= \hat{\omega}_{(1)} + \hat{\omega}_{(4)} \,, \\
  \J_u &= v_u + \omega_{(2)u} + \omega_{(5)u} \,.
 \end{split}
\end{equation}

The compatibility conditions for the dilaton structure, \eqref{eq:KPurity} and \eqref{eq:KCompatibility}, become
\begin{equation}
 \begin{split}
  \omega_{(4)} \wedge \omega_{(1)} = \hat{\omega}_{(4)} \wedge \hat{\omega_{(4)}} &= 0 \,, \\
  \omega_{(4)} \wedge \hat{\omega}_{(1)} + \hat{\omega}_{(4)} \wedge \omega_{(1)} &= \vol_{(5)} \,.
 \end{split}
\end{equation}
We solve these by taking
\begin{equation}
 \K = \omega_{(4)} \,, \qquad \hK = \hat{\omega}_{(1)} \,,
\end{equation}
i.e. we take $\omega_{(1)} = \hat{\omega}_{(4)} = 0$. In fact, this one-form and four-form are unique up to multiplication by a function: they are given by the connection one-form on $S^1$, $\sigma$ and the volume form on $K3$, $\vol_{(4)}$. Thus
\begin{equation}
 \K = \vol_{(4)} \,, \qquad \hK = \sigma \,.
\end{equation}

With this dilaton structure, the compatibility requirements for the four $\J_u$'s become
\begin{equation}
 \begin{split}
  \imath_{\tilde{\sigma}} \omega_{(2)u} &= 0 \,, \\
  \imath_{v_u} \vol_{(4)} &= 0 \,, \\
  \omega_{(2)u} \wedge \omega_{(2)v} + \frac12 \left( \imath_{v_{u}} \omega_{(5)v} + \imath_{v_{v}} \omega_{(5)u} \right) &= \delta_{uv} \vol_{(4)} \,,
 \end{split}
\end{equation}
where $\tilde{\sigma}$ denotes the vector field on the $S^1$,
\begin{equation}
 \tilde{\sigma} = \star \vol_{(4)} \,,
\end{equation}
which satisfies $\imath_{\tilde{\sigma}} \sigma = 1$. We solve these conditions by taking
\begin{equation}
 \J_U = \Omega_U \,, \qquad \J_4 = \tilde{\sigma} + \vol_{(5)} \,,
\end{equation}
where $U = 1, \ldots, 3$ and $\Omega_U$ are the hyperk\"ahler structure on $K3$.

The intrinsic torsion is now given by
\begin{equation}
 \begin{split}
  d\K &= d \vol_{(4)} \,, \\
  d\hK &= d \sigma \,, \\
  \gL_{\J_U} \J_V &= - \Omega_V \wedge d\Omega_U \,, \\
  \gL_{\J_U} \J_4 &= - \imath_{\tilde{\sigma}} d\Omega_{U} \,, \\
  \gL_{\J_U} \hK &\propto \sigma \wedge d\Omega_U \,, \\
  \gL_{\J_4} \hK &\propto L_{\tilde{\sigma}} \sigma \,, \\
  \gL_{\J_U} \kappa^4 &= 0 \,, \\
  \gL_{\J_4} \kappa^4 &= L_{\tilde{\sigma}} \vol_{(5)} \,.
 \end{split}
\end{equation}
It is easily seen that these vanish for $K3$.

\subsubsection{M-theory on $K3 \times T^2$} \label{s:K3T2}
Finally, we consider M-theory on $K3 \times T^2$, or equivalently, type II on $K3 \times S^1$. A dilaton structure is given by
\begin{equation}
 \begin{split}
  \kappa^3 &= \sqrt{g} \,, \\
  \K &= \sqrt{g} \omega_{(1)} + \omega_{(4)} + \omega_{(1)} \,, \\
  \hK &= \hat{v} + \omega_{(2)} + \omega_{(5)} \,,
 \end{split}
\end{equation}
subject to the compatibility requirements \eqref{eq:KPurity} and \eqref{eq:KCompatibility}. For $K3 \times T^2$ these are solved by
\begin{equation}
 \K = \vol_{(4)} \,, \qquad \hK = \vol_{(2)} \,,
\end{equation}
where $\vol_{(4)}$ is the volume-form of the $K3$ surface and $\vol_{(2)}$ that of the $T^2$.

With this dilaton structure, the compatibility conditions, \eqref{eq:JCompatibility}, for the five $\J_u$'s
\begin{equation}
 \J_u = v_u + \omega_{(2)u} + \omega_{(5)u} \,,
\end{equation}
become
\begin{equation}
 \begin{split}
  \vol_{(4)} \wedge \omega_{(2)u} &= 0 \,, \\
  \imath_{v_u} \vol_{(4)} &= 0 \,, \\
  \imath_{\tilde{v}_u} \vol_{(4)} &= 0 \,,
 \end{split}
\end{equation}
where $\tilde{v}_u$ is the vector field defined by $\tilde{v}_u = \star \omega_{(5)u}$, as well as
\begin{equation}
 \begin{split}
  \imath_{v_u} \omega_{(2)v} + \imath_{v_v} \omega_{(2)u} &= 0 \,, \\
  \imath_{\tilde{v}_u} \omega_{(2)v} + \imath_{v_u} \omega_{(2)v} &= 0 \,, \\
  \omega_{(2)u} \wedge \omega_{(2)v} + \imath_{v_u} \imath_{\tilde{v}_v} \vol_{(6)} + \imath_{v_v} \imath_{\tilde{v}_u} \vol_{(6)} &= \delta_{uv} \vol_{(4)} \,.
 \end{split}
\end{equation}

We solve these conditions by choosing the first three $\J$'s to be given by the holomorphic and K\"ahler 2-forms on $K3$. Thus
\begin{equation}
 \J_U = \Omega_U \,, \qquad \textrm{ for } U = 1, 2, 3 \,.
\end{equation}
The fourth and fifth $J$'s are chosen as
\begin{equation}
 \J_4 = \sigma + \tilde{\sigma} \wedge \vol_{(4)} \,, \qquad \J_5 = \sigma' + \tilde{\sigma}' \wedge \vol_{(4)} \,,
\end{equation}
where $\sigma$ and $\sigma'$ are the well-defined vector fields on the $T^2$, and $\tilde{\sigma}$ and $\tilde{\sigma}'$ their dual one-forms. The intrinsic torsion vanishes just like in the previous two examples because
\begin{equation}
 d\Omega_U = d\tilde{\sigma} = d\tilde{\sigma}' = d\vol_{(4)} = 0 \,.
\end{equation}

\section{Rewriting the action} \label{s:Reformulate}
The half-maximal structure group $\SO{d-1} \subset H_d$ is a subgroup of the maximal compact subgroup of $E_{d(d)}$ and thus implicitly defines a generalised metric of EFT. In particular, this means that it encodes all of the purely internal fields of the 10/11-dimensional supergravity. As a result, we can rewrite the EFT action, and thus that of 10/11-dimensional supergravity, in terms of the $\GH$ structure. The resulting action will have the form of a half-maximal gauged SUGRA action, although we have not yet performed a truncation. One can think of this as the half-maximal ``flux formulation'', in analogy with the maximal flux formulations \cite{Geissbuhler:2013uka,Berman:2013uda,Blair:2014zba}, where the action is rewritten using the Weitzenb\"ock connection.

\subsection{Kinetic terms}
We begin with the kinetic terms, which we determine by comparison to half-maximal gauged SUGRA upon performing a consistent truncation, see section \ref{s:ConsTruncation}. For the scalars, encoded in the $\GH$ structure, the kinetic terms are given by
\begin{equation}
 \begin{split}
  L_{kin\,s} &= \frac12 \kappa^{2-D} g^{\mu\nu} \left( D_{\mu} \J_u \wedge D_{\nu} \hJ^u + \kappa^{2-D} \left( \hJ_{u} \wedge D_\mu \J^v \right) \left( \hJ_v \wedge D_\nu \J^u \right) \right. \\
  & \quad + \left. \frac{D-2}{4} D_\mu \K \wedge D_\nu \hK \right) \,, \label{eq:ScalarKineticTerm}
 \end{split}
\end{equation}
where $D_\mu = \partial_\mu - \gL_{A_\mu}$ are the EFT-covariant external derivatives.

Similarly, we can write the kinetic terms for the gauge fields using the $\GH$ structure instead of a generalised metric. There is a universal part of the gauge kinetic terms which is given by
\begin{equation}
 L_{kin\,g}^0 = \kappa^{4-D} \left[ \frac{\kappa^{2-D}}{2} \left({\cal F}_{\mu\nu} \wedge \hJ_u\right) \left({\cal F}^{\mu\nu} \wedge \hJ^u \right) - \frac{1}{4} {\cal F}_{\mu\nu} \wedge {\cal F}^{\mu\nu} \wedge \hK \right] \,, \label{eq:UniGaugeKineticTerm}
\end{equation}
as well as a part, $L_{kin\,g}^{(D)}$ which differs from dimension to dimension. Here we will only give it in the case of $D=7$ and $D = 5$.

In $D = 7$, $L_{kin\,g}^{(7)}$ consists of a kinetic term for the two-form gauge potential ${\cal B}_{\mu\nu}$ which can be written in terms of the $\GH$ structure as
\begin{equation}
 \begin{split}
  L_{kin\,g}^{(7)} &= - \frac{1}{12} \kappa^{-6} \left( {\cal H}_{\mu\nu\rho} \wedge \hK \right) \left( {\cal H}^{\mu\nu\rho} \wedge \hK \right) \,. \label{eq:7DGaugeKineticTerm}
 \end{split}
\end{equation}
In $D=5$, there are additional one-form potentials, whose kinetic terms are given in terms of the $\GH$ structure by
\begin{equation}
 \begin{split}
  L_{kin\,g}^{(5)} &= - \frac14 \kappa^{-4} \left( {\cal F}_{\mu\nu} \wedge \K \right) \left( {\cal F}^{\mu\nu} \wedge \K \right) \,. \label{eq:5DGaugeKineticTerm}
 \end{split}
\end{equation}

In $D=6$, one could in addition to the above have further terms involving ${\cal H}_{\mu\nu\rho} \wedge \K$, since $\K \in {\cal R}_2$ in that case. The correct terms could be read off by comparison with the general $D = 6$ half-maximal gauged SUGRAs. However, since the exact form of these kinetic terms is not important for the remainder of this paper, we will not try and determine them here.

\subsection{Scalar potential} \label{s:ReformulatePot}
The scalar potential is given by
\begin{equation}
 V = - \frac14 V_0 - \frac14 V_D + 2 \J_u{}^M \J^{u\,N} \nabla_M g_{\mu\nu} \nabla_N g^{\mu\nu} \,,
\end{equation}
where $V_0$ is the same in every dimension $D \geq 5$, and thus only the universal torsion classes, $\tc_{\J\,u}$, $\tc_{\K}$, and $\tc_{C}$ appear, while $V_D$ is dimension-dependent and involves $\tc_{\hK}$. $V_0$ is given by
\begin{equation}
 \begin{split}
  V_0 &= \frac13 R_{2\,uvw} R^{2\,uvw} + \kappa^{4-D} R_{1\,uv} \wedge R^{1\,uv} \wedge \hK - 2 U_u U^u - 2 U_u T_2{}^u - 4 \kappa^{-2} \gL_{\J_u} \left( U^u \kappa \right) \\
  & \quad - T_2{}^u T_{2\,u} + \kappa^{4-D} T_1 \wedge T_1 \wedge \hK + \ldots \,, \label{eq:PotUniversal}
 \end{split}
\end{equation}
where $\ldots$ refers to terms that vanish in honestly half-maximal theories.

Up to the $T_1 \wedge T_1 \wedge \hK$ and $T_2{}^u T_{2\,u}$ terms, this potential is fixed by requiring invariance under local $\SO{d-1}_R$ symmetry as we show in appendix \ref{a:PotentialRSymmetry}. On the other hand, we have fixed the $T_1 \wedge T_1 \wedge \hK$ and $T_2{}^u T_{2\,u}$ terms by comparison with half-maximal gauged SUGRA upon performing a consistent truncation. An alternative approach would be to express the EFT scalar potential in terms of spinors, and then reformulate the resulting expression using the $\SO{d-1}$ intrinsic torsion, as was done for $D=7$ in \cite{Malek:2016bpu}.

The dimension-dependent parts are given by
\begin{equation}
 \begin{split}
  V_7 &= - \frac{1}{3} R_{2\,uvw} \epsilon^{uvw} P -  \frac14 P^2 + \ldots \,, \\
  V_6 &= - 2 P_2{}^u P_{2\,u} - \kappa^{-2} P_1{}^M P_1{}^N \left(\gamma^I\right)_{MN} \hK_I + \frac{4}{3} P_{2\,u} R_{2\,vwx} \epsilon^{uvwx} + \gamma P_2{}^u T_{2\,u} \\
  & \quad + \frac{\lambda}{2} \kappa^{-2} P_1{}^M T_1{}^N \left(\gamma^I\right)_{MN} \hK_I + \ldots \,, \\
  V_5 &= - \kappa^{-1} P_{3\,u}{}^M P_3{}^{u\,N} d_{MNP} \hK^P + \frac{\sqrt{2}}{3} P_{4\,uv} R_{2\,wxy} \epsilon^{uvwxy} + \sigma \kappa^{-1} P_1 \wedge P_1 \wedge \hK \\
  & \quad + \tau \kappa^{-2} \gL_{\hK} \left( \kappa P_1 \right) + \ldots \,, \label{eq:PotDimDep}
 \end{split}
\end{equation}
where again $\ldots$ refers to terms that vanish in a truly half-maximal theory. These have been fixed by checking that they reduce to the appropriate scalar potential of half-maximal gauged SUGRA, as we show in section \ref{s:ScalarTruncation}. Here we have not completely determined $V_6$ and $V_5$, because the general six-dimensional non-chiral half-maximal gauged SUGRA has not yet been constructed, and so the coefficients $\gamma$ and $\lambda$ are unknown, and in five dimensions $P_1 = 0$ when the trombone gauging vanishes. However, these potentials could be fixed by first expressing the EFT action in terms of spinors and then using this to rewrite it in terms of the intrinsic torsion, as was done in \cite{Malek:2016bpu}.

\section{Half-maximal consistent truncations} \label{s:ConsTruncation}
We are now in a position to discuss half-maximal consistent truncations. We begin by giving the truncation Ans\"atze for the scalars and gauge fields in sections \ref{s:ScalarAnsatz} and \ref{s:GaugeAnsatz} before discussing the conditions for consistency in section \ref{s:EmbTensor}. The truncation Ansatz is given by expanding the fields of the exceptional field theory in terms of a finite number of tensor fields, which depend only on the exceptional coordinates $Y^M$ associated to $M_d$.\footnote{Here we will focus on the bosonic sector but the fermions can be dealt with similarly by expanding them in the basis of well-defined spinor fields on $M_d$, see \cite{Malek:2016bpu}.} These fields are given by
\begin{equation}
 \left( \omega_A \,, \,\, n \,, \,\, \hat{n} \,, \,\, \rho \right) \,, \label{eq:BackgroundStructure}
\end{equation}
where $A = 1, \ldots, d-1 + N$ labels the vector representation of $\SO{d-1,N}$, and $\omega_A \in \cRY_1$, $n \in \cRY_2$, $\hat{n} \in \cRY_{D-4}$ and $\rho \in \cSY$. Note that because of the restricted coordinate dependence, the bundles $\cRY_i$ and $\cSY$ are now actually defined just over $M_d$, which we indicate by the superscript $Y$.

These tensors also satisfy a set of compatibility conditions, reminiscent of \eqref{eq:KPurity}, \eqref{eq:KCompatibility} and \eqref{eq:JCompatibility},
\begin{equation}
 \begin{split}
  \left( n \otimes n \right)\vert_{{\cal R}_c \otimes {\cal S}^4} &= 0 \,, \\
  \left( \hat{n} \otimes \hat{n} \right)\vert_{{\cal R}^*_c \otimes {\cal S}^{2D-8}} &= 0 \,, \\
  \omega_A \wedge n &= 0 \\ 
  \omega_A \wedge \omega_B &= \eta_{AB} n \,. \label{eq:BackgroundCompat}
 \end{split}
\end{equation}
where $\eta_{AB}$ is the invariant metric of $\SO{d-1,N}$, which we will use to raise/lower $A, B = 1, \ldots, d-1+N$ indices. For the following, it will also be useful to define
\begin{equation}
 \hat{\omega}_A = \omega_A \wedge \hat{n} \,.
\end{equation}

By comparison with section \ref{s:GHalfStr} we see that these tensors define a $\SO{d-1-N}$ structure on $M_d$. As we will see, $N$ determines the number of vector multiplets that are kept in the truncation. Thus, in order to keep $N \neq 0$ vector multiplets, one requires a further reduction of the exceptional generalised structure group on $M_d$ to $\SO{d-1-N} \subset \SO{d-1} \subset E_{d(d)}$. Thus, one can at most keep $N_{max} = d-1$ vector multiplets. We will refer to the set of tensors \eqref{eq:BackgroundStructure} satisfying \eqref{eq:BackgroundCompat} as the background $\SO{d-1-N}$ structure.

Let us be more precise of how the bound on the number of vector multiplets arises. Recall from section \ref{s:GHalfStr} that the third equation of \eqref{eq:BackgroundCompat} implies that at each point $\omega_A \in V_{d-1,d-1}$. The final equation requires these $\omega_A$'s to form an orthonormal basis of $V_{d-1,d-1}$. From this it is clear that one can at most keep $N \leq d -1$ vector multiplets in the consistent truncation.

This situation is to be contrasted with what typically happens when studying effective theories. In that case, one would require the compatibility conditions \eqref{eq:BackgroundCompat} to hold only when integrated over the internal space. In particular, this would mean that the $\omega_A$'s would only be required satisfy
\begin{equation}
 \int_{M_d} \omega_A \wedge \omega_B \wedge \hat{n} = \eta_{AB} \,. \label{eq:IntegralBackgroundCompat}
\end{equation}
The expression under the integral sign is a scalar density under generalised diffeomorphisms and thus really can be integrated.

The condition \eqref{eq:IntegralBackgroundCompat} makes use of an inner product on the infinite space of sections, and thus there are an infinite number of orthonormal sections with respect to this inner product. Thus, \eqref{eq:IntegralBackgroundCompat} allows one to retain an infinite number of vector multiplets. In the effective theory one instead ends up with a finite number of vector multiplets by requiring the sections to represent moduli and thus belong to certain cohomology classes. Furthermore, because the $\omega_A$'s represent cohomology classes, they need not be well-defined sections of a vector bundle, and thus are not associated with a reduced structure group.

Let us return to the general idea behind the truncation Ansatz before giving its details. As explained, we will expand all EFT fields in terms of the background $\SO{d-1-N}$ structure on $M_d$, \eqref{eq:BackgroundStructure}. We will allow the coefficients in the expansion to only depend on $x^\mu$, the coordinates on $M_D$. These then become the fields of the lower-dimensional half-maximal gauged SUGRA. Throughout, we will remove any spinor representation of $\SO{d-1-N}$, since these are related to massive gravitino multiplets, which we wish to remove in our truncation.

\subsection{Scalar Ansatz} \label{s:ScalarAnsatz}
The truncation Ansatz for the scalars involves expanding the $\GH$ structure itself in terms of the background $\SO{d-1-N}$ structure \eqref{eq:BackgroundStructure}. Thus, we let
\begin{equation}
 \begin{split}
  \langle \J_u \rangle (x,Y) &= \Sigma^{-1}(x)\, b_u{}^A(x)\, \omega_A(Y) \,, \\
  \langle \K \rangle (x,Y) &= \Sigma^{-2}(x)\, n(Y) \,, \\
  \langle \hK \rangle (x,Y) &= \Sigma^{2}(x)\, \hat{n}(Y) \,, \\
  \langle \kappa \rangle (x,Y) &= \rho(Y) \,,
  \label{eq:ScalarAnsatz}
 \end{split}
\end{equation}
and the external metric
\begin{equation}
 \langle g_{\mu\nu} \rangle(x,Y) = \hg_{\mu\nu}(x)\, \rho^2(Y) \,. \label{eq:MetricAnsatz}
\end{equation}
The $\langle \, \rangle$ brackets denote the truncation Ansatz. Here, the scalar $\Sigma(x)$ carries charge $-1/2$ under the $\mathbb{R}^+$ subgroup of $\SO{d-1,d-1} \times \mathbb{R}^+ \subset E_{d(d)}$, and appears in the Ansatz \eqref{eq:ScalarAnsatz} accordingly. The scalars $b_u{}^A(x)$ and $\hg_{\mu\nu}(x)$ are uncharged with respect to this $\mathbb{R}^+$ group. This is summarised in table \ref{t:Charges}.

One may wonder why we have not included an extra scalar degree of freedom in the Ansatz for $\kappa$ and $\hK$. The reason for this is because these amount to a global, i.e. $Y$-independent, rescaling of the $\SO{d-1-N}$ structure of the type \eqref{eq:GlobalRescaling}. Thus, these scalars cannot affect the truncation and hence do not correspond to physical degrees of freedom. The choice in \eqref{eq:ScalarAnsatz} will give us half-maximal gauged SUGRA in Einstein frame.

Using \eqref{eq:BackgroundCompat}, we see that in order for the compatibility conditions \eqref{eq:JCompatibility} to be satisfied, the scalars $b_u{}^A$ must satisfy
\begin{equation}
 b_u{}^A b_{v}{}^B \eta_{AB} = \delta_{uv} \,. \label{eq:ScalarCompatibility}
\end{equation}
Recall also that $\J_u$'s related by $\SO{d-1}_R$ rotations describe the same background. Thus, we must also identify the scalars $b_u{}^A$ which are related by $\SO{d-1}_R$ rotations. We can do this by considering the invariant combination
\begin{equation}
 P^-_A{}^B = b^u{}_A b_u{}^B \,,
\end{equation}
where from now on we will always raise / lower the $A, B$ indices by $\eta_{AB}$. Using \eqref{eq:ScalarCompatibility}, one can easily show that $P^-_A{}^B$ is a projector of rank $(d-1)$.

We can also write $P^-_A{}^B$ in the following form
\begin{equation}
 P^-_A{}^B = \frac12 \left( \delta_A^B + \gH_{AC} \eta^{BC} \right) \,,
\end{equation}
where
\begin{equation}
 \gH_{AB} = \eta_{AB} - 2 b^u{}_A b_{u\,B} \,,
\end{equation}
is satisfies
\begin{equation}
 \eta^{CD} \gH_{AC} \gH_{BD} = \eta_{AB} \,.
\end{equation}
Thus, $\gH_{AB}$ parameterises the coset space
\begin{equation}
 \gH_{AB} \in \frac{\SO{d-1,N}}{\SO{d-1}\times\SO{N}} \,.
\end{equation}
Taking into account the scalar, $\Sigma$, we see that the scalar coset space is
\begin{equation}
 M_{scalar} = \frac{\SO{d-1,N}}{\SO{d-1}\times\SO{N}} \times \mathbb{R}^+ \,,
\end{equation}
which is indeed the scalar manifold of half-maximal gauged SUGRA coupled to $N$ vector multiplets.

\subsection{Gauge fields Ansatz} \label{s:GaugeAnsatz}
Recall that the EFT contains a set of $p$-form gauge fields which are local sections of ${\cal R}_p$ with the appropriate weights. The truncation Ansatz expands these again in terms of the appropriate tensors of the background $\SO{d-1-N}$ structure.

In $D=7$ one has one-form, two-, three-, and four-form potentials, whose truncation Ans\"atze are
\begin{equation}
 \begin{split}
  \langle {\cal A}_\mu \rangle (x,Y) &= A_\mu{}^A(x)\, \omega_A(Y) \,, \\
  \langle {\cal B}_{\mu\nu} \rangle(x,Y) &= - B_{\mu\nu}(x)\, n(Y) \,, \\
  \langle {\cal C}_{\mu\nu\rho} \rangle(x,Y) &= C_{\mu\nu\rho}(x)\, \hat{n}(Y) \,, \\
  \langle {\cal D}_{\mu\nu\rho\sigma} \rangle(x,Y) &= D_{\mu\nu\rho\sigma}{}^A(x)\, \hat{\omega}_A(Y) \,. \label{eq:7DGaugeAnsatz}
 \end{split}
\end{equation}
These are the correct degrees of freedom expected in seven-dimensional half-maximal gauged SUGRA coupled to $N$ vector multiplets, up to dualisations, as usual in exceptional field theory: the one-forms and four-forms transform as vectors of $\SO{3,N}$ and satisfy a duality relation so that only half of them are propagating. This gives the correct on-shell degrees of freedom.

In $D=6$ one only has a one-form, two-form and three-form potential. However, because now $\hat{n} \in {\cal R}_2$, one obtains two two-forms in the half-maximal theory via the Ansatz
\begin{equation}
 \begin{split}
  \langle {\cal A}_\mu \rangle (x,Y) &= A_\mu{}^A(x)\, \omega_A(Y) \,, \\
  \langle {\cal B}_{\mu\nu} \rangle(x,Y) &= - B^+_{\mu\nu}(x)\, n(Y) - B^-_{\mu\nu}(x)\, \hat{n}(Y) \,, \\
  \langle {\cal C}_{\mu\nu\rho} \rangle(x,Y) &= C_{\mu\nu\rho}{}^A(x)\, \hat{\omega}_A(Y) \,. \label{eq:6DGaugeAnsatz}
 \end{split}
\end{equation}
The duality relations impose that $B_{\mu\nu}^-$ and $B_{\mu\nu}^+$ are (anti)-self-dual two-forms, while the three-forms $C_{\mu\nu\rho}{}^A$ are dual to the on-form potentials $A_\mu{}^A$. Again, we obtain the correct degrees of freedom of six-dimensional half-maximal gauged SUGRA coupled to $N$ vector multiplets.

Finally, in $D=5$ one has only the one-form and two-form potentials. However, because we now have $\hat{n} \in {\cal R}_1$, one obtains $6+N$ one-form potentials using the Ansatz
\begin{equation}
 \begin{split}
  \langle {\cal A}_\mu \rangle (x,Y) &= A_\mu{}^A(x)\, \omega_A(Y) + A_\mu{}^0(x) \, \hat{n}(Y) \,, \\
  \langle {\cal B}_{\mu\nu} \rangle(x,Y) &= - B_{\mu\nu}(x)\, n(Y) - B_{\mu\nu\,A}(x) \, \hat{\omega}^A(y) \,. \label{eq:5DGaugeAnsatz}
 \end{split}
\end{equation}
Thus, we obtain $5+N$ one-form potentials which transform in the vector of $\SO{5,N}$, as well as an additional 1-form potential $A_\mu{}^0$, which is a scalar under $\SO{5,N}$. We find the same for the two-forms potentials, which are dual to the vector fields. This corresponds to $D=5$ half-maximal gauged SUGRA coupled to $N$ vector multiplets \cite{Schon:2006kz}.

It is convenient to rewrite the five-dimensional Ansatz using indices ${\cal A} = \left(0, A\right)$, with ${\cal A} = 0, \ldots, 5+N$. These allow us to combine
\begin{equation}
 \begin{split}
  \omega_{\cal A} &= \left( \hat{n},\, \omega_A \right) \,, \\
  \hat{\omega}_{\cal A} &= \left( n,\, \hat{\omega}_A \right) \,,
 \end{split}
\end{equation}
and write the Ansatz for the gauge fields as
\begin{equation}
 \begin{split}
  \langle {\cal A}_{\mu} \rangle(x,Y) &= A_\mu{}^{\cal A} \omega_{\cal A} \,, \\
  \langle {\cal B}_{\mu\nu} \rangle(x,Y) &= - B_{\mu\nu}{}^{\cal A} \hat{\omega}_{\cal A} \,.
 \end{split}
\end{equation}

Here we have chosen not to include the scalar $\Sigma$ in the truncation Ansatz of the gauge fields. As a result, the gauge fields will be charged under the $\mathbb{R}^+$ subgroup of $\SO{d-1,d-1} \times \mathbb{R}^+ \subset E_{d(d)}$ with charges as shown in table \ref{t:Charges}. This agrees with the conventions of \cite{Schon:2006kz} for five-dimensional half-maximal gauged SUGRA.\\

\noindent\makebox[\textwidth]{
 \begin{minipage}{\textwidth}
  \begin{center}
  \begin{tabular}{|c||c|c|c|c|c|c|c|}
  \hline
  $D = 7$ & $b_u{}^A$ (0) & $\Sigma$  (-1/2) & $\bag_{\mu\nu}$ (0) & $A_\mu{}^A$ (1/2) & $B_{\mu\nu}$ (1) & $C_{\mu\nu\rho}$ (-1) & $D_{\mu\nu\rho\sigma\,A}$ (-1/2) \Tstrut\Bstrut \\
  \hline
  $D = 6$ & $b_u{}^A$ (0) & $\Sigma$  (-1/2) & $\bag_{\mu\nu}$ (0) & $A_\mu{}^A$ (1/2) & $B_{\mu\nu}^+$ (1) & $B_{\mu\nu}^-$ (-1) & $C_{\mu\nu\rho\,A}$ (-1/2) \Tstrut\Bstrut \\
  \hline
  $D = 5$ & $b_u{}^A$ (0) & $\Sigma$  (-1/2) & $\bag_{\mu\nu}$ (0) & $A_\mu{}^A$ (1/2) & $A_\mu{}^0$ (-1) & $B_{\mu\nu\,A}$ (-1/2) & $B_{\mu\nu\,0}$ (1) \Tstrut\Bstrut \\
   \hline
  \end{tabular}
 \vskip-0.5em
 \captionof{table}{\small{Charges, given in parentheses, of the fields of the half-maximal SUGRA under \\$\mathbb{R}^+ \subset \SO{d-1,d-1} \times \mathbb{R}^+ \subset \EG{d}$.}} \label{t:Charges}
 \end{center}
 \end{minipage}}

\subsection{Consistency, intrinsic torsion and embedding tensor} \label{s:EmbTensor}
As we have just shown the truncation Ansatz is given by an expansion of the EFT fields in terms of the $\SO{d-1-N}$ structure of the background. In order for the truncation to be \emph{consistent}, we now need to impose three kinds of differential constraints on the background $\SO{d-1-N}$ structure. These are most naturally formulated in terms of the intrinsic torsion of the $\SO{d-1-N}$ structure, which similar to \eqref{eq:JTorsionClasses}, can in general be written as
\begin{equation}
 \begin{split}
  dn &= \rho\, t_1 + \omega_A\, f^A + \rho^{5-D} \hat{n} \bullet t_3 \,, \\
  \gL_{\omega_{[A}} \omega_{B]} &= \rho\, r_{1\,AB} + f_{ABC} \omega^C + f_{[A} \omega_{B]} - \rho^{5-D} \hat{n} \bullet \left( L_{AB} \cdot t_3 \right) \,, \\
  \gL_{\omega_A} \hat{n} &= \rho^{D-4}\, s_{1\,A} + \rho^{D-5} \omega_A \wedge s_2 + \left( \xi_A - f_{A} \right) \hat{n} \,, \\
  \gL_{\omega_A} \rho^{D-2} &= \rho^{D-2}\, \xi_A \,,
  \label{eq:BCEmbTorsionClasses}
 \end{split}
\end{equation}
where now $\TL_{AB} = \rho^{2-D} \omega_{[A} \wedge_P \omega_{B]}$ and the irreducibles of the intrinsic torsion are now generalised tensors with no weight, in contrast to \ref{s:IntTorsion}.

The dimension-specific components of the intrinsic torsion are given by
\begin{alignat}{2}
 d\hat{n} &= \rho^2\, p_1 + \theta\, n \,, \quad&& \textrm{for } D = 7\,, \\
 d\hat{n} &= \rho\, p_1 + \rho\, \theta^A \omega_A + \rho^{-1} \hat{n} \wedge p_3 \,, \quad&& \textrm{for } D = 6 \,, \label{eq:BCD67EmbTorsionClasses}
\end{alignat}
and for $D = 5$,
\begin{equation}
 \begin{split}
  \gL_{\hat{n}} n &= \xi n + \rho\, p_2 \wedge n \,, \quad \gL_{\hat{n}} \rho^3 = \xi \rho^3 \,, \\
  \gL_{\hat{n}} \omega_A &= \rho p_{3\,A} + \xi_{AB} \omega^B + \frac12 \xi\, \omega_A - \rho^{-4} \left( \left( p_{2} \bullet \hat{n} \right) \wedge_P \omega_A \right) \cdot n \,. \label{eq:BCD5EmbTorsionClasses}
 \end{split}
\end{equation}

The first differential constraint we must impose is that the intrinsic torsion of the background $\SO{d-1-N}$ structure does not contain spinor representations of $\SO{d-1-N}$. We impose this in order to remove massive gravitino multiplets from the truncation. In particular, this implies hat
\begin{equation}
 \begin{split}
  n \wedge dn &= 0 \,, \\
  n \wedge \gL_{\omega_{A}} \omega_{B} &= 0 \,, \\
  \omega_B \wedge \gL_{\omega_A} \hat{n} &= 0 \,,
 \end{split}
\end{equation}
and for the dimension-specific parts
\begin{alignat}{2}
 d\hat{n} &= \rho^{-5} n \left( \hat{n} \wedge d\hat{n} \right) \,, \qquad && \textrm{for } D = 7 \,, \nonumber \\
 n \wedge d\hat{n} &= 0 \,, \qquad && \textrm{for } D = 6 \,, \\
 \gL_{\hat{n}} n &= \rho^{-3} n \left(\hat{n} \wedge \gL_{\hat{n}} n \right) \,, \qquad n \wedge \gL_{\hat{n}} \omega_A = 0 \,, \qquad && \textrm{for } D = 5 \,. \nonumber
\end{alignat}
As a result we have $t_3 = s_{1\,A} = s_2 = 0$, and in $D = 7$, $p_1 = 0$, in $D = 6$, $p_3 = 0$ and in $D = 5$, $p_2 = 0$. We call this the ``spinor constraint''.

Secondly, we require that we can expand the intrinsic torsion in terms of the finite number of fields which define the background $\SO{d-1-N}$ structure. This implies that any intrinsic torsion in the vector representation of $\SO{d-1-N}$ must vanish. Thus $t_1 = r_{1\,AB} = 0$, and additionally in $D = 6$, $p_1 = 0$ and in $D = 5$, $p_{3\,A} = 0$. We call this the ``closure constraint''.

Together the spinor and closure constraints imply that the intrinsic torsion of the background $\SO{d-1-N}$ structure is given by
\begin{equation}
 \begin{split}
  d n &= f^A\, \omega_A \,, \\
  \gL_{\omega_A} \omega_B &\equiv X_{ABC}\, \omega^C = f_{ABC}\, \omega^C + f_{[A}\, \omega_{B]} + \frac12 \eta_{AB} f_C\, \omega^C \,, \\
  \gL_{\omega_A} \hat{n} &= \left( \xi_A - f_A \right) \hat{n} \,, \\
  \gL_{\omega_A} \rho^{D-2} &= \rho^{D-2}\, \xi_A \,, \label{eq:EmbTensor}
 \end{split}
\end{equation}
with dimension-specific part
\begin{alignat}{2}
  d\hat{n} &= \theta\, n \,, \quad&& \textrm{for } D = 7\,, \\
  d\hat{n} &= \theta^A \omega_A \,, \quad&& \textrm{for } D = 6 \,, \label{eq:D67EmbTensor}
\end{alignat}
and for $D = 5$,
\begin{equation}
 \begin{split}
  \gL_{\hat{n}} n &= \xi n \,, \quad \gL_{\hat{n}} \rho^3 = \xi \rho^3 \,, \\
  \gL_{\hat{n}} \omega_A &= \xi_{AB} \omega^B + \frac12 \xi \omega_A \,. \label{eq:D5EmbTensor}
 \end{split}
\end{equation}
These also imply that
\begin{equation}
 \begin{split}
  \gL_{\omega_A} \hat{\omega}_B &= X_{ABC}\, \hat{\omega}^C + \left(\xi_A - f_A \right) \hat{\omega}_B \\
  &= f_{ABC}\, \hat{\omega}^C - f_{(A}\, \hat{\omega}_{B)} + \frac12 \eta_{AB} f_C\, \hat{\omega}^C + \xi_A\, \hat{\omega}_B \,, \\
  \gL_{\omega_A} n &= f_A\, n \,.
 \end{split}
\end{equation}

The final constraint is that the surviving components of the intrinsic torsion, $f_A$, $f_{ABC}$ and $\xi_A$, and in $D = 7$, $\theta$, $D = 6$, $\theta_A$ and in $D = 5$, $\xi_{AB}$ and $\xi$ must be \emph{constant}. As we will see, the action will not depend on the $\SO{d-1-N}$ structure directly, but only through the intrinsic torsion and an overall scaling via $\rho$. As a result, the dependence on the $Y^M$ coordinates on $M_d$ factorises if $f_A$, $f_{ABC}$ and $\xi_A$ and the dimension-dependent parts are constant. In this case we have a consistent truncation.

We can now identify the constants $f_A$, $f_{ABC}$ and $\xi_A$ with the universal part of the embedding tensor of half-maximal gauged SUGRA, and the dimension-dependent parts $\theta$ ($D = 7$), $\theta_A$ ($D = 6$) and $\xi_{AB}$ and $\xi$ ($D=5$) with the allowed deformations in those dimensions \cite{Schon:2006kz,Bergshoeff:2007vb}. Thus, we see that the embedding tensor obtained here corresponds to the most general solution of the linear constraint of half-maximal gauged SUGRA. Note that $\xi_A$ (and for $D = 5$, $\xi$) is the so-called trombone tensor, which must be vanishing in order to have an action principle for the gauged SUGRA. Therefore, we will in the following take $\xi_A = 0$ (and $\xi = 0$ for $D = 5$).

In order to have a consistent gSUGRA, one also needs to impose a set of quadratic constraints, which ensure that the gauge algebra closes. Just as in the case of generalised Scherk-Schwarz Ans\"atze \cite{Aldazabal:2011nj,Geissbuhler:2011mx,Grana:2012rr}, these follow from our Ansatz if we impose the section condition. However, the quadratic constraints of gauged SUGRA also allow for gaugings where the $\SO{d-1-N}$ structure violates the section condition.

\subsection{Reduction of kinetic terms} \label{s:RedKin}
In order to show that we obtain a consistent truncation, let us perform the reduction of the kinetic terms, \eqref{eq:ScalarKineticTerm}, \eqref{eq:7DGaugeKineticTerm} and \eqref{eq:5DGaugeKineticTerm}. Using \eqref{eq:ScalarAnsatz}, it is easy to show that the scalar kinetic terms become
\begin{equation}
 \begin{split}
  \langle L_{kin\,s} \rangle(x,Y) &= \rho^{-2} \hat{g}^{\mu\nu} \left[ \frac{1}{16}\, \D_\mu \gH^{AB} \D_\nu \gH_{AB} - \frac{D-2}{2}\, \Sigma^{-2}\, \D_\mu \Sigma\, \D_\nu \Sigma \right] \,,
 \end{split}
\end{equation}
where $\D_\mu \gH_{AB}$ and $\D_\mu \Sigma$ are the gauge covariant derivatives of the half-maximal theory and arise from the truncation of the EFT-covariant derivative, i.e. $\langle D_\mu \rangle \longrightarrow \D_\mu$. We see that the $Y$-dependence only appears through the conformal factor $\rho^{-2}$ and thus the equations of motions factorise.

\subsubsection{Gauge kinetic terms} \label{s:RedKinGauge}
One can use the truncation Ansatz \eqref{eq:7DGaugeAnsatz} - \eqref{eq:5DGaugeAnsatz} to find the field strengths of the half-maximal gauged SUGRA. Let us indicate how this works by working through the example of $D=5$ explicitly. The field strength ${\cal F}_{\mu\nu}$ of the 1-form potential is defined as \cite{Hohm:2013vpa}
\begin{equation}
 {\cal F}_{\mu\nu} = 2 \partial_{[\mu} A_{\nu]} - \left[ A_\mu,\, A_\nu \right]_E + dB_{\mu\nu} \,,
\end{equation}
where $\left[ \,,\, \right]_E$ denotes the antisymmetrised generalised Lie derivative. Plugging in the truncation Ansatz one finds
\begin{equation}
 \langle {\cal F}_{\mu\nu}^M \rangle(x,Y) = F_{\mu\nu}^A(x)\, \omega_A{}^M(Y) + F_{\mu\nu}^0(x)\, \hat{n}^M(Y) \,, \label{eq:5DFluxRed}
\end{equation}
where
\begin{equation}
 \begin{split}
  F_{\mu\nu}{}^A &= 2 \partial_{[\mu} A_{\nu]}^A - A_{\mu}^B A_{\nu}^C (f_{BC}{}^A + \delta^A_{[B} f_{C]}) + A_{[\mu}^0 A_{v]}^B \xi^A{}_B- B_{\mu\nu}^0 f^A \,, \\
  F_{\mu\nu}{}^0 &= 2 \partial_{[\mu} A_{\nu]}^0 - A_{[\mu}^0 A_{\nu]}^A f_A + B_{\mu\nu}^A f_A \,. \label{eq:5DFieldStrengths}
 \end{split}
\end{equation}

Plugging \eqref{eq:5DFluxRed} and \eqref{eq:5DFieldStrengths} into \eqref{eq:UniGaugeKineticTerm} and \eqref{eq:5DGaugeKineticTerm} one finds
\begin{equation}
 \begin{split}
  \langle L_{kin,g}^{(0)} \rangle &= -\frac14 \rho^{-2}(Y)\, \Sigma^{2} F_{\mu\nu}^A F^{\mu\nu\,B}\, \gH_{AB} \,, \\
  \langle L_{kin,g}^{(5)} \rangle &= - \frac14 \rho^{-2}(Y)\, \Sigma^{-4} F_{\mu\nu}^0 F^{\mu\nu\,0} \,, \label{eq:5DReducedGaugeKineticTerm}
 \end{split}
\end{equation}
where in the reduced theories the spacetime indices are raised/lowered with the metric of the half-maximal gauged SUGRA $\hat{g}_{\mu\nu}$. These are exactly the correct kinetic terms of five-dimensional half-maximal gauged SUGRA \cite{Schon:2006kz}. Note, in particular, that it is a non-trivial check that we obtain the correct powers of the scalar $\Sigma$ in \eqref{eq:5DReducedGaugeKineticTerm}. Finally, as required, the dependence on the $Y$ coordinates factorises thanks to our truncation Ansatz, so that we obtain a consistent truncation.

\subsection{Reduction of scalar potential} \label{s:ScalarTruncation}
Let us now calculate the reduction of the scalar potential using the truncation Ansatz \eqref{eq:ScalarAnsatz} and \eqref{eq:EmbTensor}. We begin by calculating the intrinsic $\GH$ torsion using the truncation Ansatz. We find that its universal part becomes
\begin{equation}
 \begin{split}
  \langle T_1 \rangle &= \rho^{-2}\, \omega_A\, \Sigma^{-2} P_+^{AB} f_B \,, \\
  \langle T_2{}^u \rangle &= \rho^{-1}\, \Sigma^{-1} b^{u\,A} f_A \,, \\
  \langle R_{2\,uvw} \rangle &= \rho^{-1}\, \Sigma^{-1} b_u{}^A b_v{}^B b_w{}^C f_{ABC} \,, \\
  \langle R_{1\,uv} \rangle &= \rho^{-2}\, \omega_A \, \Sigma^{-2} b_u{}^C b_v{}^D P_{+}{}^{AB} f_{BCD} \,, \\
  \langle U_u \rangle &= \rho^{-1}\, \Sigma^{-1} b_u{}^A \xi_A \,, \label{eq:IntrinsicTorsionReduction}
 \end{split}
\end{equation}
with the other components vanishing. Recall that gauged SUGRAs with non-vanishing trombone tensor, $\xi_A \neq 0$, do not admit an action principle. Thus we will here also take $\xi_A = 0$.

Using \eqref{eq:IntrinsicTorsionReduction} one immediately finds that the universal part of the scalar potential becomes
\begin{equation}
 \begin{split}
  \langle |e| V_0 \rangle &= \rho^{D-2} |\bae| \left[ \Sigma^{-2} f_{ABC} f_{DEF} \left( \frac{1}{12} \gH^{AD} \gH^{BE} \gH^{CF} - \frac14 \gH^{AD} \eta^{BE} \eta^{CF} + \frac16 \eta^{AD} \eta^{BE} \eta^{CF} \right) \right. \\
  & \quad \left. + \Sigma^{-2} \gH^{AB} f_A f_B \right] \,. \label{eq:UniPotentialReduction}
 \end{split}
\end{equation}
Here we made use of the identity
\begin{equation}
 \begin{split}
  0 &= \left[ P_{-}^{AD} P_-^{BE} \left( \frac13 P_-^{CF} + P_+^{CF} \right) - \frac1{12} \gH^{AD} \gH^{BE} \gH^{CF} + \frac14 \gH^{AD} \eta^{BE} \eta^{CF} \right. \\
  & \quad \left. - \frac16 \eta^{AD} \eta^{BE} \eta^{CF} \right] X_{ABC} X_{DEF} \,, \label{eq:f2Identity}
 \end{split}
\end{equation}
for any totally antisymmetric $X_{ABC} = X_{[ABC]}$.

The scalar potential \eqref{eq:UniPotentialReduction} agrees with the universal part of the scalar potential of half-maximal gauged SUGRA \cite{Townsend:1983kk,Bergshoeff:1985mr,Romans:1985tw,Schon:2006kz,Dibitetto:2015bia}. Note in particular, that because of our truncation Ansatz, the potential only depends on $Y$ through the conformal factor $\rho$. This factorisation ensures that we have a consistent truncation.

One can also use the truncation Ansatz to calculate the dimension-specific part of the intrinsic torsion. Using \eqref{eq:D67TorsionClasses}, \eqref{eq:ScalarAnsatz} and \eqref{eq:D67EmbTensor} one finds that in $D = 7$ there is only one additional non-vanishing component of the intrinsic torsion
\begin{equation}
 \langle P_1 \rangle = \rho^{-1} \, \theta \Sigma^{4} \,. \label{eq:ITD7Reduction}
\end{equation}
For $D = 6$, one finds the only one extra non-vanishing component given by
\begin{equation}
 \begin{split}
  \langle P_1 \rangle &= \rho^{-2} \omega_A \Sigma^2\, P_+^{AB} \theta_B \,, \\
  \langle P_2{}^u \rangle &= \rho^{-1} \Sigma^3\, b_u{}^A \theta_A \,. \label{eq:ITD6Reduction}
 \end{split}
\end{equation}
Finally, for $D = 5$ one finds, using \eqref{eq:D5TorsionClasses}, \eqref{eq:ScalarAnsatz} and \eqref{eq:D5EmbTensor},
\begin{equation}
 \begin{split}
  \langle P_1 \rangle &= \rho^{-1} \Sigma^2\, \xi \,, \\
  \langle P_{3\,u} \rangle &= - \rho^{-2} \omega_A \Sigma\, P_+^{AB} b_u{}^C \xi_{BC} \,, \\
  \langle P_{4\,uv} \rangle &= \rho^{-1} \Sigma^2\, b_u{}^A b_v{}^B \xi_{AB} \,. \label{eq:ITD5Reduction}
 \end{split}
\end{equation}
We will take $\xi = 0$ in the following, since otherwise the gauged SUGRA does not admit an action principle.

Using these, the dimension-dependent part of the scalar potential, equation \eqref{eq:PotDimDep} reduces to
\begin{equation}
 \begin{split}
  \langle |e| V_7 \rangle &= \rho^{5} |\bae| \left( -\frac{1}{3} \Sigma^3\, \gH^{ABC} f_{ABC} \theta - \frac14 \Sigma^8 \theta^2 \right) \,, \\
  \langle |e| V_6 \rangle &= \rho^4 |\bae| \left( -2 \Sigma^6 \eta^{AB} \theta_A \theta_B + \frac{4}{3} \Sigma^2\, \gH^{ABCD} \theta_A f_{BCD} \right. \\
  & \quad \left. + \gamma \Sigma^2\, P_{-}^{AB} f_A \theta_B  + \lambda \Sigma^2 P_+^{AB} f_A \theta_B \right) \,, \\
  \langle |e| V_5 \rangle &= \rho^3 |\bae| \left[ \frac14 \Sigma^4\, \left( \gH^{AB} \gH^{CD} - \eta^{AB} \eta^{CD} \right) \xi_{AC} \xi_{BD} + \frac{\sqrt{2}}{3} \Sigma\, \gH^{ABCDE} \xi_{AB} f_{CDE} \right] \,.
 \end{split} \label{eq:PotDimDepTruncation}
\end{equation}
Here we have defined
\begin{equation}
 \begin{split}
  \gH^{ABC} &= \epsilon^{uvw} b_u{}^A b_v{}^B b_w{}^C \,, \qquad \qquad \quad \textrm{for } D = 7 \,, \\
  \gH^{ABCD} &= \epsilon^{uvwx} b_u{}^A b_v{}^B b_w{}^C b_x{}^D \,, \qquad \quad \textrm{for } D = 6 \,, \\
  \gH^{ABCDE} &= \epsilon^{uvwxy} b_x{}^A b_v{}^B b_w{}^C b_x{}^D b_y{}^E \,, \quad \textrm{for } D = 5 \,.
 \end{split}
\end{equation}
This gives the correct scalar potential of half-maximal gauged SUGRA \cite{Townsend:1983kk,Bergshoeff:1985mr,Romans:1985tw,DAuria:2000afl,Andrianopoli:2001rs,Bergshoeff:2005pq,Schon:2006kz,Bergshoeff:2007vb,Dibitetto:2015bia}, up to the correct choice of the coefficients $\lambda$ and $\gamma$ when $D = 6$. Again, we see that our Ansatz ensures that the only dependence on $Y$ appears through the conformal factor $\rho$ and thus we have a consistent truncation.

\subsection{Proof of consistency}
Let us now show how that a reduction Ansatz satisfying the above conditions gives a consistent truncation. We will do this by showing that the dependence on the $Y$-coordinates in the equations of motion factorises, where, crucially, we do not impose the truncation Ansatz on the variations of the fields. As a result, the full equations of motion are satisfied if those of the half-maximal gauged SUGRA, which correspond to the $x$-dependent expression in the equations of motion, are satisfied.

Recall that EFT has fields in the tensor hierarchy with ``external indices'', and scalar fields (from the external $D$-dimensional point of view) which parameterise the coset space $E_{d(d)} / H_d$. The equations of motion of the tensor hierarchy fields are easily dealt with. As discussed in \cite{Hohm:2013vpa,Hohm:2013uia,Wang:2015hca,Abzalov:2015ega,Musaev:2015ces,Berman:2015rcc,Bosque:2016fpi}, the variations of the field strengths, ${\cal F}_{\mu\nu}$, ${\cal H}_{\mu\nu\rho}$, \ldots, are given by external covariant derivatives and nilpotent derivatives, $d$, of the variations of the tensor hierarchy fields themselves. Thus, after integrating by parts, the variation of the gauge kinetic terms and topological terms involve external covariant derivatives and nilpotent derivatives of the field-strengths and scalar fields. As we have shown in the preceding sections, our truncation Ansatz implies that for all these terms the $Y$-dependence factorises and appears only through the background $\SO{d-1-N}$ structure multiplying an $x$-dependent expression. Which of the tensor fields defining the $\SO{d-1-N}$ structure appear depends on the EFT index structure of the full expression.

One must also consider what happens to the scalar kinetic terms \eqref{eq:ScalarKineticTerm} under variations of the external 1-form gauge field $\delta A_\mu$. For example, one would have to consider terms such as
\begin{equation}
 \begin{split}
  \delta_A \left( D_\mu \K \wedge D_\nu \hK \right) &= - \gL_{\delta A_\mu} \K \wedge D_\nu \hK - D_\mu \K\wedge \gL_{\delta A_\nu} \hK \,, \\
  \delta_A \left( D_\mu \J_u \wedge D_\nu \hJ^u \right) &= - \gL_{\delta A_\mu} \J_u \wedge D_\nu \hJ^u - D_\mu \J_u \wedge \gL_{\delta A_\nu} \hJ^u \,. \label{eq:VariationScalar}
 \end{split}
\end{equation}
However, one can rewrite these equations by integrating by parts those terms involving derivatives of $\delta A_\mu$ to obtain an expression with only derivatives on the scalar and the external covariant derivatives of the scalars. These expressions must necessarily be tensorial and thus expressible in terms of the generalised Lie derivative or the nilpotent derivative, $d$, of the scalar fields (and their external covariant derivatives).

To be more explicit, when $D \geq 6$ one can use the identity
\begin{equation}
 \gL_{\delta A_\mu} \K = \delta A_\mu \wedge d\K + d \left( \delta A_\mu \wedge K \right) \,,
\end{equation}
and integration by parts to rewrite the first equation of \eqref{eq:VariationScalar} in terms of the nilpotent derivatives, $d$, acting on $K$, $\hat{K}$ and their external covariant derivatives. Similarly, when $D \geq 5$ one can easily prove the identities\footnote{These can easily be shown by writing the generalised Lie derivative in terms of the ``$Y$-tensor'' \cite{Berman:2012vc} as in \eqref{eq:GenLieDerivativeY}.}
\begin{equation}
 \begin{split}
  \gL_{\Lambda} V \wedge W &= \Lambda \wedge \left( \gL_V W - V \wedge d W \right) + \ldots \,, \\
  V \wedge \gL_\Lambda W &= \Lambda \wedge \left( V \wedge dW - \gL_V W \right) + \ldots \,,
 \end{split}
\end{equation}
where $\ldots$ denote total derivative terms and $V, \Lambda \in \Gamma\left({\cal R}_1\right)$ and $W \in \Gamma\left({\cal R}_{D-3}\right)$. These identities can be used to rewrite the second equation of \eqref{eq:VariationScalar} to only involve the generalised Lie derivative of $\J_u$ and $D_\mu \J_u$ acting on $\hJ_u$ and $D_\nu \hJ_u$, as well as terms involving the nilpotent derivative, $d$, of $\hJ_u$ and its external covariant derivative, and similarly for the first equation when $D = 5$ with $\hK$ and $\K$ instead of $\J_u$ and $\hJ_u$. Using these results it is straightforward to show that the $Y$-dependence also factorises in the variation of the scalar kinetic term with respect to $A_\mu$.

Similarly, the full EFT Lagrangian has an external Einstein-Hilbert term, $L_{EH}$, where all external derivatives have been replaced with external covariant derivatives, $D_\mu$, see e.g. \cite{Berman:2015rcc}. Thus, we must also consider the variation of these terms under variations $\delta A_\mu$. After integration by parts so there are no derivatives on $\delta A_\mu$, one finds
\begin{equation}
 \begin{split}
  \delta_A L_{EH} &= \delta A_\mu{}^M \left( g^{\mu\nu} \partial_M D_\nu \ln |g| + \partial_M D_\nu g^{\mu\nu} - \frac12 g^{\mu\lambda} \partial_M g^{\nu\rho} D_\lambda g_{\nu\rho} + g^{\mu\lambda} \partial_M g^{\nu\rho} D_\nu g_{\rho\lambda} \right. \\
  & \quad \left. + \frac12 \partial_M g^{\mu\nu} D_\nu \ln g \right) + \ldots \,,
 \end{split} \label{eq:EHAVariation}
\end{equation}
where $\ldots$ refer to the total derivative terms.\footnote{The author thanks Chris Blair for sharing this result.} Once again, the expression in brackets must be a generalised tensor. However, since $g_{\mu\nu}$ is a generalised scalar density and there is no way of obtaining a generalised tensor from $\partial_M$ derivatives of generalised scalar densities, it follows that in the variation \eqref{eq:EHAVariation} one must be able to replace
\begin{equation}
 \partial_M g_{\mu\nu} \longrightarrow |g|^{-1/D} \partial_M \left( |g|^{1/D} g_{\mu\nu} \right) \,, \qquad \partial_M D_{\mu} g_{\nu\rho} \longrightarrow |g|^{-1/D} \partial_M \left( |g|^{1/D} D_{\mu} g_{\nu\rho} \right) \,.
\end{equation}
Indeed, an explicit calculation shows that this can be done. As a result, \eqref{eq:EHAVariation} vanishes when imposing the truncation Ansatz of the metric, \eqref{eq:MetricAnsatz}.

It remains to show that the $Y$-dependence in the equations of motion of the scalar fields also factorises. Again, the terms involving external covariant derivatives factorise as required, but now there are additional terms which involve only internal derivatives. To show that the $Y$-dependence factorises for these terms, we will show that they can be rewritten in terms of the intrinsic torsion of the $\GH$ structure.

The terms only containing internal derivatives are given by the ``generalised Ricci tensor''. As shown in equation (4.20) of \cite{Coimbra:2012af}, one can rewrite the generalised Ricci tensor in terms of covariant derivatives of a spinor. Schematically,
\begin{equation}
 \begin{split}
  R \cdot \epsilon &\sim \nabla^2 \epsilon \,, \label{eq:ScalarEoM}
 \end{split}
\end{equation}
where $R$ denotes the generalised Ricci tensor, $\epsilon$ some $H_d$ spinor, $\cdot$ a particular $H_d$ action, and $\nabla$ is a torsion-free $H_d$ connection. The crucial property for us is that the combination on the right-hand side is independent of the choice of torsion-free $H_d$ connection. Note that there is no unique torsion-free $H_d$ connection \cite{Coimbra:2011ky,Coimbra:2012af} unlike in the case of $O(d)$ structures in differential geometry.

Consider now the covariant derivative of a $\GH$ spinor on the internal space, $\varepsilon$, using a $\GH$ connection, $\tilde{\nabla}$, which in general will not be torsion-free. By definition,
\begin{equation}
 \tilde{\nabla} \varepsilon = 0 \,.
\end{equation}
However, since $\GH \subset H_d$, $\tilde{\nabla}$ is a torsion-full $H_d$ connection, and thus
\begin{equation}
 \tilde{\nabla} \varepsilon = \nabla \varepsilon + T \cdot \varepsilon \,,
\end{equation}
where $\nabla$ is a torsion-free $H_d$ connection, in general depending on the choice of $\tilde{\nabla}$, and $T$ denotes the torsion of $\tilde{\nabla}$. This implies that we can write
\begin{equation}
 \nabla \varepsilon = - T \cdot \varepsilon \,,
\end{equation}
where $T$ depends on $\nabla$ and thus implicitly on the choice of $\GH$ connection, $\tilde{\nabla}$. This implies that $T$ is not necessarily entirely intrinsic.

Now consider \eqref{eq:ScalarEoM} evaluated on a $\GH$ spinor. We have
\begin{equation}
 R \cdot \varepsilon = T^2 \cdot \varepsilon \,,
\end{equation}
independent of which $H_d$ connection we use. Thus, this is also clearly independent of the choice of $\GH$ connection and we see that $T^2$ is necessarily projected onto its intrinsic part. This implies that the generalised Ricci tensor can be expressed as the square ot the intrinsic torsion of the $\GH$ structure. As we showed in section \ref{s:ScalarTruncation}, the truncation Ansatz ensures that the $Y$-dependence of the intrinsic torsion factorises. One can use this to show that it also necessarily factorises in the equations of motion, expressed as the square of the intrinsic torsion. This suffices to show that the truncation Ansatz outlined in section \ref{s:ScalarAnsatz} subject to the conditions discussed in \ref{s:EmbTensor} leads to a consistent truncation of the original theory.

Note that this proof relied only on the fact that we have a supersymmetric truncation Ansatz. The amount of preserved supersymmetry was irrelevant. This means that our proof can be extended to other amounts of SUSY, using an analogous truncation Ansatz, and imposing the analogous conditions of section \ref{s:EmbTensor}, i.e. the closure condition, spinor condition and that the resulting embedding tensor is constant. It also extends to four-dimensional and chiral six-dimensional truncations which we will discuss in sections \ref{s:D=4} and \ref{s:D=6b}.

\subsection{Examples} \label{s:ExampleTruncation}
Let us now give an example of such a half-maximal consistent truncation, which comes from the truncation of M-theory on $K3 \times T^n$, where $K3$ need not be compact. We have already discussed how these backgrounds define a $\GH$ structure with vanishing torsion, and thus it is easy to define a consistent truncation on them.

\subsubsection{M-theory on $K3$} \label{s:K3Trunc}
The consistent truncation of M-theory on $K3$ is obtained by using the $\SO{3}$ structure defined by $K3$ in the truncation Ansatz \eqref{eq:ScalarAnsatz}. In particular, we take
\begin{equation}
 \begin{split}
  \rho^5 &= \sqrt{g} \,, \qquad n = \vol_{(4)} \,, \qquad \hat{n} = 1 \,, \qquad \omega_A = \Omega_A \,,
 \end{split}
\end{equation}
where $A = 1, 2, 3$. As we showed in section \ref{s:K3Example}, these objects have vanishing intrinsic torsion and thus we obtain a half-maximal seven-dimensional with only a gravitational supermultiplet.

\subsubsection{M-theory on $K3 \times S^1$} \label{s:K3S1Trunc}
We have already described in section \ref{s:K3S1Example} that $K3 \times S^1$ has a $\SO{4}$ structure. However, because the background is a trivial product of $K3$ with $S^1$, the structure group can be further reduced to $\SO{3}$. The $\SO{3}$ structure is given by
\begin{equation}
 \begin{split}
  \rho^4 &= \sqrt{g} \,, \qquad n = \vol_{(4)} \,, \qquad \hat{n} = \sigma \,, \qquad \omega_{U} = \Omega_U \,, \\
  \omega_4 &= \tilde{\sigma} + \vol_{(5)} \,, \qquad \omega_5 = \tilde{\sigma} - \vol_{(5)} \,,
 \end{split}
\end{equation}
where $U = 1, 2, 3$ and we use the same conventions as in section \ref{s:K3S1Example}.

Following section \ref{s:K3S1Example}, one can see that the $\omega_A$'s now satisfy
\begin{equation}
 \begin{split}
  \omega_A \wedge n &= 0 \,, \\
  \omega_A \wedge \omega_B &= \eta_{AB} n \,,
 \end{split}
\end{equation}
where $\eta_{AB} = \mathrm{diag}\left(1,1,1,1,-1\right)$ and thus they define a $\SO{3}$ structure. However, one can easily check that the intrinsic torsion of this $\SO{3}$ structure still vanishes, i.e.
\begin{equation}
 dn = d\hat{n} = \gL_{\omega_A} \omega_B = \gL_{\omega_A} \hat{n} = \gL_{\omega_A} \rho = 0 \,.
\end{equation}

This means that one can keep one vector multiplets in the consistent truncation on $K3 \times S^1$ and obtain a six-dimensional half-maximal non-chiral supergravity. This is not surprising: if we had first performed a consistent truncation on $K3$ as in section \ref{s:K3Trunc} and then performed a consistent truncation of this theory on $S^1$ we would have obtained a six-dimensional half-maximal SUGRA coupled to one vector multiplet.

\subsubsection{M-theory on $K3 \times T^2$} \label{s:K3T2STrunc}
In section \ref{s:K3T2}, we have discussed how $K3 \times T^2$ defines a $\SO{5}$ structure. Let us explicitly show that this background actually defines a $\SO{3}$ structure, since it is a product manifold of $K3$ with a generalised parallelisable space. The $\SO{3}$ structure is defined by
\begin{equation}
 \begin{split}
  \rho^3 &= \sqrt{g} \,, \qquad n = \vol_{(4)} \,, \qquad \hat{n} = \vol_{(2)} \,, \qquad \omega_U = \Omega_U \,, \\
  \omega_4 &= \sigma + \tilde{\sigma} \wedge \vol_{(4)} \,, \qquad \omega_5 = \sigma' + \tilde{\sigma}' \wedge \vol_{(4)} \,, \\
  \omega_6 &= \sigma - \tilde{\sigma} \wedge \vol_{(4)} \,, \qquad \omega_7 = \sigma' - \tilde{\sigma}' \wedge \vol_{(4)} \,,
 \end{split}
\end{equation}
where $U = 1, 2, 3$ and we use the same conventions as in section \ref{s:K3T2}. These sections indeed satisfy \eqref{eq:BackgroundCompat} with $\eta_{AB} = \mathrm{diag}\left(1,1,1,1,1,-1,-1\right)$.

Finally, it is easy to see that the intrinsic torsion of this $\SO{3}$ structure still vanishes. Thus, we obtain a consistent truncation to five-dimensional half-maximal SUGRA coupled to two vector multiplets. The fact that we have a product manifold of a hyperk\"ahler manifold with a generalised parallelisable manifold gives us the extra structure needed to keep the extra vector multiplets.

\subsection[Universal consistent truncations for half-maximal warped AdS and Mink vacua]{Universal consistent truncations for half-maximal warped AdS and \\Minkowski vacua} \label{s:UniTruncation}
Using the technology developed here we can prove that for any  half-maximal warped AdS$_D$ or Mink$_D$ solution of type II or 11-dimensional supergravity, there is a consistent truncation to a half-maximal gauged SUGRA in $D$ dimensions keeping only the gravitational supermultiplet. This proves a particular case of a conjecture of \cite{Gauntlett:2007ma} (the case where $D \geq 4$ and we have half-maximal supersymmetry), as well as a particular case of the conjecture of \cite{Duff:1985jd}. The case of supersymmetric seven-dimensional AdS vacua has been proven in \cite{Passias:2015gya} by explicit construction.

The proof is straightforward and here we will present it for $D \geq 5$, excluding chiral supersymmetry in six dimensions. For $D = 4$ we give the proof in section \ref{s:4DUniTruncation} and for chiral six-dimensional vacua in \ref{s:6DChiralConsTruncation}. As we have already discussed in section \ref{s:Vacua}, a half-maximal $AdS_D$ or Minkowski vacuum must have a ``weakly integrable'' $\GH$ structure. This means that its intrinsic torsion satisfies
\begin{equation}
 \begin{split}
  d\K &= \gL_{\J_u} \hK = \gL_{\J_u} \kappa^{D-2} = 0 \,, \\
  \gL_{\J_u} \J_v &= \bR_{uvw} \J^w \,, \label{eq:WeakIntegrabilityUni}
 \end{split}
\end{equation}
as well as
\begin{alignat}{2}
  d\hK &- \frac1{4} \epsilon_{uvw} \bR^{uvw} \K \,, \qquad &&\textrm{for } D = 7 \,, \nonumber \\
  d\hK &= - \frac1{18} \epsilon_{uvwx} \J^u \bR^{vwy} \,, \qquad &&\textrm{for } D = 6 \,, \label{eq:WeakIntegrabilityD67} \\
  \gL_{\hK} \J_u &= -\frac1{3\sqrt{2}} \epsilon_{uvwxy} \J^{y} \bR^{wxy} \,, \qquad \gL_{\hK} \K = \gL_{\hK} \kappa^3 = 0 \,, \qquad &&\textrm{for } D = 5 \,. \nonumber
\end{alignat}
Here $\bR_{uvw} = \kappa R_{2\,uvw}$, in terms of the intrinsic torsion given in \eqref{eq:JTorsionClasses}, must be \emph{constant}. It encodes the cosmological constant of the $AdS_D$ vacuum in a way that breaks the R-symmetry of the half-maximal SUGRA to match that of the appropriate lower-dimensional superconformal algebra. The case $\bR_{uvw} = 0$ corresponds to a Minkowski vacuum, as follows from \cite{Coimbra:2014uxa} and \cite{Coimbra:2016ydd}.

As we have already discussed the $\GH$ structure, and thus the sections $\K$, $\hK$, $\J_u$ and $\kappa$ only depend on the $Y^M$ coordinates and not those of the external space. In other words, they are in fact sections of the $\cRY_i$ and $\cSY$ bundles, and thus they define a $\SO{d-1}$ structure on $M_d$. As a result, we can use them to define a truncation around this vacuum, by taking $\rho = \kappa$, $n = \K$, $\hat{n} = \hK$ and $\omega_u = \J_u$ in \eqref{eq:BackgroundStructure}, \eqref{eq:BackgroundCompat} and \eqref{eq:ScalarAnsatz}, with $u = 1, \ldots, d-1$ and $\eta_{uv} = \delta_{uv}$.

From the weak integrability conditions \eqref{eq:WeakIntegrabilityUni}, \eqref{eq:WeakIntegrabilityD67} with \eqref{eq:EmbTensor}, we see that we fulfil the conditions necessary to have a consistent truncation. In particular, we obtain a half-maximal gauged SUGRA with embedding tensor
\begin{equation}
 f_{uvw} = \bR_{uvw} \,,
\end{equation}
and
\begin{alignat}{2}
 \theta &= - \frac14 \epsilon_{uvw} \bR^{uvw} \,, \qquad &&\textrm{for } D = 7 \,, \nonumber \\
 \theta_u &= - \frac1{18} \epsilon_{uvwx} \bR^{vwx} \,, \qquad &&\textrm{for } D = 6 \,, \\
 \xi_{uv} &= - \frac1{3\sqrt{2}} \epsilon_{uvwxy} \bR^{wxy} \,, \qquad &&\textrm{for } D = 5 \nonumber \,.
\end{alignat}

Note that a general half-maximal AdS and Minkowski vacuum will only give us enough structure to keep the gravitational multiplet, as we are doing here. Only when the vacuum admits more structure, can additional vector multiplets be kept, as we saw in the examples \ref{s:ExampleTruncation}.

\section{Relation to heterotic DFT}\label{s:HetConsTruncation}
We will now show how the technology developed here can be used to obtain $\SO{d-1,N}$ heterotic double field theory, with $N < d - 1$, from exceptional field theory. This will in particular allow us to see which consistent truncations of type II, or 11-dimensional, SUGRA can also be obtained from heterotic SUGRA. Because consistent truncations of type II and 11-dimensional SUGRA only keep $N < d-1$ vector multiplets, as we have shown in section \ref{s:ConsTruncation}, we can only obtain heterotic DFT with $N < d-1$ gauge fields. One may expect to be able to see the full number of gauge fields of the heterotic theory by studying effective theories, rather than consistent truncations. We leave this question to further work.

\subsection{The heterotic DFT reduction Ansatz}
To obtain heterotic double field theory from exceptional field theory, we clearly need to break half the supersymmetries, and thus our internal space must have a $\GH$ structure. If we assume that the internal space actually has a $\SO{d-1-N}$ structure, then we will obtain $N$ vector multiplets. These $\SO{d-1-N}$ structure is defined as in section \ref{s:ConsTruncation}, i.e. we have
\begin{equation}
 \left( \omega_A \,, \,\, n \,, \,\, \hat{n} \,, \,\, \rho \right) \,, \label{eq:BackgroundHetStructure}
\end{equation}
where $A = 1, \ldots, d-1 + N$ labels the vector representation of $\SO{d-1,N}$, and $\omega_A \in \cRY_1$, $n \in \cRY_2$, $\hat{n} \in \cRY_{D-4}$ and $\rho \in \cSY$. These bundles are defined over the background and thus \eqref{eq:BackgroundHetStructure} only depend on the $Y^M$ coordinates. These sections must satisfy the compatibility conditions
\begin{equation}
 \begin{split}
  \left( n \otimes n \right)\vert_{{\cal R}_c \otimes {\cal S}^4} &= 0 \,, \\
  \left( \hat{n} \otimes \hat{n} \right)\vert_{{\cal R}^*_c \otimes {\cal S}^{2D-8}} &= 0 \,, \\
  \omega_A \wedge n &= 0 \\ 
  \omega_A \wedge \omega_B &= \eta_{AB} n \,, \label{eq:HetBackgroundCompat}
 \end{split}
\end{equation}
where $\eta_{AB}$ is the invariant metric of $\SO{d-1,N}$. We further require the intrinsic torsion of this $\SO{d-1-N}$ structure to satisfy the same conditions as for consistent truncations \eqref{eq:EmbTensor}. Thus,
\begin{equation}
 \begin{split}
  d n &= f^A\, \omega_A \,, \\
  \gL_{\omega_A} \omega_B &\equiv X_{ABC}\, \omega^C = f_{ABC}\, \omega^C + f_{[A}\, \omega_{B]} + \frac12 \eta_{AB} f_C\, \omega^C \,, \\
  \gL_{\omega_A} \hat{n} &= \left( \xi_A - f_A \right) \hat{n} \,, \\
  \gL_{\omega_A} \rho^{D-2} &= \rho^{D-2}\, \xi_A \,, \label{eq:HetEmbTensor}
 \end{split}
\end{equation}
and the dimension-specific part
\begin{alignat}{2}
  d\hat{n} &= \theta\, n \,, \qquad && \qquad \textrm{for } D = 7\,, \nonumber \\
  d\hat{n} &= \theta^A \omega_A \,, && \qquad \textrm{for } D = 6 \,, \label{eq:D67HetEmbTensor}
\end{alignat}
and for $D = 5$,
\begin{equation}
 \begin{split}
  \gL_{\hat{n}} n &= \xi n \,, \quad \gL_{\hat{n}} \rho^3 = \xi \rho^3 \,, \\
  \gL_{\hat{n}} \omega_A &= \xi_{AB} \omega^B + \frac12 \xi \omega_A \,. \label{eq:D5HetEmbTensor}
 \end{split}
\end{equation}
with $f_{ABC}$, $f_A$, $\xi_A$, $\xi$, $\xi_{AB}$, $\theta_A$ and $\theta$ constant. While this is the most general set-up one can consider, any theory with $\xi_A \neq 0$ or $\xi \neq 0$ will not admit an action principle, just like for consistent truncations, while the deformations and gaugings $\xi_{AB}$, $\theta_A$ and $\theta$ do not have an obvious higher-dimensional origin in terms of heterotic string theory. Thus, it seems natural to remove these and hence require the $\SO{d-1,d-1}$ structure, $\rho$, $n$, $\hat{n}$ to be integrable.

We now use an Ansatz for the scalar fields of the exceptional field theory which is similar to that in used in section \ref{s:ConsTruncation},
\begin{equation}
 \begin{split}
  \langle \J_u{}^M \rangle (x,Y) &= b_u{}^A(x,Y)\, \omega_A{}^M(Y) \,, \\
  \langle \K \rangle (x,Y) &= n(Y) \,, \\
  \langle \hK \rangle (x,Y) &= e^{-2d}(x,Y)\, \hat{n}(Y) \,, \\
  \langle \kappa \rangle (x,Y) &= e^{-2d/(D-2)}(x,Y)\, \rho(Y) \,, \\
  \langle g_{\mu\nu} \rangle(x,Y) &= \hat{g}_{\mu\nu}(x,Y)\, e^{-4d/(D-2)}(x,Y)\, \rho^2(Y) \,, \label{eq:HetAnsatz}
 \end{split}
\end{equation}
with one of the key differences being that the scalar fields can now depend on both $x^\mu$ and $Y^M$. The above Ansatz will give an string-frame metric and make comparison easier with heterotic DFT. As we will see, the fields $e^{-2d}$ and $b_u{}^A$ become the generalised dilaton of the heterotic DFT and the left-moving frame fields.

For the gauge fields, we use the analogous Ans\"atze, i.e. \eqref{eq:7DGaugeAnsatz}, \eqref{eq:6DGaugeAnsatz} and \eqref{eq:5DGaugeAnsatz}, with the coefficients of the $\SO{d-1-N}$ structure again being allowed to depend on both $x^\mu$ and $Y^M$.  The analogous construction preserving maximal supersymmetry, i.e. based on a generalised Scherk-Schwartz Ansatz, has recently been used to obtain massive IIA \cite{Ciceri:2016dmd} and generalised IIB \cite{Baguet:2016prz} from exceptional field theory.

An arbitrary dependence on $Y^M$, however, is incompatible with half-maximal SUSY, and thus we need to impose further restrictions on the $Y^M$ dependence. In order to obtain a half-maximally supersymmetric theory, we need to ensure that the internal derivatives do not source spinors of $\SO{d-1}$, analogous to the condition required for consistent truncations. Further making use of the analogy with consistent truncations, we want to be able to expand the internal derivatives in terms of the $\SO{d-1-N}$ structure. This implies that we require
\begin{equation}
 \begin{split}
  \partial_M &= \rho^{D-2} \hat{\omega}^A{}_M \partial_A \,, \\
 \end{split} \label{eq:HetCoordDep1}
\end{equation}
where
\begin{equation}
 \begin{split}
  \partial_A = \omega_A{}^M \partial_M \,. \label{eq:HetCoordDep2}
 \end{split}
\end{equation}
Thus, the half-maximal theory can only depend on $d-1+N$ of the internal $Y^M$ coordinates.

While we have written these ``twisted'' derivatives, $\partial_A = \omega_A{}^M \partial_M$, as partial derivatives, they will in general not commute. However, we want to interpret them as coordinate derivatives and thus we must impose that
\begin{equation}
 \left[ \partial_A ,\, \partial_B \right] = 0 \,.
\end{equation}
Using the definition of the generalised Lie derivative in terms of the ``Y-tensor'' \cite{Berman:2012vc},
\begin{equation}
 \begin{split}
  \gL_{\omega_A} \omega_B{}^M &= \omega_A{}^N \partial_N \omega_B{}^M - \omega_B{}^N \partial_N \omega_A{}^M + Y^{MN}_{PQ} \omega_B{}^P \partial_N \omega_A{}^Q \,,
 \end{split}
\end{equation}
we can write
\begin{equation}
 \left[ \partial_A,\, \partial_B \right] = \gL_{\omega_A} \omega_B{}^M \partial_M - Y^{MN}_{PQ} \omega_B{}^P \partial_M \omega_A{}^Q \partial_N \,.
\end{equation}
Thus, we see that in order for the twisted derivatives, $\partial_A$, to commute, we require
\begin{equation}
 \gL_{\omega_A} \omega_B{}^M \partial_M = X_{AB}{}^C \partial_C = f_{AB}{}^C \partial_C + f_{[A} \partial_{B]} + \frac12 \eta_{AB} f^C \partial_C = 0 \,, \label{eq:HetfConstraint}
\end{equation}
as well as
\begin{equation}
 Y^{MN}_{PQ} \omega_B{}^P \partial_M \omega_A{}^Q \partial_N = 0 \,. \label{eq:HetMixedConstraint}
\end{equation}
To obtain the first condition we used \eqref{eq:EmbTensor}. These conditions should as usual be understood as acting on any of the field of the reduced theory that we thus obtain. If we take $f_A = 0$, as usual for the heterotic theory, then the first condition \eqref{eq:HetfConstraint} reduces to exactly that of the heterotic DFT \cite{Hohm:2011ex}
\begin{equation}
 f_{AB}{}^C \partial_C = 0 \,.
\end{equation}
The condition \eqref{eq:HetMixedConstraint} requires the section condition to hold when mixed between objects of the reduced theory and the ``twists'' $\omega_A{}^M$.

\subsection{Heterotic DFT generalised Lie derivative}
Let us see how the generalised Lie derivative reduces. Consider
\begin{equation}
 \begin{split}
  \langle V \rangle(x,Y) &= V^A(x,Y) \omega_A(Y) \,, \\
  \langle W \rangle(x,Y) &= W^A(x,Y) \omega_A(Y) \,,
 \end{split}
\end{equation}
with $V^A$ and $W^A$ satisfying \eqref{eq:HetCoordDep1} and \eqref{eq:HetCoordDep2}. Then we find
\begin{equation}
 \langle \gL_{V} W{}^M \rangle = \omega_A{}^M L^{(f)}_V W^A \,,
\end{equation}
where we have defined
\begin{equation}
 \begin{split}
  L_V^{(f)} W^A &= L_V W^A + f_{BC}{}^A V^B W^C + f_B V^{[B} W^{A]} + \frac12 V^B W_V f^A \,, \\
  L_V W^A &= V^B \partial_B W^A - W^B \partial_B V^A + W^B \partial^A V_B \,,
 \end{split} \label{eq:HetGenDiffeo}
\end{equation}
where we raise/lower the $A, B = 1, \ldots, d-1+N$ indices with $\eta_{AB}$. This is the gauged $\SO{d-1,N}$ double field theory generalised Lie derivative, where the gauge group is encoded in $f_{ABC}$ and $f_A$. This means that the intrinsic torsion of the $\GH$ structure on the internal space now defines the gauge group of the heterotic DFT rather than that of the lower-dimensional gauged SUGRA. It is also easy to check that the section condition of exceptional field theory reduces to that of the heterotic double field theory
\begin{equation}
 \eta^{AB} \partial_A \partial_B = 0 \,.
\end{equation}

We have thus obtained the heterotic double field theory. What is happening here is analogous to what happens when considering a Scherk-Schwarz reduction of double field theory, see for example \cite{Grana:2012rr}. The key difference is that here we obtain heterotic DFT from exceptional field theory, and thus must break half the supersymmetry. This difference is important if one wants to understand dualities between type II and heterotic theory.

It is easy to see that the fields $e^{-2d}$ and $b_u{}^A$ correspond to the generalised dilaton and the left-moving vielbeine. For example, let us find their transformation properties under the DFT generalised diffeomorphism. These can be found by acting with the generalised Lie derivative on the $\SO{d-1}$ structure. One finds
\begin{equation}
 \begin{split}
  \langle \gL_{V} \J_u \rangle &= \left( L^{(f)}_{\tilde{V}} b_u{}^A \right) \omega_A \,, \qquad
  \langle \hK \rangle = \left( L^{(\xi)}_{\tilde{V}} e^{-2d} \right) \hat{n} \,,
 \end{split}
\end{equation}
where $V = \tilde{V}^A \omega_A$, $L^{(f)}$ is given in \eqref{eq:HetGenDiffeo} and
\begin{equation}
 L^{(\xi)}_V e^{-2d} = \partial_A \left( e^{-2d} V^A \right) + e^{-2d} \xi_A V^A \,,
\end{equation}
is indeed the action of the heterotic generalised Lie derivative on the generalised dilaton. Furthermore, the $\SO{d-1}_R$ symmetry becomes part of the generalised Lorentz symmetry of the heterotic DFT. The $\SO{d-1}_R$-invariant combination
\begin{equation}
 \gH_{AB} = \eta_{AB} - 2 b^u{}_A b_{u\,B} \,,
\end{equation}
is the generalised metric of the heterotic DFT. This shows that the $b_u{}^A$ are the generalised frame fields of the heterotic theory \cite{Hohm:2010xe,Hohm:2011ex}.

\subsection{Intrinsic torsion and scalar potential}
Using the half-maximal reformulation of exceptional field theory that we developed in section \ref{s:4DConsTruncations} one can now proceed to calculate the reduction of the entire action. This yields the heterotic DFT action. Let us exemplify this by calculating the reduction of the scalar potential, which gives the heterotic generalised Ricci scalar. The reduction of the kinetic and topological terms are so similar to what happened in section \ref{s:ConsTruncation} when applying a consistent truncations that we will not go through the details here. An example of how this is done for $D =7$ can be found in \cite{Malek:2016vsh}.

We begin by calculating the reduction of the intrinsic torsion of the $\GH$ structure. Its universal components become
\begin{equation}
 \begin{split}
  \langle T_1 \rangle &= \rho^{-2} e^{4d/(D-2)} P_{+}^{AB} f_A \omega_B \,, \\
  \langle T_2{}^u \rangle &= \rho^{-1} e^{2d/(D-2)} b_u{}^A f_A \,, \\
  \langle R_{1\,uv} \rangle &= \rho^{-2} e^{4d/(D-2)} P_+^{AB} \Omega^{(f)}_{uv\,A} \omega_B \,, \\
  \langle R_{2\,uvw} \rangle &= \rho^{-1} e^{2d/(D-2)} \Omega^{(f)}_{uvw} \,, \\
  \langle U_u \rangle &= \rho^{-1} e^{2d/(D-2)} \Omega_u \,, \\
  \langle T_3 \rangle &= \langle S_{1\,u} \rangle = \langle S_2 \rangle = 0 \,, \label{eq:HetITUni}
 \end{split}
\end{equation}
where we have defined
\begin{equation}
 \begin{split}
  \Omega^{(f)}_{uv\,A} &= L^{(f)}_{b_[u} b_{v]A} \,, \\
  \Omega^{(f)}_{uvw} &= b_w{}^A \Omega^{(f)}_{uv\,A} = b_{w\,A} L^{(f)}_{b_{[u}} b_{v]}{}^A \,, \\
  \Omega^{(\xi)}_u &= e^{2d} L^{(\xi)}_{b_u} e^{-2d} \,.
 \end{split}
\end{equation}
Note that the generalised Lie derivative acts as usual on a scalar density as
\begin{equation}
 \begin{split}
  L_V e^{-2d} &= \partial_A \left(V^A e^{-2d} \right) \,.
 \end{split}
\end{equation}

For the dimension-dependent parts of the intrinsic torsion one finds
\begin{alignat}{3}
 \langle P_1 \rangle &= e^{8d/5} \theta \,, \quad \langle P_2 \rangle &&= 0 \,, \quad&& \textrm{ for } D = 7 \,, \\
 \langle P_1 \rangle &= \rho^{-2} e^{-d} P_+^{AB} \theta_A \omega_B \,, \quad \langle P_2{}^u \rangle &&= \rho^{-1} e^{-3d/2} b^{u\,A} \theta_A \,, \quad&& \textrm{ for } D = 6 \,,
\end{alignat}
and for $D=5$
\begin{equation}
 \begin{split}
  \langle P_{3\,u} \rangle &= e^{-2d/3} \rho^{-2}b_u{}^A P_+^{BC} \xi_{AC} \omega_B \,, \qquad \langle P_{4\,uv} \rangle = \rho^{-1} e^{-4d/3} b_u{}^A b_v{}^B \xi_{AB} \,, \\
  \langle P_1 \rangle &= 0 \,, \qquad \qquad \langle P_2 \rangle = 0 \,.
 \end{split}
\end{equation}
We have here taken $\xi = 0$ as this is a trombone gauging for which one does not obtain an action principle. We can already see that the gaugings $\theta$, $\theta_A$ and $\xi_{AB}$ are problematic. Comparing with \eqref{eq:HetITUni} we see that these gaugings must have a non-trivial weight under generalised diffeomorphisms, and thus cannot be viewed as parameters. This once again highlights that their interpretation in the heterotic theory is troublesome.

Using these results we find that the universal part of the scalar potential, \eqref{eq:PotUniversal}, reduces to
\begin{equation}
  \begin{split}
   \langle |e| V_0 \rangle &= \rho^{D-2} |\bae| e^{-2d} \left( \frac13 \Omega^{(f)}_{uvw} \Omega^{(f)uvw} + P_+^{AB} \Omega^{(f)}_{uv\,A} \Omega^{(f)uv}{}_B - 2 \Omega_u \Omega^{u} \right. \\
   & \left. - 4 b_u{}^A \partial_A \Omega^{u} - 2 \Omega_u b^{u\,A} f_A + \gH^{AB} f_A f_B \right) \,, \label{eq:HetPotUniTruncation}
  \end{split}
\end{equation}
where we have taken $\xi_A = 0$, as otherwise the theory does not admit an action. The first line in \eqref{eq:HetPotUniTruncation} corresponds to the usual generalised Ricci scalar of heterotic DFT, i.e. where only the gaugings $f_{ABC} \neq 0$, are considered, in the frame formulation \cite{Siegel:1993th,Siegel:1993xq,Hohm:2011ex}.

We can also easily calculate the dimension-dependent parts. These are, up to a conformal transformation of the external metric, unmodified from the case when considered a consistent truncation, equation \eqref{eq:PotDimDepTruncation}. Explicitly, we find
\begin{equation}
 \small
 \begin{split}
  \langle |e| V_7 \rangle &= \rho^{5} |\bae| \left( \frac1{16} e^{-4d/5} \theta \epsilon^{uvw} \Omega^{(f)}_{uvw} - \frac1{16} e^{2d/5} \theta^2 \right) \,, \\
  \langle |e| V_6 \rangle &= \rho^4 |\bae| \left( -8 e^{-6d} \eta^{AB} \theta_A \theta_B + \frac{16}{3} e^{-4d}\, b_u{}^A \Omega^{(f)}_{vwx} \epsilon^{uvwx} \theta_A + e^{-4d} \left( \gamma P_{-}^{AB} + \lambda P_+^{AB} \right) f_A \theta_B \right) \,, \\
  \langle |e| V_5 \rangle &= \rho^3 |\bae| \left[ \frac14 e^{2d/3}\, \left( \gH^{AB} \gH^{CD} - \eta^{AB} \eta^{CD} \right) \xi_{AC} \xi_{BD} + \frac{\sqrt{2}}{3} e^{8d/3}\, b_u{}^A b_v{}^B \Omega^{(f)}_{wxy} \epsilon^{uvwxy} \xi_{AB} \right] \,.
 \end{split}
 \normalsize
\end{equation}
Once again we see that the gaugings $\theta$, $\theta_A$ and $\xi_{AB}$ must have non-vanishing weights under the generalised Lie derivative and thus it is not clear whether they can be interpreted as gaugings or deformations of the theory. This is perhaps not surprising as these gaugings and deformations do not have an obvious higher-dimensional origin in terms of the heterotic theory.

\subsection{M-theory / heterotic duality} \label{s:HetDuality}
We showed that EFT can be reduced to heterotic DFT by expanding the EFT fields in terms of a background $\SO{d-1-N}$ structure, just like for consistent truncations. However, the would-be lower-dimensional fields are now still allowed to depend on the internal coordinates $Y^M$ subject to the constraints
\begin{equation}
 \begin{split}
  \omega_M &= \rho^{D-2} \omega^A_M \omega_A{}^N \partial_N \,, \\
  X_{AB}{}^C \partial_C &= 0 \,, \\
  Y^{MN}_{PQ} \omega_A{}^P \partial_M \omega_A{}^Q \partial_N &= 0 \,,
 \end{split} \label{eq:HetConditions1}
\end{equation}
as well as the section condition
\begin{equation}
 \eta^{AB} \partial_A \otimes \partial_B = \eta^{AB} \partial_A \partial_B = 0 \,, \label{eq:HetConditions2}
\end{equation}
when acting on any of the fields of the heterotic DFT. Furthermore, the intrinsic torsion of the $\SO{d-1-N}$ structure, $\rho$, $n$, $\hat{n}$ and $\omega_A$, becomes the gauge group of the heterotic DFT. We already indicated that one should also require the $\SO{d-1,d-1}$ structure, defined by $\rho$, $n$, $\hat{n}$, to be integrable, as otherwise the deformations do not have an obvious interpretation in terms of the heterotic string.

Consider now a consistent truncation of M-theory, or type II string theory, defined by some $\SO{d-1-N}$ structure, that is by the tensors $\rho(Y)$, $n(Y)$, $\hat{n}(Y)$ and $\omega_A(Y)$, and its intrinsic torsion. Next, consider the heterotic theory obtained by the above procedure with the fields satisfying \eqref{eq:HetConditions1} and \eqref{eq:HetConditions2}. If these conditions can be solved by allowing the fields to depend on $d-1$ coordinates, then the consistent truncation of M-theory / type II also arises as a truncation of the heterotic theory, and thus these theories are dual on such a space. This happens, for example, when considering the truncation on $K3$ as in section \ref{s:ExampleTruncation}.

It should be noted that because here we are working with consistent truncations, we are only seeing the 10-dimensional heterotic theory with a small number of vector multiplets. This guarantees that both theories give the same lower-dimensional half-maximal SUGRA. For instance, in the above example of M-theory on $K3 \times T^{d-4}$, one obtains an ungauged $\left(11-d\right)$-dimensional SUGRA with scalar coset space $\frac{\SO{d-1,d-4}}{\SO{d-1}\times\SO{d-4}}$ and thus the dual heterotic theory on $T^{d-1}$ is already truncated to have only $d-4$ vector multiplets in ten dimensions. This answers some of the questions raised in \cite{Lu:1998xt} which studied the M-theory / heterotic duality in the context of consistent truncations.

It would be interesting to try and investigate the relation between EFT and heterotic DFT from an effective viewpoint where one would hope to see the duality between the low-energy theories of the M-theory and heterotic compactifications. In particular, this should allow one to obtain the full heterotic theory with $16$ vector multiplets.

\subsection{Modified $\SO{5,N}$ double field theory} \label{s:ModDFT}
Recall that in five dimensions, there is an extra vector field in the half-maximal spectrum \cite{Schon:2006kz}. In the consistent truncation this arises because the dilaton structure contains a generalised vector field, $\hK$, equivalently $\hat{n}$ in the consistent truncation. As we have shown in \ref{s:ConsTruncation}, this allowed us to expand generalised vectors of the exceptional field theory in terms of $\hat{n}$, in addition to the usual $\omega_A$ vector fields from the $\SO{5-N}$ structure.

This leads to the question if we can also associate an extra coordinate with this vector field in a modified half-maximal $\SO{5,N}$ double field theory. From the set-up outlined in the previous section, this arises if we slightly relax the constraint \eqref{eq:HetCoordDep1} and \eqref{eq:HetCoordDep2} to become
\begin{equation}
 \partial_M = \rho^{-3} n_M \partial_0 + \rho^{-3} \hat{\omega}^A{}_M \partial_A \,, \label{eq:ModHetDependence}
\end{equation}
when acting on any field of our half-maximal theory. In other words, we are now allowing for the possibility that
\begin{equation}
 \partial_0 \equiv \rho\, \hat{n}^M \partial_M \neq 0 \,,
\end{equation}
when acting on any field of this half-maximal theory. In particular, this relaxation is compatible with half-maximal supersymmetry. Note that we can take $N$ arbitrary large, if we ignore its origin as a consistent truncation of $\EG{6}$.

Let us now follow the procedure outlined above with the modified constraint \eqref{eq:ModHetDependence}. We will denote a generalised vector field of this theory, which has $6+N$ components, by ${\cal V}^{\cal A} = \left(v,\, V^A \right)$ with $A = 1, \ldots, 5+N$. From the truncation of the exceptional field theory we now obtain the generalised Lie derivative
\begin{equation}
 \gL_{{\cal V}} {\cal W}^{\cal A} = \left( \gL_{{\cal V}} w ,\, \gL_{{\cal V}} W^A \right) \,,
\end{equation}
given by
\begin{equation}
 \begin{split}
  \gL_{{\cal V}} w &= L^\xi_{V} v \equiv \partial_A \left( V^A v \right) + V^A v \xi_A \,, \\
  \gL_{{\cal V}} W^A &= L^f_V W^A + \partial_0\left( v W^A \right) + v W_B \xi^{BA} + \frac12 v W^A \xi \,, \label{eq:ModifiedGeneralisedLieDerivative}
 \end{split}
\end{equation}
where
\begin{equation}
 L^f_V W^A = V^B \partial_B W^A - W^B \partial_B V^A + W^B \partial^A V_B + f_{BC}{}^A V^B W^C \,,
\end{equation}
is the usual generalised Lie derivative of heterotic double field theory with gauge group defined by $f_{BC}{}^A$. Note that the singlet components, $v$, of the modified generalised vector field transform as scalar densities under the usual DFT generalised Lie derivative.

Let us for the remainder focus solely on the ``undeformed theory'' where $f_{ABC} = f_A = \xi_A = \xi_{AB} = \xi = 0$. The section condition for this theory is obtained from
\begin{equation}
 d^{MNP} \partial_M \otimes \partial_N = 0 \,,
\end{equation}
which gives
\begin{equation}
 \eta^{AB} \partial_A \otimes \partial_B = \partial_A \otimes \partial_0 + \partial_0 \otimes \partial_A = 0 \,,
\end{equation}
where as usual we are taking the derivatives to act on arbitrary products of fields, or as double derivatives on any one field of our theory. We see that there are two distinct solutions:
\begin{equation}
 \begin{split}
 (a) & \quad \partial_0 = 0 \,, \qquad \eta^{AB} \partial_A \otimes \partial_B = 0 \,, \\
 (b) & \quad \partial_A = 0 \,, \qquad \partial_0 \neq 0 \,.
 \end{split}
\end{equation}
Solutions of type (a) correspond to those of the usual $\SO{5,N}$ heterotic DFT coming from ten dimensions, while those of type (b) correspond to a 5+1 split of half-maximal six-dimensional SUGRA. In this case the $\SO{5,N}$ symmetry is unbroken in six dimensions and thus we are led to identify this theory with the six-dimensional ${\cal N}=\left(2,0\right)$ SUGRA coupled to $N$ tensor multiplets. The tensor fields have in this description been dualised into five-dimensional vector fields.

We see that this modified $\SO{5,N}$ double field theory, which is contained inside the $\EG{6}$ EFT, unifies 10-dimensional half-maximal SUGRA with ${\cal N}=\left(2,0\right)$ SUGRA in six dimensions. This is analogous to what happens in double field theory at $\SL{2}$ angles \cite{Ciceri:2016hup}. It would be interesting to investigate the role of the deformations $f_{ABC}$, $f_A$, $\xi_A$, $\xi_{AB}$ and $\xi$ in this theory. One might imagine that when using solution (b) of the section condition, they are related to the A-D-E gauge group of the self-dual strings of the chiral six-dimensional theory \cite{Witten:1995ex}.

As outlined in \ref{s:HetConsTruncation}, one can determine the action of this theory from the $\EG{6}$ EFT action. We leave this, and further study of this theory for future work. Let us end this section by mentioning that the analogous construction in four dimensions would allow dependence on $\SL{2}$ copies of $6+N$ coordinates, would reduce $\EG{7}$ exceptional field theory to ``double field theory at $\SL{2}$ angles'' \cite{Ciceri:2016hup}.

\section{Four-dimensional half-maximal SUGRA} \label{s:D=4}
Let us now turn our attention to four-dimensional half-maximal supergravities obtained from exceptional field theory. The necessary exceptional generalised structure group is in this case
\begin{equation}
 \GH = \SU{4} \simeq \SO{6} \,.
\end{equation}
We will now describe this structure group bosonically in $\EG{7}$ EFT. As we already mentioned, this will be similar to the pattern discussed in section \ref{s:GHalfDgeq5}, but not identical to it.

In particular, we will make use of the ${\cal R}_1$, ${\cal R}_2$ and ${\cal R}_3$ bundles, of weight $\frac12$, $1$ and $\frac32$, respectively, under the generalised Lie derivative, whose fibres are the vector spaces listed in table \ref{t:TensorHierarchy}. We will also use a wedge product which maps
\begin{equation}
 \begin{split}
  R_1 \wedge R_1 &\longrightarrow R_2 \,, \\
  R_1 \wedge_S R_1 &\longrightarrow \mbf{1} \,, \\
  R_1 \wedge R_2 &\longrightarrow R_3 \,,
 \end{split}
\end{equation}
and similarly for the corresponding bundle maps. This is similar to the notation in \ref{s:DiffForms}, except that we denote the wedge product onto the adjoint without the subscript $P$ and instead add a subscript $S$ for the product onto the singlet. Explicitly we have for $A_1, A_2 \in R_1$ and $B \in R_2$,
\begin{equation}
 \begin{split}
  \left( A_1 \wedge A_1 \right)^\alpha &= A_1{}^M A_2{}^N \left(t^\alpha\right)_{MN} \,, \\
  \left( A_1 \wedge_S A_1 \right) &= A_1{}^M A_2{}^N \Omega_{MN} \,, \\
  \left( A \wedge B \right)^{M\,\alpha} &= \left(\mathbb{P}_{\mbf{912}}\right)^{M\,\alpha}{}_{N\,\beta} A{}^N B^\beta \,. \label{eq:4DWedge}
 \end{split}
\end{equation}
Here $M, N = 1, \ldots, 56$ labels the fundamental representation of $\EG{7}$, $\alpha = 1, \ldots, 133$ labels the adjoint representation, $t_{\alpha}$ are the generators of $\EG{7}$, $\Omega_{MN}$ is the symplectic invariant of $\EG{7}$ and
\begin{equation}
 \left(\mathbb{P}_{\mbf{912}}\right)^{M\,\alpha}{}_{N\,\beta} = - \frac{12}{7} \left(t_\beta\right)^M{}_K \left(t^\alpha\right)_N{}^K + \frac47 \left(t_\beta\right)_N{}^K \left(t^\alpha\right)_K{}^M + \frac17 \delta^M_N \delta^\alpha_\beta \,,
\end{equation}
is the projector onto the $\mbf{912}$, as given in \cite{deWit:2002vt}.

The $\EG{7}$ generalised Lie derivative is given by \cite{Berman:2012vc,Coimbra:2012af,Hohm:2013uia}
\begin{equation}
 \begin{split}
  \gL_{\Lambda} A^M &= \Lambda^N \partial_N A^M - 12\, \left(\mathbb{P}_{\mbf{133}}\right)^M{}_N{}^K{}_L A^N \partial_K \Lambda^L + \frac12 A^M \partial_N \Lambda^N \,, \\
  \gL_{\Lambda} B_\alpha &= \Lambda^N \partial_N B_\alpha + 12 f_{\alpha\beta}{}^\gamma \left(t_\beta\right)_L{}^K B_\gamma \partial_K \Lambda^L + B_\alpha \partial_N \Lambda^N \,,
 \end{split}
\end{equation}
where $A \in \Gamma\left({\cal R}_1\right)$ and $B \in \Gamma\left({\cal R}_2\right)$, $f_{\alpha\beta}{}^\gamma$ are the $\EG{7}$ structure constants and
\begin{equation}
 \begin{split}
  \left(\mathbb{P}_{\mbf{133}}\right)^M{}_N{}^K{}_L &= \left(t_\alpha\right)^M{}_N \left(t^\alpha\right)^K{}_L \\
  &= \frac1{24} \delta^M_N \delta^K_L + \frac1{12} \delta^M_L \delta^K_N + \left(t_\alpha\right)^{MK} \left(t_\alpha\right)_{NL} - \frac1{24} \Omega^{MK} \Omega_{NL} \,,
 \end{split}
\end{equation}
is the projector onto the adjoint.

Throughout we will be raising and lowering fundamental $\EG{7}$ indices with $\Omega_{MN}$ and using a north-west south-east convention, i.e.
\begin{equation}
 V^M = \Omega^{MN} V_N \,, \qquad V_M = V^N \Omega_{NM} \,,
\end{equation}
with
\begin{equation}
 \Omega^{MK} \Omega_{NK} = \delta^M_K \,.
\end{equation}
The adjoint indices $\alpha, \beta = 1,\, \ldots, \, 133$ are raised/lowered with the Killing metric
\begin{equation}
 \kappa_{\alpha\beta} = \left(t_\alpha\right)^M{}_N \left(t_\beta\right)^N{}_M \,.
\end{equation}

\subsection{$\SO{6}$ structure and intrinsic torsion}
\subsubsection{Axio-dilaton structure} \label{s:4DAxioDilStructure}
We begin by observing that
\begin{equation}
 \SO{6} \subset \SO{6,6} \subset E_{7(7)} \times \mathbb{R}^+ \,,
\end{equation}
and thus we begin by describing a $\SO{6,6}$ structure. Here we see the first difference to the case $D \geq 5$ since the maximal commutant of $\SO{6,6} \subset \EG{7}$ is $\SL{2}$ not $\UO$. In order to break the $\SL{2}$ we now need a $\SL{2}$ \emph{triplet} of sections of the ${\cal R}_2$ bundle, satisfying certain compatibility conditions. From table \ref{t:TensorHierarchy} we see that the ${\cal R}_2$ bundle is in this case the adjoint bundle, of weight $1$ under the generalised Lie derivative.

Thus, a $\SO{6,6}$ structure is defined by a triplet of adjoint fields $K_{ij}{}^\alpha$, with $K_{[ij]}{}^\alpha = 0$ and where $\alpha = 1, \ldots, 133$ labels the adjoint of $\EG{7}$ and $i, j = 1, 2$ are fundamental $\SL{2}$ indices. The compatibility conditions that these must satisfy are
\begin{equation}
 \begin{split}
  \left( \K_{ij} \otimes \K_{kl} \right)\vert_{1539} &= 0 \,, \\
  \left[ \K_{ij} ,\, \K_{kl} \right] &= -2 \kappa^2 \left[ \epsilon_{i(k} \K_{l)j} + \epsilon_{j(k} \K_{l)i} \right] \,, \\
  \tr \left( \K_{ij} \K_{kl} \right) &= \K_{ij}{}^\alpha \K_{kl\,\alpha} = 12 \kappa^4 \epsilon_{i(k} \epsilon_{l)j} \,, \label{eq:4DAxioCompat}
 \end{split}
\end{equation}
where $\kappa$ is a scalar density of weight $\frac12$, and $\epsilon_{ij}$ is the $\SL{2}$ invariant antisymmetric tensor. We will use this to raise/lower $\SL{2}$ indices according to the north-west south-east convention:
\begin{equation}
 v^i = \epsilon^{ij} v_j \,, \qquad v_i = v^j \epsilon_{ji} \,,
\end{equation}
with
\begin{equation}
 \epsilon^{ik} \epsilon_{jk} = \delta^i_k \,.
\end{equation}
Under the decomposition $\EG{7} \longrightarrow \SO{6,6} \times \SL{2}$, the adjoint branches as
\begin{equation}
 \mbf{133} \longrightarrow \dbf{66}{1} \oplus \dbf{32}{2} \oplus \dbf{1}{3} \,.
\end{equation}
The conditions \eqref{eq:4DAxioCompat} imply that the $\K_{ij}$'s correspond to the $\dbf{1}{3}$. It is easy to check that this breaks $\EG{7} \longrightarrow \SO{6,6}$.

As we will see, upon reduction to half-maximal supergravity, the $\SO{6,6}$ structure contains the degrees of freedom of the four-dimensional axio-dilaton. Thus we will also call the $\SO{6,6}$ structure an axio-dilaton structure.

Let us at this stage also mention that four-dimensional half-maximal supergravities have $\SU{4} \times \UO$ R-symmetry. The $\UO$ generator is contained inside the axio-dilaton structure as
\begin{equation}
 \hK = \delta^{ij} \K_{ij} \,.
\end{equation}
From \eqref{eq:4DAxioCompat} one can see that $\hK$ is anti-Hermitian and thus must correspond to the $\UO \subset \SL{2}$ generator.

The axio-dilaton structure here is very similar to the ${\cal N}=2$ hypermultiplet structure \cite{Ashmore:2015joa}: the two cases define $\mathrm{SO}^*(12)$ and $\SO{6,6}$ structure groups, respectively. These are different real forms of $\SO{12}$, and are thus related by analytic continuation.\footnote{When comparing our formulae to \cite{Ashmore:2015joa} it is important to note that we are following the conventions of \cite{Hohm:2013uia} and hence our traces, in particular, differ.} However, one should not be fooled into thinking that under the embedding ${\cal N}=2 \rightarrow {\cal N}=4$, the axio-dilaton structure reduces to the hypermultiplet structure.

\subsubsection{$\SO{6}$ structure} \label{s:4DSO6Structure}
In order to further break the structure group to $\SO{6} \subset \SO{6,6} \subset \EG{7} \times \mathbb{R}^+$ we need to introduce a further twelve generalised vector fields $\J_{u\,i} \in \Gamma\left({\cal R}_1\right)$ where $u = 1, \ldots, 6$ labels the vector representation of $\SO{6}_R$, the $\SO{6}$ R-symmetry group, while $i = 1, 2$ transforms under the $\SL{2}$ group generated by the axio-dilaton structure $\K_{ij}$. We will throughout be raising/lowering the $\SO{6}_R$ indices with $\delta_{uv}$.

The generalised vector fields are subject to a compatibility requirement which is very similar to that in section \ref{s:GHalfDgeq5}. We require that
\begin{equation}
 \begin{split}
  \J_{u\,i} \wedge \K_{jk} &= 0 \,, \\
  \J_{u\,i} \wedge \J_{v\,j} &= \delta_{uv} \K_{ij} + \epsilon_{ij} \J_{uv} \,, \label{eq:4DSO6Compat}
 \end{split}
\end{equation}
where $\wedge$ has been defined in \eqref{eq:4DWedge}, and the $\J_{uv}$ are $\SO{6}_R$ generators, similar to what we found for $D \geq 5$. These conditions can be understood as follows. When decomposing $\EG{7} \longrightarrow \SO{6,6} \times \SL{2}$, the fundamental representation branches as
\begin{equation}
 \mbf{56} \longrightarrow \dbf{12}{2} \oplus \dbf{32'}{1} \,.
\end{equation}
The first condition in \eqref{eq:4DSO6Compat} implies that $\J_{u\,i} \in \dbf{12}{2}$. Decomposing further under
\begin{equation}
 \SO{6}_S \times \SO{6}_R \times \SL{2} \subset \SO{6,6} \times \SL{2} \subset \EG{7} \,,
\end{equation}
where the subscripts $S / R$ stands for the structure and R-symmetry group, respectively, one finds that
\begin{equation}
 \dbf{12}{2} \longrightarrow \tbf{6}{1}{2} \oplus \tbf{1}{6}{2} \,.
\end{equation}
These two representations will appear in the second equation of \eqref{eq:4DSO6Compat} with opposite signs and thus we now find that
\begin{equation}
 \J_{u\,i} \in \tbf{1}{6}{2} \,.
\end{equation}
Twelve such vectors, together with the axio-dilaton structure break $\EG{7} \times \mathbb{R}^+ \longrightarrow \SO{6}_S$.

From the compatibility conditions \eqref{eq:4DAxioCompat} and \eqref{eq:4DSO6Compat} it follows that the $\J_{u\,i}$ completely determine the $\K_{ij}$, as well as $\J_{uv}$, via
\begin{equation}
 \begin{split}
  \K_{ij} &= \frac16 \J^u{}_i \wedge \J_{u\,j} \,, \\
  \J_{uv} &= \frac12 \J_{u\,i} \wedge \J_{v}{}^{i} \,,
 \end{split}
\end{equation}
and that $\K_{ij}$ and $\J_{uv}$ act as $\SL{2}$ and $\SO{6}_R$ transformations on the $\J_{u\,i}$, i.e.
\begin{equation}
 \begin{split}
  \K_{ij} \cdot \J_{u\,k} &= 2 \kappa^2 \epsilon_{k(i} \J_{|u|j)} \,, \\
  \J_{uv} \cdot \J_{w\,i} &= - \kappa^2 \delta_{w[u} \J_{v]i} \,,
 \end{split}
\end{equation}
where we use the north-west south-east convention for raising/lowering indices, e.g.
\begin{equation}
 \left( \K_{ij} \cdot \J_{u\,k} \right)^M = \K_{ij}{}^\alpha \left(t_\alpha\right)^{MN} \J_{u\,k\,N} \,.
\end{equation}
The compatibility requirements also imply that
\begin{equation}
 \J_{u\,i} \wedge_S \J_{v\,j} = - 6 \kappa^2 \delta_{uv} \epsilon_{ij} \,.
\end{equation}

It is worth pointing out that the twelve $\J_{u\,i}$ completely determine all three $\K_{ij}$'s. This is different to the situation for $D \geq 5$ (see section \ref{s:GHalfDgeq5}), where the $\J_u$'s only determined $\K$ but not $\hK$.

\subsubsection{Intrinsic torsion of the axio-dilaton structure} \label{s:4DAxioDilTorsion}
Just as for $D \geq 5$ we now first find the intrinsic torsion of the axio-dilaton structure. Let us first calculate what representations are expected in the intrinsic torsion, following \cite{Coimbra:2014uxa}. The space of torsions is given by
\begin{equation}
 \begin{split}
  W &= \mbf{912} \oplus \mbf{56} \\
  &= \dbf{352'}{1} \oplus \dbf{220}{2} \oplus 2 \cdot \dbf{12}{2} \oplus \dbf{32'}{3} \oplus \dbf{32'}{1} \,,
 \end{split}
\end{equation}
where we have decomposed $\EG{7} \longrightarrow \SO{6,6} \times \SL{2}$. On the other hand, the space of $\SO{6,6}$ connections is
\begin{equation}
 \begin{split}
  K_{\SO{6,6}} &= \left( \dbf{12}{2} \oplus \dbf{32'}{1} \right) \otimes \dbf{66}{1} \\
  &= \dbf{1728'}{1} \oplus \dbf{560}{2} \oplus \dbf{352'}{1} \oplus \dbf{220}{2} \oplus \dbf{32'}{1} \oplus \dbf{12}{2} \,.
 \end{split}
\end{equation}
As a result, the image of the torsion map $\tau_{\SO{6,6}}: K_{\SO{6,6}} \longrightarrow W$ is
\begin{equation}
 \mathrm{Im} \tau_{\SO{6,6}} = \dbf{352'}{1} \oplus \dbf{220}{2} \oplus \dbf{32'}{1} \oplus \dbf{12}{2} \,.
\end{equation}
Finally, we find that the intrinsic torsion lies in
\begin{equation}
 \begin{split}
  W_{\SO{6,6}} &= W / \mathrm{Im}\tau_{\SO{6,6}} \,, \\
  &= \dbf{12}{2} \oplus \dbf{32'}{3} \,. \label{eq:4DAxioWInt}
 \end{split}
\end{equation}

We now want to find explicit expressions for the intrinsic torsion. Just like in section \ref{s:IntTorsion}, it will be given by tensors built out of one derivative of the axio-dilaton structure $\left(\K_{ij},\kappa\right)$. To do this, we will use the combination
\begin{equation}
 \left( d\K_{ij} \right)^M = - 12 \left(t_\alpha\right)^{MN} \partial_N \K_{ij}{}^\alpha - \frac12 \Omega^{MN} W_{ij,N} \,, \label{eq:4DdK}
\end{equation}
where $W_{ij,N}$ is a set of three compensator fields, as introduced in \cite{Hohm:2013uia}: For a tensor $T \in \Gamma\left({\cal R}_2\right)$, one can construct a covariant derivative by taking
\begin{equation}
 \left( dT \right)^M = - 12 \left(t_\alpha\right)^{MN} \partial_N T^\alpha - \frac12 \Omega^{MN} W_N \,, \label{eq:4DdAdjoint}
\end{equation}
where the compensator field, $W$, must satisfy
\begin{equation}
 \left(t_\alpha\right)^{MN} W_{M} \partial_N = \omega^{MN} W_{M} \partial_N = \left( t_\alpha \right)^{MN} W_{M} W_{N} = 0 \,,
\end{equation}
and must have the following anomalous transformation under the generalised Lie derivative
\begin{equation}
 \Delta_\Lambda W_{M} = - 24 \left(t^\alpha\right)_P{}^N T_\alpha \partial_M \partial_N \Lambda^P \,.
\end{equation}
As shown in \cite{Hohm:2013uia}, the combination \eqref{eq:4DdK} is then a generalised vector field.

For a general tensor in $T \in \Gamma\left({\cal R}_2\right)$ it is not clear how to construct an appropriate compensator field $W_M$. However, here there is a way to construct the compensator fields because we have a triplet of $\K_{ij}$ related by the compatibility requirements \eqref{eq:4DAxioCompat}. The appropriate compensator fields are given by
\begin{equation}
 W_{ij,M} = - \frac1{2\kappa^2} \K_{k(i}{}^\alpha \partial_M K^k{}_{j)\alpha} \,.
\end{equation}
Thus we find that the combination
\begin{equation}
 \left( d\K_{ij} \right)^M = - 12 \left(t_\alpha\right)^{MN} \partial_N \K_{ij}{}^\alpha + \frac1{4\kappa^2} \Omega^{MN} \K_{k(i}{}^\alpha \partial_N K^k{}_{j)\alpha} \,,
\end{equation}
transforms as a tensor.

In general, one would expect $\left(d\K_{ij}\right)^M$ to contain the following representations of $\SO{6,6} \times \SL{2} \subset \EG{7}$
\begin{equation}
 \dbf{1}{3} \otimes \left[ \dbf{12}{2} \oplus \dbf{32'}{1} \right] = \dbf{12}{2} \oplus \dbf{12}{4} \oplus \dbf{32'}{3} \,.
\end{equation}
However, from \eqref{eq:4DAxioWInt}, we know that there is no component transforming in the $\dbf{12}{4}$. Thus, we find that
\begin{equation}
 \left(d\K_{ij}\right)^M = \K_{ij}{}^\alpha \left(t_\alpha\right)^M{}_N \tilde{T}^N + \ldots \,, \label{eq:4DAxioDilTorsion}
\end{equation}
where $\tilde{T}^M$ is the components transforming in the $\dbf{12}{2}$ and $\ldots$ denotes the $\dbf{32'}{3}$. We do not need its explicit form since it only contains spinorial representations under $\SO{6}_S$ and thus it will not play a role in truly half-maximal theories, as we are considering here.

Finally, let us mention that the definition of the intrinsic torsion of the $\SO{6,6}$ structure is compatible with \cite{Ashmore:2015joa}. Up to an unimportant change of signature, we can use the same formula \eqref{eq:4DAxioDilTorsion} to define the intrinsic torsion of a hypermultiplet structure, as relevant in \cite{Ashmore:2015joa}. It is easy to show that the vanishing of the intrinsic torsion as given in \eqref{eq:4DAxioDilTorsion} is \emph{equivalent}, up to integration by parts, to the vanishing of the moment map for the hypermultiplet structure in equations (4.6) and (4.7) of \cite{Ashmore:2015joa}. However, for the study of consistent truncations, it will be crucial to have a local expression for the intrinsic torsion as opposed to the integral one in \cite{Ashmore:2015joa}.

\subsubsection{Intrinsic torsion of the $\SO{6}$ structure} \label{s:4DSO6Torsion}
We now turn to the intrinsic torsion of the $\SO{6}$ structure. The representation theory analysis gives
\begin{equation}
 \begin{split}
  W &= \mbf{912} \oplus \mbf{56} \\
  &= \tbf{15}{6}{2} \oplus \tbf{6}{15}{2} \oplus \tbf{10}{1}{2} \oplus \tbf{\bar{10}}{1}{2} \oplus \tbf{1}{10}{2} \\
  & \quad \oplus \tbf{1}{\bar{10}}{2} \oplus 2 \cdot \mbf{\left(6,1,2\right)} \oplus 2 \cdot \mbf{\left(1,6,2\right)} \oplus \ldots \,, \\
  K_{\SO{6}} &= \tbf{15}{1}{1} \otimes \left[ \tbf{6}{1}{2} \oplus \tbf{1}{6}{2} \oplus \ldots \right] \\
  &= \tbf{64}{1}{2} \oplus \tbf{15}{6}{2} \oplus \tbf{10}{1}{2} \oplus \tbf{\bar{10}}{1}{2} \oplus \tbf{6}{1}{2} \oplus \ldots \,, \\
  \mathrm{Im}\tau_{\SO{6}} &= \tbf{15}{6}{2} \oplus \tbf{10}{1}{2} \oplus \tbf{\bar{10}}{1}{2} \oplus \tbf{6}{1}{2} \oplus \ldots \,, \label{eq:WSO6}
 \end{split}
\end{equation}
and hence the intrinsic torsion has components in the representations
\begin{equation}
 W_{\SO{6}} = \tbf{6}{15}{2} \oplus \tbf{1}{10}{2} \oplus \tbf{1}{\bar{10}}{2} \oplus \tbf{6}{1}{2} \oplus 2 \cdot \tbf{1}{6}{2} \oplus \ldots \,, \label{eq:WIntSO6}
\end{equation}
where all representations here are of $\SO{6}_S \times \SO{6}_R \times \SL{2}$ (apart from the first line of \eqref{eq:WSO6} where we are referring to $\EG{7}$ representations) and $\ldots$ refers to components of the intrinsic torsion which transform in spinorial representations of $\SO{6}_S$. We will ignore these components as they will vanish in truly half-maximal backgrounds, and thus in all applications relevant to us here.

Explicitly, the intrinsic torsion is as usual given by tensorial combinations of derivatives of the $\SO{6}$ structure. We can make use of $d\K_{ij}$ as already discussed in section \ref{s:4DAxioDilTorsion}, as well as use $\J_{u\,i}$ to generate generalised diffeomorphisms. However, we can also introduce a derivative of the $\SO{6}_R$ generators $\J_{uv} \in {\cal R}_{2}$. This is defined as
\begin{equation}
 d\J_{uv}{}^M = - 12 \left(t_\alpha\right)^{MN} \partial_N \J_{uv}{}^\alpha - \frac12 \Omega^{MN} \Omega_{KL} \J_{[u}{}^{i\,K} \partial_N \J_{v]i}{}^L \,, \label{eq:4DdJ}
\end{equation}
which can be understood in the same way as $d\K_{ij}{}^M$ in \eqref{eq:4DdK}. In particular, the second term is a compensator field, constructed out of the twelve generalised vector fields $\J_{u\,i}$.

We can now give the intrinsic torsion explicitly. The only independent combinations of derivatives of the $\SO{6}$ structure are
\begin{equation}
 \begin{split}
  d\K_{ij} &= - \frac12 \K_{ij} \cdot T_1 - \kappa \J_{u(i} T_2{}^u{}_{j)} + \ldots \,, \\
  d\J_{uv} &= 2 \kappa^2 R_{1\,uv} - \kappa R_{2\,uvw\,k} \J^{w\,k} - \kappa T_{2[u}{}^k \J_{v]k} + \ldots \,, \\
  \gL_{\J_{u\,i}} \J_{v\,j}{} - \gL_{\J_{v\,j}} \J_{u\,i}{} &= - 2 \kappa R_{2\,uvw(i} \J^w{}_{j)} + \kappa \K_{ij} \cdot R_{1\,uv} + \kappa T_{2\,v(i} \J_{|u|j)} - \kappa T_{2\,u(i} \J_{|v|j)} \\
  & \quad - 2 \kappa \epsilon_{ij} T_{2(u}{}^k \J_{v)k} - \frac12 \kappa \delta_{uv} \epsilon_{ij} T_{2\,w\,k} \J^{w\,k} + \frac12 \kappa \delta_{uv} \epsilon_{ij} \kappa^2 T_1 + \ldots \,, \\
  \gL_{\J_{u\,i}} \kappa^2 &= 0 \,, \label{eq:4DIntTorsion}
 \end{split}
\end{equation}
where again the $\ldots$ stand for components which are $\SO{6}_S$ spinors and hence will vanish in all applications considered. We have also assumed that the following component of the intrinsic torsion vanishes
\begin{equation}
 \gL_{\J_{u\,i}} \kappa = 0 \,.
\end{equation}
As  we will discuss further in section \ref{s:4DConsTruncations} this will have to vanish for four-dimensional supergravities with an action principle.

These components of the intrinsic torsion transform in the following representations of $\SO{6}_S \times \SO{6}_R \times \SL{2}$.
\begin{equation}
 \begin{split}
  T_1^M &\in \tbf{6}{1}{2} \,, \\
  T_{2\,u\,i} &\in \tbf{1}{6}{2} \,, \\
  R_{1\,uv}{}^M &\in \tbf{6}{15}{2} \,, \\
  R_{2\,uvw\,i} &\in \tbf{1}{10}{2} \oplus \tbf{1}{\bar{10}}{2} \,,
 \end{split}
\end{equation}
which in particular implies that
\begin{equation}
 \begin{split}
  T_1{}^M \J_{u\,M} &= R_{1\,uv}{}^M \J_{u\,M} = 0 \,.
 \end{split}
\end{equation}
We have assumed that the other $\tbf{1}{6}{2}$ representation of \eqref{eq:WIntSO6} vanishes by taking $\gL_{\J_{u\,i}} \kappa^2 = 0$.

To see that the above are the only independent combinations possible, note that
\begin{equation}
 \gL_{\J_{u\,i}} \J_{v\,j} + \gL_{\J_{v\,j}} \J_{u\,i} = \delta_{uv} d\K_{ij} + \epsilon_{ij} d\J_{uv} \,,
\end{equation}
which follows from the compatibility conditions \eqref{eq:4DAxioCompat}, \eqref{eq:4DSO6Compat} and the definitions \eqref{eq:4DdK} and \eqref{eq:4DdJ}. Furthermore, as we have already noted, the $\K_{ij}$ can all be expressed in terms of $\J_{u\,i}$ and thus $\gL_{\J_{u\,i}} \K_{jk}$ will not be independent of \eqref{eq:4DIntTorsion}. Finally, one can show using the compatibility conditions \eqref{eq:4DAxioCompat}, \eqref{eq:4DSO6Compat} that the most general form of the intrinsic torsion is as given in \eqref{eq:4DIntTorsion}. The calculation is similar in spirit to that given in section \ref{s:IntTorsion} and appendix \ref{A:TorsionClassesDecomp} and thus we will not repeat it here.

\subsection{Half-maximal flux vacua} \label{s:4DVacua}
We can again use the technology introduced to determine the conditions to have generic half-maximal warped Mink$_4$ and AdS$_4$ vacua of type II or 11-dimensional SUGRA. Lorentz and AdS requires the vector field $A_\mu{}^M$ of EFT to vanish. Also the $\SO{6}$ structure must be independent of the external space.

As shown in \cite{Coimbra:2014uxa,Coimbra:2016ydd}, Minkowski vacua require that the intrinsic torsion of the $\SO{6}$ structure vanishes, i.e.
\begin{equation}
 \begin{split}
  d\K_{ij} &= d\J_{uv} = \gL_{\J_{u\,i}} \J_{v\,j} = \gL_{\J_{u\,i}} \kappa^2 = 0 \,.
 \end{split}
\end{equation}
We call such $\SO{6}$ structures integrable	.

One can also determine the conditions in order to have a half-maximal $AdS_4$ vacuum. This can be found from the supersymmetry variations, or by comparison with four-dimensional half-maximal gauged SUGRA \cite{Louis:2014gxa}. From there one sees that AdS$_4$ vacua have only the $\tbf{1}{10}{2}$ component of the intrinsic torsion non-vanishing. Thus, we must have
\begin{equation}
 \begin{split}
  d\K_{ij} &= \gL_{\J_{u\,i}} \kappa^2 = 0 \,, \\
  d\J_{uv} &= - \bR_{uvw\,i} \J^{w\,i} \,, \\
  \gL_{\J_{u\,i}} \J_{v\,j} &= - \bR_{uvw\,i} \J^w{}_j \,, \label{eq:4DAdS}
 \end{split}
\end{equation}
where $\bR_{uvw\,i}$ is constant and satisfies
\begin{equation}
 \bR_{uvw\,i} = - \frac1{3!} \epsilon_{uvwxyz} \delta_{ij} \bR^{xyz\,j} \,, \label{eq:4DAdS10}
\end{equation}
where we have raised/lowered the $\SL{2}$ indices with $\epsilon_{ij}$ and the $\SO{6}_R$ indices with $\delta_{uv}$. We call a $\SO{6}$ structure satisfying \eqref{eq:4DAdS} a weakly integrable $\SO{6}$ structure. Note that $\bR_{uvw\,i} = \kappa R_{2\,uvw\,i}$ in terms of the intrinsic torsion components appearing in \eqref{eq:4DIntTorsion}.

\subsection{Reformulating the $\EG{7}$ EFT}
In order to study four-dimensional half-maximal consistent truncations, it is useful to first reformulate the $\EG{7}$ EFT in terms of the $\SO{6}$ structure, rather than the generalised metric. This makes ${\cal N}=4$ supersymmetries manifest at the level of the full exceptional field theory without any truncation. One can do this in the same way as we did in section \ref{s:Reformulate}.

Here we will exclusively focus on the scalar potential, since this is the most interesting part. It is more or less straightforward to write the appropriate kinetic terms for the scalars as we did in section \ref{s:Reformulate} for $D \geq 5$. One way to rewrite the scalar potential is to express it in terms of spinors and then to re-express those in terms of the intrinsic torsion \eqref{eq:4DIntTorsion}. This was the approach taken in \cite{Malek:2016bpu}.

Here we have instead determined the scalar potential by comparison with four-dimensional half-maximal gauged SUGRA. We find that it is given by
\begin{equation}
 \begin{split}
  V &= - \frac14 \left[\frac34 T_2^{u\,i} T_{2\,u}{}^j \delta_{ij} - \frac{1}{16} \kappa^{-2} T_1^M T_1^N \K_{ij}{}^\alpha \left(t_\alpha\right)_{MN} \delta^{ij} + \frac13 R_{2\,uvw\,i} R_2^{uvw}{}_j \delta^{ij} \right. \\
  & \quad \left. - \frac1{12} \kappa^{-2} R_{1\,uv}{}^M R_1^{uv\,N} \left(t_\alpha\right)_{MN} \K^\alpha_{ij} \delta^{ij} - \frac19 R_{2\,uvw\,i} R_{2\,xyz\,j} \epsilon^{uvwxyz} \epsilon^{ij} \right] + \ldots \,,
 \end{split}
\end{equation}
where again $\ldots$ refer to terms which vanish in a half-maximal consistent truncation. It is worth emphasising that not any potential with global $\SO{6,N} \times \SL{2}$ invariance can be obtained in this way, and in particular, it is a non-trivial result that the four-dimensional half-maximal gauged SUGRA potential can be obtained this way.

\subsection{Consistent truncations} \label{s:4DConsTruncations}

\subsubsection{Truncation Ansatz}
We can now define consistent truncations to four-dimensional half-maximal gauged SUGRA by expanding all fields of the EFT in terms of a background $\SO{6-N}$ structure. This is defined by the following sections of exceptional generalised bundles over the background
\begin{equation}
 n_{IJ} \in \Gamma\left(\cRY_1\right) \,, \qquad \omega_{A\,I} \in \Gamma\left(\cRY_2\right) \,, \qquad \rho \in \Gamma\left(\cSY\right) \,, \label{eq:TruncationsSections}
\end{equation}
where $I = 1, 2$ are $\SL{2}$ indices, and $A = 1, \ldots, 6 + N$ are $\SO{6+N}$ indices. Here $\cRY_i$, and $\cSY$ denote the bundles defined over the background we are truncating on. Thus, the tensors in \eqref{eq:TruncationsSections} only depend on the $Y^M$ coordinates, not the ``external'' four-dimensional $x^\mu$ coordinates.

The sections \eqref{eq:TruncationsSections} are subject to the compatibility conditions
\begin{equation}
 \begin{split}
  \left( n_{IJ} \otimes n_{KL} \right)\vert_{1539} &= 0 \,, \\
  \left[ n_{IJ},\, n_{KL} \right] &= - 2 \rho^2 \left[ \epsilon_{I(K} n_{L)J} + \epsilon_{J(K} n_{L)I} \right] \,, \\
  \mathrm{tr} \left( n_{IJ} n_{KL} \right) &= 12 \rho^4 \epsilon_{I(K} \epsilon_{L)J} \,, \\
  \omega_{A\,I} \wedge n_{JK} &= 0 \,, \\
  \omega_{A\,I} \wedge \omega_{B\,J} &= \eta_{AB} n_{IJ} + \epsilon_{IJ} \LO_{AB} \,. \label{eq:TruncCompat}
 \end{split}
\end{equation}
These imply that the background has a $\SO{6-N}$ structure. Equations \eqref{eq:TruncCompat} also lead to
\begin{equation}
 \begin{split}
  n_{IJ} \cdot \omega_{A\,K} &= 2 \rho^2 \epsilon_{K(I} \omega_{|A|J)} \,, \\
  \omega_{AB} \cdot \omega_{C\,I} &= - \rho^2 \eta_{C[A} \omega_{B]I} \,,
 \end{split}
\end{equation}

We can now give the scalar truncation Ansatz. This comes from the expansion of the $\SO{6}$ structure in terms of the background $\SO{6-N}$ structure, with the coefficients becoming scalar fields of the four-dimensional SUGRA.
\begin{equation}
 \begin{split}
  \langle \J_{u\,i} \rangle(x,Y) &= b_u{}^A(x)\, a_i{}^I(x)\, \omega_{A\,I}(Y) \,, \\
  \langle \K_{ij} \rangle(x,Y) &= a_i{}^I(x)\, a_j{}^J(x)\, n_{IJ}(Y) \,, \\
  \langle \kappa \rangle(x,Y) &= \rho(Y)\,, \\
  \langle g_{\mu\nu} \rangle(x,Y) &= \bag_{\mu\nu}(x) \rho(Y) \,. \label{eq:4DScalarAnsatz}
 \end{split}
\end{equation}
Note that here we have not included a scalar in the expansion of $\kappa$ for the same reason as in \ref{s:ConsTruncation}. Such a scalar is just a global, i.e. $Y$-independent, rescaling of $\rho$ from the perspective of the background $\SO{6-N}$ structure. Thus, it will leave the EFT background invariant.

In order for the compatibility conditions, \eqref{eq:4DAxioCompat} and \eqref{eq:4DSO6Compat}, to be fulfilled, the scalars must satisfy
\begin{equation}
 \begin{split}
  b_u{}^A b_{v}{}^B \eta_{AB} &= \delta_{uv} \,, \\
  a_i{}^I a_j{}^J \epsilon_{IJ} &= \epsilon_{ij} \,. \label{eq:4DScalarCompat}
 \end{split}
\end{equation}
We also identify any configuration of scalars related by $\SO{6}_R \times \UO_R$ symmetries. As a result, the following R-symmetry invariant combinations are useful
\begin{equation}
 P_{-}^{AB} = b_u{}^A b^{u\,A} = \frac12 \left( \eta^{AB} - \gH^{AB} \right) \,, \qquad \gH^{IJ} = a_i{}^I a_j{}^J \delta^{ij} \,,
\end{equation}
where we will from now onwards always raise/lower $u, v = 1, \ldots, 6$ indices with $\delta_{uv}$. Note that the conditions \eqref{eq:4DScalarCompat} imply that $\gH_{AB}$ and $\gH_{IJ}$ parameterise the coset spaces
\begin{equation}
 \gH_{AB} \in \frac{\SO{6,n}}{\SO{6}\times\SO{n}} \,, \qquad \gH_{IJ} \in \frac{\SL{2}}{\UO} \,.
\end{equation}
Thus the scalar manifold is that expected of half-maximal gauged SUGRA coupled to $n$ vector multiplets,
\begin{equation}
 {\cal M}_{scalar} = \frac{\SO{6,n}}{\SO{6}\times\SO{n}} \times \frac{\SL{2}}{\UO} \,.
\end{equation}

The truncation Ansatz for the gauge fields is similarly given by
\begin{equation}
 \begin{split}
  \langle A_\mu \rangle(x,Y) &= A_\mu{}^{A\,I}(x)\, \omega_{A\,I}(Y) \,, \\
  \langle B_{\mu\nu} \rangle(x,Y) &= - B_{\mu\nu}{}^{IJ}\, n_{IJ} + B_{\mu\nu,AB}\, \omega^{AB} \,,
 \end{split}
\end{equation}
leading to $12+2N$ (electric plus magnetic) vector fields, and a two-form potential in the adjoint of the global symmetry group $\SL{2} \times \SO{6,N}$.

\subsubsection{Consistency conditions, intrinsic torsion and embedding tensor}
Just like in section \ref{s:ConsTruncation}, we need to impose three conditions on the intrinsic torsion of the $\SO{6-N}$ structure in order to have a consistent truncation. These are that the intrinsic torsion does not contain any spinor representations of $\SO{6-N}$, that it can be expanded only in terms of $\omega_{A\,I}$ and $n_{IJ}$ and that the coefficients are constant. The second condition means that the components of the intrinsic torsion transforming in the vector representation of $\SO{6-N}$ must vanish. With the first two conditions fulfilled, the $\SO{6-N}$ intrinsic torsion is given by
\begin{equation}
 \begin{split}
  dn_{IJ} &= - \omega_{A(I} f^A{}_{J)} \,, \\
  d\omega_{AB} &= - f_{ABC\,I} \omega^{C\,I} - f_{[A}{}^I \omega_{B]I} \,, \\
  \gL_{\omega_A} \rho^2 &= 0 \,, \\
  \gL_{\omega_{AI}} \omega_{BJ} &= - f_{ABC(I} \omega^C{}_{J)} + \frac12 \left[ f_{B(I} \omega_{|A|J)} - f_{A(I} \omega_{|B|J)} \right] - \frac{1}{2} \eta_{AB} \omega_{C(I} f^C{}_{J)} \\
  & \quad - \epsilon_{IJ} \left( f_{(A}{}^K \omega_{B)K} + \frac{1}{4} \eta_{AB} f_{C\,K} \omega^{C\,K} - \frac{1}{2} f_{ABC\,K} \omega^{C\,K} - \frac{1}{2} f_{[A}{}^K \omega_{B]K}
  \right)
  \,. \label{eq:4DEmbTensor}
 \end{split}
\end{equation}

The final condition then is that $f_{ABC\,I}$ and $f_{A\,I}$ are constant. These become the embedding tensor of the four-dimensional half-maximal gauged SUGRA, and we see that they are automatically of the most general form possible, i.e. satisfying the linear constraint of four-dimensional half-maximal gauged SUGRA \cite{Schon:2006kz}. Note that we have also restricted ourselves to those gaugings which have vanishing trombone, i.e. $\gL_{\omega_A} \rho^2 = 0$. This ensures that the gauged SUGRAs we obtain admit an action principle \cite{LeDiffon:2008sh}.

The gaugings of half-maximal gauged SUGRA must also satisfy certain quadratic constraints \cite{Schon:2006kz}. Here these follow from closure of the generalised Lie derivative. Thus, they are satisfied automatically, if the background $\SO{6-N}$ structure satisfies the section condition.

\subsubsection{Reduction of scalar potential}
First we calculate the reduction of the intrinsic torsion and find
\begin{equation}
 \begin{split}
  \langle T_{1} \rangle &= \rho^{-2} P_+^{AB} f_A{}^I \omega_{B\,I} \,, \\
  \langle T_{2\,u\,i} \rangle &= \rho^{-1} a_i{}^I b_u{}^A f_{A\,I} \,, \\
  \langle R_{uv}{}^M \rangle &= \rho^{-2} b_u{}^A b_v{}^B P_+{}^{CD} f_{ABD}{}^{I} \omega_{D\,I} \,, \\
  \langle R_{uvw,i} \rangle &= \rho^{-1} b_u{}^A b_v{}^B b_w{}^C a_i{}^I f_{ABC\,I} \,, \label{eq:4DITTrunc}
 \end{split}
\end{equation}
with all other components, i.e. the spinorial ones, vanishing.

From this we immediately obtain
\begin{equation}
 \begin{split}
  \langle V |e| \rangle &= -\frac14 |e| \rho^{4} \left[ f_{ABC\,I} f_{DEF\,J} \gH^{IJ} \left( \frac1{12} \gH^{AD} \gH^{BE} \gH^{CF} - \frac14 \gH^{AD} \eta^{BE} \eta^{CF} + \frac16 \eta^{AD} \eta^{BE} \eta^{CF} \right) \right. \\
  & \quad \left. - \frac19 f_{ABC\,I} f_{DEF\,J} \epsilon^{IJ} \gH^{ABCDEF} + \frac34 f_A{}^I f_B{}^J \gH_{IJ} \gH^{AB} \right] \,,
 \end{split}
\end{equation}
where we have made use of the identity \eqref{eq:f2Identity} and also defined
\begin{equation}
 \gH^{ABCDEF} = \epsilon^{uvwxyz} b_u{}^A b_v{}^B b_w{}^C b_x{}^D b_y{}^E b_z{}^F \,.
\end{equation}
This is the four-dimensional half-maximal gauged SUGRA, including $\SL{2}$ angles, with general gaugings and $N$ vector multiplets \cite{Schon:2006kz}. Note in particular that the our Ansatz ensures that the only dependence on $Y$ appears in the conformal factors. This guarantees that the truncations is consistent.

\subsection{Universal consistent truncations for half-maximal AdS and Mink vacua}\label{s:4DUniTruncation}
We can now prove the conjecture of \cite{Gauntlett:2007ma} for the case of half-maximal $AdS_4$ vacua, and the corresponding statement for $Mink_4$ vacua. That is, we will prove that for any warped half-maximal $AdS_4$ or $Mink_4$ vacuum of type II or 11-dimensional SUGRA, there exists a consistent truncation keeping only the gravitational supermultiplet.

The proof is completely analogous to that presented in section \ref{s:UniTruncation}. As we have shown in section \ref{s:4DVacua}, a half-maximal AdS$_4$ or Mink$_4$ vacuum of type II or 11-dimensional SUGRA has a (weakly) integrable $\SO{6}$ structure, and the tensors defining the $\SO{6}$ structure, i.e. $\J_{u\,i}$, $\K_{ij}$ and $\kappa$, depend only on the internal space, i.e. only on the $Y$ coordinates. This means that we can use these tensors as our background $\SO{6}$ structure in the truncation Ansatz \eqref{eq:4DScalarAnsatz}, i.e. we take
\begin{equation}
 \rho(Y) = \kappa(Y) \,, \qquad n_{ij}(Y) = \K_{ij}(Y) \,, \qquad \omega_{u\,i}(Y) = \J_{u\,i}(Y) \,.
\end{equation}

Furthermore, the weak integrability of the $\SO{6}$ structure implies that
\begin{equation}
 \begin{split}
  dn_{ij} &= \gL_{\omega_{u\,i}} \rho^2 = 0 \,, \\
  d\omega_{uv} &= - \bR_{uvw\,i} \omega^{w\,i} \,, \\
  \gL_{\omega_{u\,i}} \J_{\omega\,j} &= - \bR_{uvw\,i} \omega^w{}_j \,, \label{eq:4DWeak}
 \end{split}
\end{equation}
where $\bR_{uvw\,i}$ is constant and satisfies \eqref{eq:4DAdS10}. Comparing with \eqref{eq:4DEmbTensor} we see that we fulfil all the conditions to a consistent truncation. The four-dimensional half-maximal gauged SUGRA that we obtain has an embedding tensor given by
\begin{equation}
 f_{uvw\,i} = \bR_{uvw\,i} \,.
\end{equation}

\subsection{Relation to DFT at $\SL{2}$ angles}
One can follow the same procedure as in section \ref{s:HetConsTruncation} to reduce the $\EG{7}$ EFT to the recently constructed double field theory at $\SL{2}$ angles \cite{Ciceri:2016hup}. Since the procedure is fairly straightforward we will not do this here. Nonetheless, this could be interesting, as this would, for example, show which four-dimensional backgrounds can be obtained by a consistent truncation of both type II (or 11-dimensional) and type I SUGRA. 

Similar to our method, \cite{Ciceri:2016hup} shows how to obtain four-dimensional half-maximal gauged SUGRAs with gaugings at $\SL{2}$ angles. These are obtained as generalised Scherk-Schwarz reductions of their double field theory at $\SL{2}$ angles. However, exactly as for generalised Scherk-Schwarz reductions of DFT and EFT \cite{Aldazabal:2011nj,Geissbuhler:2011mx,Grana:2012rr,Dibitetto:2012rk,Aldazabal:2013sca,Berman:2012uy,Geissbuhler:2013uka,Berman:2013cli,Godazgar:2013dma,Godazgar:2013oba,Godazgar:2013pfa,Hohm:2014qga,Lee:2014mla,Lee:2015xga}, the procedure in \cite{Ciceri:2016hup} requires a generalised parallelisable background in order to perform a consistent truncation \cite{Lee:2014mla}, since the higher-dimensional theory only has ${\cal N}=1$. There may be four-dimensional gaugings which come from truncations on backgrounds with exceptional generalised $\SO{6-N}$ structure, and thus these can only be captured by the methods developed here. For example, these would be any four-dimensional half-maximal gauged SUGRAs coming from 11-dimensional SUGRA \cite{Malek:2016vsh}.

\section{Six-dimensional chiral half-maximal supergravity}\label{s:D=6b}
In six dimensions, there are two different half-maximal supergravities. One is non-chiral and we have already described how this can be obtained from exceptional field theory in \ref{s:GHalfDgeq5}. The chiral half-maximal supergravity can also be obtained from exceptional field theory, and in particular this allows us to study consistent truncations of type II and 11-dimensional supergravity to the six-dimensional ${\cal N}=\left(2,0\right)$ supergravity.

\subsection{$\SO{5}$ structure}
We begin by describing the appropriate exceptional generalised $\GH \subset \SO{5,5} \times \mathbb{R}^+$ structure. From table \ref{t:GHalf}, we see that
\begin{equation}
 \GH = \SO{5} \simeq \USp{4} \subset \USp{4} \times \USp{4} \subset \SO{5,5} \times \mathbb{R}^+ \,.
\end{equation}
This structure can again be defined by a set of well-defined tensors which behave like differential forms, exactly like in section \ref{s:GHalfDgeq5}. A $\SO{5} \subset \SO{5,5} \times \mathbb{R}^+$ is defined by five globally well-defined nowhere vanishing tensors $\J_u \in \Gamma\left({\cal R}_2\right)$ and a scalar density $\kappa \in \Gamma\left({\cal S}\right)$, of weight $\frac14$, which satisfy
\begin{equation}
 \J_u \wedge \J_v = \delta_{uv} \kappa^4 \,.
\end{equation}
Here $u = 1, \ldots, 5$ transforms under the $\SO{5}_R$ R-symmetry. Using the indices $I, J = 1, \ldots, 10$ to denote the fundamental representation of $\SO{5,5}$ with $\eta_{IJ}$ the $\SO{5,5}$ metric, we can write that above as
\begin{equation}
 \J_u{}^I \J_v{}^J \eta_{IJ} = \delta_{uv} \kappa^4 \,. \label{eq:6DChiralCompat}
\end{equation}
Under $\SO{5}_S \times \SO{5}_R \subset \SO{5,5}$, where the subscripts $S / R$ denote the structure and R-symmetry group, respectively, the $\mbf{10}$ representation of $\SO{5,5}$ branches as
\begin{equation}
 \mbf{10} \longrightarrow \dbf{5}{1} \oplus \dbf{1}{5} \,.
\end{equation}
The compatibility condition \eqref{eq:6DChiralCompat} implies that $\J_u$'s live in $\dbf{1}{5}$ representation. It is straightforward to see that this breaks $\SO{5,5} \longrightarrow \SO{5}$.

Comparing to the half-maximal structures of \ref{s:GHalfDgeq5} we see that we do not have an analogue of the ``dilaton'' structure since $\SO{5}$ is not a subgroup of $\SO{4,4}$. This is a manifestation of the fact that the chiral six-dimensional supergravity does not have a scalar in its gravitational supermultiplet.

\subsection{Intrinsic torsion}
The intrinsic torsion of the $\SO{5}$ structure can be determined in the same way as we have done in sections \ref{s:IntTorsion} and \ref{s:D=4}. First, let us use representation theory to see which components we are expecting to find. Following \cite{Coimbra:2014uxa}, and using the same notation as in sections \ref{s:4DAxioDilTorsion} and\ref{s:4DSO6Torsion} and appendices \ref{s:D=7} -- \ref{s:D=5}, we have
\begin{equation}
 \begin{split}
  W &= \mbf{144} \oplus \obf{16} \\
  &= 2 \cdot \dbf{4}{4} \oplus \dbf{4}{16} \oplus \dbf{16}{4} \,, \\
  K_{\USp{4}} &= \dbf{4}{4} \otimes \dbf{10}{1} = \dbf{4}{4} \oplus \dbf{16}{4} \oplus \dbf{20}{4} \,, \\
  W_{int} &= \dbf{4}{4} \oplus \dbf{4}{16} \,. \label{eq:6DChiralWInt}
 \end{split}
\end{equation}
The first line refers to $\SO{5,5}$ representations while all other lines refer to representation under $\SO{5}_S \times \SO{5}_R$, where $S / R$ denote the structure and R-symmetry group, respectively.

To find explicit expressions for these components of the intrinsic torsion, we make use of the exterior derivative that we introduced in section \ref{s:DiffForms},
\begin{equation}
 d: \Gamma\left({\cal R}_2\right) \longrightarrow \Gamma\left({\cal R}_1\right) \,.
\end{equation}
We can use this to differentiate $\J_u$ and find the intrinsic torsion
\begin{equation}
 d\J_u = \J_u \wedge T + S_u \,.
\end{equation}
Explicitly, following our conventions outlined in appendix \ref{s:D=6}, we have
\begin{equation}
 \left(d\J_u\right)^M \equiv \left(\gamma_I\right)^{MN} \partial_N \J_u{}^I = \frac12 \kappa \left(\gamma_I\right)^{MN} J_u{}^I T_N + \kappa^2 S_u{}^M \,. \label{eq:6DChiralTorsion}
\end{equation}

$S_u{}^M$ here satisfies
\begin{equation}
 S_u \wedge \J^u = 0 \,,
\end{equation}
or, more explicitly,
\begin{equation}
 S_u{}^M \left(\gamma^I\right)_{MN} \J^u{}_I \,.
\end{equation}
This implies that $S_u{}^M \in \dbf{4}{16}$ of $\SO{5}_S \times \SO{5}_R$. To derive \eqref{eq:6DChiralTorsion}, note that $d\J_u$, for fixed $u = 1, \ldots, 5$, is at each point valued in $\mbf{16}$ of $\SO{5,5}$. Decomposing this under $\USp{4}_S \times \USp{4}_R$ and taking into account the $u = 1, \ldots, 5$ index, we see that at each point $d\J_u$ takes values in
\begin{equation}
 \dbf{1}{5} \otimes \dbf{4}{4} = \dbf{4}{4} \oplus \dbf{4}{16} \,.
\end{equation}
These two representations correspond to the intrinsic torsion components $T_M$ and $S_u{}^M$. The factors of $\kappa$ in \eqref{eq:6DChiralTorsion} are conventions so that $T_M$ and $S_u{}^M$ have weight $-1$ under the generalised Lie derivative. We also see that $T_M$ and $S_u{}^M$ correspond to the representations expected from \ref{eq:6DChiralWInt}.

Finally, we can use the intrinsic torsion to formulate the conditions for six-dimensional ${\cal N}=\left(2,0\right)$ warped Minkowski vacua. Lorentz symmetry of the external space requires the gauge fields of the EFT tensor hierarchy to vanish. Furthermore, the $\SO{5}$ structure must be independent of the external space. Finally, as shown in \cite{Coimbra:2014uxa,Coimbra:2016ydd}, its intrinsic torsion must vanish, i.e.
\begin{equation}
 d\J_u^M \equiv \left(\gamma_I\right)^{MN} \partial_N \J_u{}^I = 0 \,.
\end{equation}
We call these ``integrable $\SO{5} \subset \SO{5,5} \times \mathbb{R}^+$ structures''. There can be no ${\cal N}=\left(2,0\right)$ AdS$_6$ vacua.

\subsection{Consistent truncations} \label{s:6DChiralConsTruncation}
Let us show how to obtain consistent truncations of exceptional field theory which lead to the chiral six-dimensional half-maximal supergravity. To do this, we require a background which admits an exceptional generalised $\SO{5-N}$ structure which is defined by $5+N$ sections of the $\cRY_2$ bundle and a scalar density $\rho \in \cSY$ of weight $\frac14$. The $Y$ superscript in the bundles $\cSY$ and $\cRY_2$ denotes that these bundles are defined solely over the internal space, $M_d$, excluding the external six-dimensional spacetime. Thus $\omega_A(Y)$ and $\rho(Y)$ only depend on the $Y^M$ coordinates. These sections must satisfy the compatibility condition
\begin{equation}
 \omega_A \wedge \omega_B \equiv \omega_A{}^I \omega_B{}^J \eta_{IJ} = \eta_{AB} \rho^4 \,,
\end{equation}
where $\eta_{AB}$ is the $\SO{5,N}$ invariant metric. We will see that $N$ corresponds to the number of tensor multiplets kept in the truncation.

The truncation Ansatz is to expand the $\SO{5}$ structure in terms of this background $\SO{5-N}$ structure. Thus, we take
\begin{equation}
 \begin{split}
  \langle \J_u \rangle (x,Y) &= b_u{}^A(x) \omega_A(Y) \,, \\
  \langle \kappa \rangle(x,Y) &= \rho(Y) \,. \label{eq:6DChiralScalarAnsatz}
 \end{split}
\end{equation}
In order for the $\SO{5}$ structure to satisfy the compatibility conditions \eqref{eq:6DChiralCompat}, the scalars must satisfy
\begin{equation}
 b_u{}^A b_v{}^B \eta_{AB} = \delta_{uv} \,. \label{eq:6DChiralScalarCompat}
\end{equation}
Scalars related by $\SO{5}_R$ rotations define the same supergravity background and thus we want to identify such configurations. This can be achieved by using the $\SO{5}_R$-invariant combination
\begin{equation}
 P_-^{AB} = b_u{}^A b^{u\,B} \,,
\end{equation}
where we raise/lower the $\SO{5}_R$ indices with $\delta_{uv}$. From \eqref{eq:6DChiralScalarCompat}, we see that $P_-^{AB}$ is a projector of rank $5$ and can also be expressed as
\begin{equation}
 P_-^{AB} = \frac12 \left( \eta^{AB} - \gH^{AB} \right) \,,
\end{equation}
with $\gH^{AB}$ satisfying
\begin{equation}
 \gH^{AC} \gH^{BD} \eta_{CD} = \eta^{AB} \,.
\end{equation}
From this we see that the scalars parameterise the coset space
\begin{equation}
 {\cal M}_{scalar} = \frac{\SO{5,N}}{\SO{5} \times \SO{N}} \,,
\end{equation}
which is the scalar manifold of chiral half-maximal gauged SUGRA coupled to $N$ tensor multiplets.

The truncation Ansatz for the other fields of the exceptional field theory is obtained by similarly expanding them in terms of the $\SO{5-N}$ structure. However, because the $\SO{5-N}$ structure is defined only by sections of $\cRY_2$ and $\cSY$, we have no objects in which to expand the generalised vector fields of exceptional field theory. This means that the truncation Ansatz for the gauge fields is
\begin{equation}
 \begin{split}
  \langle A_\mu{}^M \rangle(x,Y) &= 0 \,, \\
  \langle B_{\mu\nu}{}^I \rangle(x,Y) &= B_{\mu\nu}{}^A(x) \, \omega_A{}^I(Y) \,, \\
  \langle C_{\mu\nu\rho\,M} \rangle(x,Y) &= 0 \,,
 \end{split}
\end{equation}
and hence we will not have any 1-form or 3-form potentials in the chiral six-dimensional theory. We do, however, obtain $5+N$ tensor fields $B_{\mu\nu}{}^A$. This gives the correct matter content of six-dimensional chiral SUGRA coupled to $N$ tensor multiplets.

Finally, we need to impose a set of differential constraints on $\omega_A$ and $\rho$ in order to have a consistent truncation. From \eqref{eq:6DChiralTorsion} we see that in general the intrinsic torsion of the $\SO{5-N}$ structure is given by
\begin{equation}
 d\omega_A = \omega_A \wedge t + s_A \,.
\end{equation}
However, $t$ and $s_A$ will source the intrinsic torsion $T$ and $S_u$ of the $\SO{5}$ structure \eqref{eq:6DChiralTorsion} and these transform in representations which are spinors under $\SO{5}_S$. Such representations are related to massive gravitino multiplets which we wish to truncate from our theory. Thus, we require
\begin{equation}
 d\omega_A = 0 \,.
\end{equation}
This corresponds to the fact that chiral six-dimensional supergravity does not admit any gaugings.

We can also trivially extend the proof of sections \ref{s:UniTruncation} and \ref{s:4DUniTruncation} to show that for any warped ${\cal N} = \left(2,0\right)$ Minkowski vacuum there is a consistent truncation keeping only the gravitational supermultiplet.

\section{Half-maximal structures in DFT} \label{s:HalfDFT}
So far we have described how to obtain general half-maximal gauged SUGRAs from EFT. One may wonder how to obtain such half-maximal gauged SUGRAs from DFT. DFT is compatible both with ${\cal N}=1$ \cite{Jeon:2011sq,Hohm:2011nu}, as well as ${\cal N} = 2$ SUSY \cite{Hohm:2011zr,Hohm:2011dv,Jeon:2011vx,Jeon:2012hp} in ten dimensions.\footnote{The corresponding ${\cal N}=2$ generalised geometry is discussed in \cite{Coimbra:2011nw}.} If one starts with ${\cal N}=1$ SUSY then the consistent truncation must not break any further supersymmetry. Thus, one must perform a generalised Scherk-Schwarz Ansatz \cite{Aldazabal:2011nj} on a generalised parallelisable space \cite{Lee:2014mla}.

However, if one started with ${\cal N}=2$ SUSY, one can consider truncations on backgrounds which break half the supersymmetry. These thus have non-trivial generalised structure group. There are however several inequivalent half-maximal structure groups in this case. Fortunately, all these cases arise as special cases of the unique EFT set-up we just discussed. The different scenarios arise by changing the relative embedding of the $\SO{d-1,d-1} \mathbb{R}^+$ structure group of DFT relative to the half-maximal $\SO{d-1} \subset E_{(d)d} \times \mathbb{R}^+$ structure group.

Despite the fact that all the half-maximal cases in DFT arise as special cases of what we have considered so far, we will now discuss the particular choice of half-maximal structure group $\ON{d} \subset \ON{d} \times \ON{d} \times \mathbb{R}^+$. We hope that this may clarify some aspects of the previous sections and the relation of the half-maximal EFT structure groups and DFT.

\subsection{DFT generalised Lie derivative and section condition}
In the following we will require the generalised Lie derivative
\begin{equation}
 \gL_{V} W^M = V^N \partial_N W^M - W^N \partial_N V^M + W^N \partial^M V_N \,,
\end{equation}
where $V$ and $W$ are generalised vector fields and we raise/lower indices with the $\ON{d,d}$ metric $\eta_{MN}$. On a scalar density $e^{-2d}$, the generalised Lie derivative acts as
\begin{equation}
 \gL_{V} e^{-2d} = \partial_M \left( V^M e^{-2d} \right) \,.
\end{equation}

In order for the algebra of generalised Lie derivatives to close, we must impose the so-called ``section condition''
\begin{equation}
 \eta^{MN} \partial_M \otimes \partial_N = 0 \,, \label{eq:DFTSectionCondition}
\end{equation}
where the derivatives are taken to act on any pair of fields or on any one field twice.

\subsection{$\ON{d} \subset \ON{d,d} \times \mathbb{R}^+$ structures in DFT}
Let us begin dy defining an $\ON{d} \subset \ON{d,d} \times \mathbb{R}^+$ structure in DFT. This is equivalent to having a well-defined nowhere-vanishing generalised tensor density $e^{-2d}$ and $d$ well-defined nowhere-vanishing generalised vector fields $\J_u{}^M$, $u = 1, \ldots, d$, satisfying
\begin{equation}
 \J_u{}^M \J_v{}^N \eta_{MN} = \delta_{uv} \,. \label{eq:DFTCompat}
\end{equation}
With respect to the $\ON{d}_S \times \ON{d}_R \subset \ON{d,d}$ subgroup, where the subscripts $S/R$ denote the structure and R-symmetry groups, the $\J_u$'s transform in
\begin{equation}
 \J_u \in \dbf{1}{d} \,.
\end{equation}
It is straightforward to see that these break $\ON{d,d} \times \mathbb{R}^+ \longrightarrow \ON{d}$ and thus define a generalised $\ON{d}$ structure. Note that on any background one can find $d$ generalised vector fields satisfying \eqref{eq:DFTCompat} -- these are just the left-moving set of generalised vielbeine \cite{Hohm:2010xe,Coimbra:2011nw} --, but in general these are only well-defined up to $\ON{d}$ rotations. Here we require the $\J_u$'s, satisfying \eqref{eq:DFTCompat}, to be globally well-defined, and only then do we have a $\ON{d} \subset \ODD \times \mathbb{R}^+$ structure.

Not all degrees of freedom of $\J_u$ are physical. In particular, those related by $\ON{d}_R$ transformations define the same background. Thus, it is natural to use the $\ON{d}_R$-invariant combination
\begin{equation}
 \gH^{MN} = \eta^{MN} - 2 \J_u{}^M J^{v\,N} \,, \label{eq:ODDGenMetric}
\end{equation}
where we raise/lower the $u = 1,\ldots, d$ label by $\delta_{uv}$ and the $\ON{d,d}$ indices $M, N = 1, \ldots, 2d$ with $\eta_{MN}$. From \eqref{eq:DFTCompat}, it follows that
\begin{equation}
 \gH^{MP} \gH^{NQ} \eta_{PQ} = \eta^{MN} \,.
\end{equation}
Thus, $\gH_{MN}$ parameterises the coset space
\begin{equation}
 \gH_{MN} \in \frac{\ON{d,d}}{\ON{d} \times \ON{d}} \,.
\end{equation}
At this point, it is clear that the $\J_u{}^M$ are just the left-moving frame fields of the frame-formalism of double field theory \cite{Hohm:2010xe}.

\subsection{Intrinsic torsion} \label{s:DFTIntTorsion}
Let us now find the intrinsic torsion of the $\ON{d}$ structure. First, we find what representations we expect, following \cite{Coimbra:2014uxa}. The space of torsions is given by \cite{Coimbra:2011nw}
\begin{equation}
 W = E \oplus \Lambda^3 E \,,
\end{equation}
where $E$ denotes the representation of the generalised tangent bundle. Let us decompose this under $\ON{d}_S \times \ON{d}_R$. We find
\begin{equation}
 E \longrightarrow V_S \oplus V_R \,,
\end{equation}
where $V_{S/R}$ denotes the vector representation of $\ON{d}_{S/R}$. Thus,
\begin{equation}
 W = \Lambda^3 V_R \oplus \Lambda^3 V_S \oplus \left( V_S \otimes \Lambda^2 V_R \right) \oplus \left( \Lambda^2 V_S \otimes V_R \right) \oplus V_R \oplus V_S \,.
\end{equation}

Now consider, the space of generalised $\ON{d}$-connections. The difference between any two $\ON{d}$-connections is a tensor, necessarily valued in $E^* \otimes \mathbf{adj}(\ON{d}) \simeq E \otimes \Lambda^2 V_S$, and thus the space of generalised $\ON{d}$-connections is given by
\begin{equation}
 \begin{split}
  K_{\ON{d}} &= E^* \otimes \Lambda^2 V_S \\
  &= \left( \Lambda^2 V_S \otimes V_R \right) \oplus \Lambda^3 V_S \oplus \mathrm{Sym}\left(\Lambda^2 V_S, V_S\right) \oplus V_S \,,
 \end{split}
\end{equation}
where $\mathrm{Sym}\left(\Lambda^2 V_S, V_S\right)$ is defined as the traceless and not totally antisymmetric part of $V_S \otimes \Lambda^2 V_S$, such that
\begin{equation}
 V_S \otimes \Lambda^2 V_S = \Lambda^3 V_S \oplus V_S \oplus \mathrm{Sym}\left(\Lambda^2 V_S, V_S \right) \,.
\end{equation}
Finally, we can now calculate the space of intrinsic torsion which is given by
\begin{equation}
 W_{int} = W / K_{\ON{d}} = \Lambda^3 V_R \oplus \left(V_S \otimes \Lambda^2 V_R \right) \oplus V_R \,. \label{eq:DFTWInt}
\end{equation}

Let us now find explicit expressions for the intrinsic torsion. We are looking for tensors formed from one derivative of the $\ON{d}$ structure. The only combinations we can use are
\begin{equation}
 \gL_{\J_u} \J_v = {\cal W}_{uv} \,, \qquad \gL_{\J_u} e^{-2d} = {\cal W}_u \,. \label{eq:DFTIntTorsionPre}
\end{equation}
As we show in appendix \ref{A:DFTTorsion} these can, in general, be written as
\begin{equation}
 \begin{split}
  \gL_{\J_{u}} \J_v &= R_{uv}{}^M + R_{uvw} \J^{w\,M} \,, \\
  \gL_{\J_u} e^{-2d} &= U_u e^{-2d} \,, \label{eq:DFTIntTorsion}
 \end{split}
\end{equation}
where $R_{uv}{}^M = R_{[uv]}{}^M$, $R_{uvw} = R_{[uvw]}$ and
\begin{equation}
 R_{uv}{}^M \J_{w\,M} = 0 \,.
\end{equation}
$R_{uv}{}^M$, $R_{uvw}$ and $U_u$ are the components of the intrinsic torsion and transform in the following representations of $\ON{d}_S \times \ON{d}_R \subset \Odd$:
\begin{equation}
 \begin{split}
  R_{uv}{}^M &\in V_S \otimes \Lambda^2 V_R \,,\\
  R_{uvw} &\in \Lambda^3 V_R \,, \\
  U_u & \in V_R \,,
 \end{split} \label{eq:DFTIntrinsicTorsion}
\end{equation}
These are precisely the representations appearing in \eqref{eq:DFTWInt}. By comparison with section \ref{s:IntTorsion}, we see that one can say that DFT always has an integrable $\ON{d,d}$ structure. As we will see this means that we cannot obtain the most general half-maximal gauged SUGRA from DFT.

\subsection{Reformulating double field theory}
We will now show how to reformulate the NS-NS part of double field theory, which makes use of the generalised metric, in terms of the $\ON{d}$ structure. We will focus on rewriting the so-called scalar potential where the generalised metric appears.

We can rewrite the scalar potential in terms of the intrinsic torsion, in an analogous way to the so-called ``flux formulation''  \cite{Geissbuhler:2013uka,Berman:2013uda,Blair:2014zba} which rewrites the scalar potential in terms of the torsion of the Weitzenb\"ock connection, thus making the identity structure manifest. We begin by writing the most general linear combination of squares of the intrinsic torsion which only makes use of the $\ON{d}$ structure, and also include a term involving the derivative of the intrinsic torsion. Thus we write
\begin{equation}
 {\cal R} = a_1 R_{uvw} R^{uvw} + a_2 R_{uv}{}^M R^{uv,N} \eta_{MN} + a_3 U_u U^u + a_4 \gL_{\J_u} U^u \,,
\end{equation}
where the final term is just
\begin{equation}
 \gL_{\J_u} U^u = \J_u{}^M \partial_M U^u \,,
\end{equation}
but we have written it in terms of the generalised Lie derivative to highlight that it is a tensor.

We can fix the coefficients appearing here, up to an overall scale which can always be absorbed into $e^{-2d}$, by requiring ${\cal R}$ to be invariant under local $\ON{d}_R$ transformations. This is analogous to \cite{Berman:2013uda,Blair:2014zba} where the scalar potential was rewritten in terms of the Weitzenb\"ock torsion by requiring invariance under local $\ON{d} \times \ON{d}$ or $\SO{5}$ transformations, in the case of DFT and $\SL{5}$ EFT, respectively. Here we begin by calculating the anomalous transformation of the intrinsic torsion under $\ON{d}_R$ and find
\begin{equation}
 \begin{split}
  \Delta_\lambda R_{uvw} &= 3 \J_{[u}{}^M \partial_{|M|} \lambda_{vw]} \,, \\
  \Delta_\lambda R_{uv}{}^M &= \left( \eta^{MN} - \J_w{}^M \J^{w,N} \right) \partial_N \lambda_{uv} \,, \\
  \Delta_\lambda U_u &= \J^{v,M} \partial_M \lambda_{uv} \,, \\
  \Delta_\lambda \gL_{\J_u} U^u &= \J_u{}^M \partial_M \J_v{}^N \partial_N \lambda^{uv} - U^u \J^{v,M} \partial_M \lambda_{uv} \,. \label{eq:ODDAnomalous}
 \end{split}
\end{equation}
Note that we can also write
\begin{equation}
 \Delta_\lambda \gL_{\J_u} U^u = \frac12 \left( \gL_{\J_u} \J_v{}^M \right) \partial_M \lambda_{uv} - U^u \J^{v,M} \partial_M \lambda_{uv} - \frac12 \J_v{}^N \partial^M \J_{u,N} \partial_M \lambda_{uv} \,, \label{eq:ODDAnomalous2}
\end{equation}
where the final term vanishes by the section condition \eqref{eq:DFTSectionCondition}.

Using \eqref{eq:ODDAnomalous} and \eqref{eq:ODDAnomalous2} one finds, up to the section condition,
\begin{equation}
 \begin{split}
  \Delta_\lambda {\cal R} &= \left(3a_1 - a_2\right) \J_w{}^M \J^{w}{}_N \gL_{\J_u} \J_{v}{}^{N} \partial_M \lambda^{uv} + \left(a_2 + \frac{a_4}{2} \right)\gL_{\J_u} \J_v{}^M \partial_M \lambda^{uv} \\
  & \quad + \left(a_3-a_4\right) U_u \J_v{}^M \partial_M \lambda^{uv} \,,
 \end{split}
\end{equation}
and thus we find that
\begin{equation}
 a_4 = a_3 = - 2a_2 = - 6a_1 \,.
\end{equation}
For later convenience we choose the overall scale by setting $a_1 = \frac{1}{3}$ so that
\begin{equation}
 {\cal R} =  \frac13 R_{uvw} R^{uvw} + R_{uv}{}^M R^{uv}{}_M - 2 U_u U^u - 2 \gL_{\J_u} U^u \,. \label{eq:ODDGenRicciScalar}
\end{equation}

The full scalar potential includes a term involving internal derivatives of the external metric \cite{Hohm:2013nja}. This can also be rewritten using \eqref{eq:ODDGenMetric} in terms of $\J_u{}^M$ so that the scalar potential, up the section condition, is given by
\begin{equation}
 V = - \frac14 {\cal R} + \frac12 \J_u{}^M \J^{u,N} \partial_M g_{\mu\nu} \partial_N g^{\mu\nu} \,, \label{eq:ODDV}
\end{equation}
with ${\cal R}$ given in \eqref{eq:ODDGenRicciScalar}.

\subsection{Consistent truncation} \label{s:ODDConsTruncation}
We will now show how to obtain half-maximal consistent truncation using the formalism described above. These will yield half-maximal gauged supergravities in $D = 10 - d$ dimensions, but these are not the most general half-maximal gauged supergravities. The most general half-maximal gauged SUGRAs are obtained by considering EFT as we have shown in sections \ref{s:ConsTruncation}, \ref{s:4DConsTruncations} and \ref{s:6DChiralConsTruncation}.

Our truncation Ansatz is based on a factorisation Ansatz for the $\ON{d}$ structure in terms of a background $\ON{d-N}$ structure. This is defined by a generalised scalar density $e^{-2\lambda}$ and $d+N$ generalised vector fields $\omega_A{}^M$ satisfying
\begin{equation}
 \omega_A{}^M \omega_B{}^N \eta_{MN} = \eta_{AB} \,, \label{eq:ODDOmegaCompat}
\end{equation}
where $\eta_{AB}$ is a constant $\ON{d,N}$ metric, and $A = 1, \ldots, d + N$ labels the number of vector multiplets kept in the truncation, as we will see. This means that out of the $d+N$ generalised vector fields, $d$ are sections of the $V_R$ vector bundle, while $N$ are sections of the $V_S$ vector bundle. 

The scalar Ansatz is given by
\begin{equation}
 \begin{split}
  \langle \J_u{}^M \rangle(x,Y) &= \omega_A{}^M(Y) \, b_u{}^A(x) \,, \\
  \langle d \rangle(x,Y) &= \lambda(Y) + \phi(x) \,. \label{eq:ODDScalarTruncationAnsatz}
 \end{split}
\end{equation}
In order for $\J_u{}^M$ to satisfy the compatibility requirement \eqref{eq:DFTCompat} the scalar fields $b_u{}^M(x)$ are subject to the constraint
\begin{equation}
 b_u{}^A b_{v}{}^B \eta_{AB} = \delta_{uv} \,. \label{eq:ODDbCompat}
\end{equation}

The truncation Ansatz for the other fields, the external metric $g_{\mu\nu}$, the vector potential $A_\mu{}^M$ and the two-form potential $B_{\mu\nu}$, are given by
\begin{equation}
 \begin{split}
  \langle g_{\mu\nu} \rangle(x,Y) &= \bag_{\mu\nu}(x) \,, \\
  \langle A_\mu{}^M \rangle(x,Y) &= A_\mu{}^{A}(x)\, \omega_{A}{}^M(Y) \,, \\
  \langle B_{\mu\nu} \rangle(x,Y) &= B_{\mu\nu}(x) \,. \label{eq:ODDOtherTruncationAnsatz}
 \end{split}
\end{equation}

The truncation procedure here is very similar to the generalised Scherk-Schwarz Ansatz used to obtain maximally consistent truncation. However, there one uses the full generalised vielbein $E_{\bar{M}}{}^M$, where $\bar{M} = 1, \ldots, 2d$, and the twist matrix used in the truncation Ansatz is $\ON{d,d}$-group valued. Here the analogue of the twist matrix are the $d+n$ generalised vectors $\omega_A{}^M$ where $A = 1, \ldots, d+N$ and subject to \eqref{eq:ODDOmegaCompat}. This gives more freedom than the generalised Scherk-Schwarz Ansatz. For example, a background may admit a generalised $\ON{d}$ structure but not be generalised parallelisable.

\subsubsection{Consistency conditions and embedding tensor}
In order to have a consistent truncation we need to impose certain differential constraints. These can be expressed in terms of the intrinsic torsion of the background $\ON{d-N}$ structure. Completely analogously to section \ref{s:DFTIntTorsion}, the $\ON{d-N}$ intrinsic torsion is in general given by
\begin{equation}
 \begin{split}
  \gL_{\omega_A} \omega_B{}^M &= r_{AB}{}^M + f_{ABC} \omega^{C\,M} \,, \\
  \gL_{\omega_A} e^{-2\lambda} &= \xi_A e^{-2\lambda} \,. \label{eq:DFTEmbeddingTensor}
 \end{split}
\end{equation}
In order to have a consistent truncation we must impose that $r_{AB}{}^M = 0$ and $f_{ABC}$ and $\xi_A$ are constant. The first condition ensures that the modes kept in the truncation do not source other modes. The second condition implies that all the $Y$-dependence in the action factorises. The constants $f_{ABC}$ and $\xi_A$ can then be identified as the embedding tensor of the half-maximal gauged SUGRA. In order for the lower-dimensional gauged SUGRA to have an action principle we must take $\xi_A = 0$.

\subsubsection{Reduction of scalar potential}
Given the reduction Ansatz \eqref{eq:ODDScalarTruncationAnsatz} and the constraints \eqref{eq:DFTEmbeddingTensor}, the intrinsic torsion becomes
\begin{equation}
 \begin{split}
  \langle R_{uv}{}^M \rangle &= b_u{}^A b_v{}^B f_{ABC} P_+^{CD} \omega_{D}{}^M \,, \\
  \langle R_{uvw} \rangle &= b_u{}^A b_v{}^B b_w{}^C f_{ABC} \,, \\
  \langle U_u \rangle &= e^{-2\lambda} b_u{}^A \xi_A \,,
 \end{split}
\end{equation}
where $P_+^{AB} = \frac12 \left( \eta^{AB} + \gH^{AB} \right)$ is the right-moving projector. However, recall that we must take $\xi_A = 0$ in order for the lower-dimensional gauged SUGRA to have an action principle. Thus, we will in the following take $\xi_A = 0$ and hence $\langle U_u \rangle = 0$.

Using the identity \eqref{eq:f2Identity}, we find that the scalar potential thus becomes
\begin{equation}
 \langle V \rangle = - \frac14 e^{-2d} e^{-2\lambda} f_{ABC} f_{DEF} \left( \frac{1}{12} \gH^{AD} \gH^{BE} \gH^{CF} - \frac14 \gH^{AD} \eta^{BE} \eta^{CF} + \frac16 \eta^{AD} \eta^{BE} \eta^{CF} \right) \,.
\end{equation}
We see that the only dependence on the internal, $Y^M$, coordinates appears through the conformal factor $e^{-2d}$. This ensures that we have a consistent truncation. Furthermore, the scalar potential that we have thus obtained is that of half-maximal gauged SUGRA coupled to $N$ vector multiplets and with only the gaugings $f_{ABC} \neq 0$.

In sections \ref{s:ConsTruncation} and \ref{s:4DConsTruncations}, we have seen how to generate more general gaugings. In particular, by comparison with the situation here, we see that the other gaugings arise from the intrinsic torsion of the $\ON{d,d}$ structure, which in DFT always vanishes.

\section{Discussion}
In this paper we showed how to describe half-maximal supersymmetry in exceptional field theory in $D \geq 4$ dimensions using the language of exceptional generalised $G$ structures. In particular, we showed that half-maximal EFT backgrounds admit a set of globally well-defined nowhere vanishing tensors, which can be thought of as the exceptional field theory analogue of differential forms. We also gave explicit expressions for the intrinsic torsion of the $\GH$ structures in terms of derivatives of these tensors. This allowed us to write down the (weak) integrability conditions which implies that we have a half-maximal warped Minkowski or AdS vacuum.

One of the main applications considered in this paper were consistent half-maximal truncations of 10- and 11-dimensional SUGRA. We showed how to construct such truncations and that these lead to the most general half-maximal gauged SUGRAs, including the complete set of allowed gaugings and deformations of the half-maximal gauged SUGRAs. In particular, we also obtained those gaugings and deformations which are not accessible via generalised Scherk-Schwarz reductions of double field theory, such as the de Roo-Wagemans angles in four dimensions \cite{deRoo:1985jh}, but which are typically required for interesting phenomenology. We leave it for future work to use this formalism to try and find new consistent truncations, for example uplifting the gauged SUGRA in seven dimensions that contains a stable deSitter vacuum \cite{Dibitetto:2015bia}.

Using our set-up we showed that any warped half-maximal AdS$_{D}$ or Mink$_D$ vacuum, for $D \geq 4$, of type II or 11-dimensional SUGRA admits a consistent truncation keeping only the gravitational supermultiplet. This proves the conjecture of \cite{Gauntlett:2007ma} in the case of half-maximal supersymmetry in $D \geq 4$ dimensions, and extends it to include Minkowski vacua, proving a special case of \cite{Duff:1985jd}. Another interesting feature we found was that it is not possible to keep more than $d-1$ vector multiplets in a consistent truncation of type II or 11-dimensional SUGRA, where $d$ is the rank of the relevant exceptional group, $\EG{d}$. To obtain more vector multiplets one needs to presumably go to an effective theory. One immediate consequence of this is that there exists a consistent truncation on $K3$ but with only a small number of vector multiplets.

We also showed how to reduce exceptional field theory to heterotic double field theory. This makes use of an Ansatz like for consistent truncations but the would-be lower-dimensional fields are still allowed to depend on the exceptional $Y$ coordinates, subject to certain restrictions. These are required for consistency and compatibility with half-maximal supersymmetry and imply that the half-maximal theory has a $\left(d-1+N\right)$-dimensional ``extended space''. The generalised Lie derivative then becomes that of heterotic DFT with the gauge group defined by the intrinsic torsion of the $\GH$ structure that was used in the reduction.

This relationship between EFT and heterotic DFT makes it easy to see which lower-dimensional theories can be obtained from truncations of both type II strings / M-theory and the heterotic theory. For example, one can equally interpret the M-theory truncations on $K3$ as a heterotic truncation on a torus, where the $\SO{3}$ structure of $K3$ now instead defines vector multiplets of the 10-dimensional heterotic theory. In this sense it makes dualities between these theories manifest.

In five dimensions, we showed that the $\EG{6}$ EFT reduces to a new $\SO{5,N}$ DFT which has a $\left(6+N\right)$-dimensional extended space. The section condition for this theory allows two inequivalent solutions. The first allows dependence on five of the $6+N$ coordinates, breaking the $\SO{5,N}$ symmetry. This corresponds to ten dimensional ${\cal N}=1$ SUGRA in a $5+5$ split. The second solution allows dependence on a single coordinate, preserving the global $\SO{5,N}$ symmetry, and corresponds to a $5+1$ split of ${\cal N}=\left(2,0\right)$ SUGRA. It would be interesting to further study this theory. For example, one should try and understand the allowed gaugings and deformations of this modified DFT from the perspective of the six-dimensional ${\cal N}=\left(2,0\right)$ theory.

Our results lead to several avenues for future study. For example, it would be interesting to use the (weak) integrability conditions to study half-maximal AdS and Minkowski vacua and their moduli, as was done in \cite{Ashmore:2015joa,Ashmore:2016oug,Ashmore:2016qvs} for the case of ${\cal N}=2$ vacua. One may ask what the appropriate notion of cohomology is that controls the deformation problem of the (weakly) integrable $\GH$ structures. One may also wonder whether one can use the integrability conditions presented here to find new AdS vacua. Such results would have interesting applications in phenomenology and holography.

It should also be straightforward to generalise our method of finding consistent truncations to cases where different amounts of supersymmetry are preserved. This should allow one to prove the conjecture of \cite{Gauntlett:2007ma} that all supersymmetric warped AdS vacua of 10- and 11-dimensional supergravity admit a consistent truncation keeping only the gravitational supermultiplet.

The relationship between EFT and heterotic DFT also deserves further investigation. For example, recent work on incorporating $\alpha$' corrections of the heterotic string in generalised geometry \cite{Coimbra:2014qaa} lend themselves to a natural interpretation as a dimensional reduction, see also \cite{Andriot:2011iw,Grana:2012rr}. It would be interesting to see to what extent these arise from the formalism presented here.

A drawback of the method of obtaining heterotic DFT presented here is that it only allows one to keep a small number of vector multiplets, bounded by the rank of the relevant exceptional group. This is because it is based on a consistent truncation Ansatz. It would be interesting to see whether one can generalise this to an effective approach, in which one would hope to see the full type II / heterotic dualities. Related to this, and given the recent work on gauge enhancement at self-dual tori \cite{Aldazabal:2015yna,Aldazabal:2017wbk,Cagnacci:2017ulc} in DFT, one might wonder whether the formalism here can capture gauge enhancements of half-maximal string and M-theory compactifications.

\section*{Acknowledgements}
The author thanks Chris Blair, Davide Cassani, Marina Gra\~{n}a, Gianluca Inverso, Diego Marqu\'{e}s, Felix Rudolph, George Papadopoulos, Chris Pope, Henning Samtleben and Daniel Waldram for helpful discussions. The author would also like to thank the organisers of the Banff International Research Station workshop on ``String and M-theory geometries: Double Field Theory, Exceptional Field Theory and their Applications'' for hospitality while part of this work was completed. This work is supported by the ERC Advanced Grant ``Strings and Gravity" (Grant No. 320045).

\appendix

\section{Seven-dimensional half-maximal SUGRA} \label{s:D=7}

\subsection{Conventions}
For $D=7$ the relevant exceptional group is $\SL{5}$, whose generalised Lie derivative is given by
\begin{equation}
 \gL_{\Lambda} V^{a} = \frac12 \Lambda^{bc} \partial_{bc} V^a - V^b \partial_{bc} \Lambda^{ac} + \left(\frac15 + \frac{\lambda}{2} \right) V^a \partial_{bc} \Lambda^{bc} \,,
\end{equation}
where the indices $a = 1, \ldots, 5$ label the fundamental representation of $\SL{5}$ and $V^a$ has weight $\lambda$.

For a generalised vector field, $V^{ab}$, (recall that our conventions are that these carry weight $\frac15$) the generalised Lie derivative is thus
\begin{equation}
 \gL_{\Lambda} V^{ab} = \frac12 \Lambda^{cd} \partial_{cd} V^{ab} - 2 V^{c[b} \partial_{cd} \Lambda^{a]d} + \frac12 V^{ab} \partial_{cd} \Lambda^{cd} \,.
\end{equation}
From this, the $Y$-tensor can be read off to be
\begin{equation}
 Y^{ab,cd}_{ef,gh} = 4! \delta^{abcd}_{efgh} = \epsilon^{abcdi} \epsilon_{efghi} \,.
\end{equation}

We follow similar conventions to \cite{Wang:2015hca} for the wedge products and nilpotent derivative. We let $A_{1,2} \in \Gamma\left({\cal R}_1\right)$, $B_{1,2} \in \Gamma\left({\cal R}_2\right)$, $C_{1,2} \in \Gamma\left({\cal R}_3\right)$ and $D \in \Gamma\left({\cal R}_4\right)$. The wedge product is as follows. 
\begin{equation}
 \begin{split}
  \left( A_1 \wedge A_2 \right)_a &= \frac14 A_1{}^{bc} A_2{}^{de} \epsilon_{abcde} \,, \\
  \left( A \wedge B \right)^a &= A^{ab} B_b \,, \\
  \left( A \wedge C \right)_{ab} &= \frac14 \epsilon_{abcde} A^{cd} C^e \,, \\
  A \wedge D &= A_{ab} D^{ab} \,, \\
  \left( A \wedge_P D \right)^a{}_b &= A_{bc} D^{ac} - \frac15 \delta^a{}_b A_{cd} D^{cd} \,, \\
  \left( B_1 \wedge B_2 \right)_{ab} &= B_{2[a} B_{|1|b]} \,, \\
  B \wedge C &= B_a C^a \,, \\
  \left(B \wedge_P C\right)^a{}_b &= B_b C^a - \frac15 \delta^a{}_b B_c C^c \,.
 \end{split}
\end{equation}
Similarly, the bullet products we need are given by
\begin{alignat}{2}
  \bullet: R_3 \otimes R_3 &\longrightarrow R_1 \,, \qquad \left( C_1 \bullet C_2 \right)^{ab} &&= C_2^{[a} C_{1}^{b]} \,, \\
  \bullet: R_2 \otimes R_3 &\longrightarrow \mbf{1} \,, \qquad B \bullet C &&= B_a C^a \,, \\
  \bullet: R_3 \otimes R_4 &\longrightarrow R_2 \,, \qquad \left( C \bullet D \right)_a &&= C^b D_{ba} \,,
\end{alignat}
and the nilpotent derivatives by
\begin{equation}
 \begin{split}
  \left( dB \right)^{ab} &= \frac12 \epsilon^{abcde} \partial_{cd} B_e \,, \\
  \left( dC \right)_a &= \partial_{ba} C^b \,, \\
  \left( dD \right)^a &= \frac12 \epsilon^{abcde} \partial_{bc} D_{de} \,.
 \end{split}
\end{equation}

\subsection{Dilaton and $\SO{3}$ structures}
With these conventions a dilaton structure is defined by
\begin{equation}
 \left( \K_a,\, \hK^a,\, \kappa \right) \,, \label{eq:SL5Dilaton}
\end{equation}
of weight $\frac25$, $\frac35$ and $\frac15$, satisfying the compatibility conditions \eqref{eq:KPurity}, \eqref{eq:KCompatibility}
\begin{equation}
 \K_a \hK^a = \kappa^5 \,. \label{eq:SL5DilCompat}
\end{equation}
Here $\kappa$ is a scalar density of weight $\frac15$. Under the branching, $\SL{5} \longrightarrow \SL{4} \times \mathbb{R}^+$  we find that the $\mbf{5}$ and $\obf{5}$ representations decompose as
\begin{equation}
 \begin{split}
  \mbf{5} &\longrightarrow \mbf{4}_1 \oplus \mbf{1}_{-4} \,, \\
  \obf{5} &\longrightarrow \obf{4}_{-1} \oplus \mbf{1}_4 \,.
 \end{split}
\end{equation}
One can see that the compatibility equations \eqref{eq:SL5DilCompat} imply that
\begin{equation}
 \K \in \mbf{1}_4 \,, \qquad \hK \in \mbf{1}_{-4} \,,
\end{equation}
and are thus stabilised by $\SO{3,3} \simeq \SL{4} \subset \SL{5}$.

An $\SO{3}$ structure is defined by a dilaton structure with three additional generalised vector fields $\J_u{}^{ab}$ satisfying
\begin{equation}
 \begin{split}
  \J_u{}^{ab} \K_b &= 0 \,, \\
  \frac14 \epsilon_{abcde} J_u{}^{bc} J_v{}^{de} &= \delta_{uv} \K_a \,. \label{eq:SL5SO3Compat}
 \end{split}
\end{equation}
Under $\SL{5} \longrightarrow \SL{4} \times \mathbb{R}^+$, the $\mbf{10}$ representation decomposes as
\begin{equation}
 \begin{split}
  \mbf{10} &\longrightarrow \mbf{4}_{-3} \oplus \mbf{6}_{2} \,.
 \end{split}
\end{equation}
The first equation of \eqref{eq:SL5SO3Compat} can be viewed as a map from $\mbf{10} \longrightarrow \mbf{5}$. From the $\mathbb{R}^+$ charges one can see that it implies that the $\J_u \in \mbf{6}_{2}$ only.

Decomposing further under $\SO{3}_S \times \SO{3}_R \simeq \SU{2}_S \times \SU{2}_R \subset \SL{4}$, where the $_{S/R}$ subscripts denote the structure and R-symmetry group, the $\mbf{6}$ branches as
\begin{equation}
 \mbf{6} \longrightarrow \left(\mbf{3},\mbf{1}\right) \oplus \left(\mbf{1},\mbf{3}\right) \,.
\end{equation}
The triplet of $\SU{2}_{S/R}$ would appear in the second equation of \eqref{eq:SL5SO3Compat} with opposite signs. Thus we find that the triplet of generalised vectors, $\J_u$, must live in $\left(\mbf{1},\mbf{3}\right)$, i.e. are a triplet of the $\SU{2}_R$ R-symmetry group. This breaks $\SL{5} \times \mathbb{R}^+ \longrightarrow \SU{2}_S \simeq \SO{3}_S$ and thus defines a $\SO{3}$ structure.

\subsection{Intrinsic torsion}

\subsubsection{Intrinsic torsion of dilaton structure} \label{A:7DDilIT}
We begin by calculating the representations expected in the dilaton structure, before giving their explicit expressions. Recall from \eqref{eq:WInt} that the intrinsic torsion is given by
\begin{equation}
 W_{\SL{4}} = W/\mathrm{Im} \tau_{\SL{4}} \,,
\end{equation}
where $W = \mbf{15} \oplus \mbf{40} \oplus \mbf{10}$ of $\SL{5}$. Decomposing these under $\SL{4}$ one finds the representations
\begin{equation}
 W = \mbf{20} \oplus \mbf{10} \oplus \obf{10} \oplus 2 \cdot \mbf{6} \oplus 2 \cdot \mbf{4} \oplus \obf{4} \oplus \mbf{1} \,.
\end{equation}
On the other hand, $\tau: K_{\SL{4}} \longrightarrow W$ maps from the space of $\SL{4}$ connections, $K_{\SL{4}}$ to the space of torsions. $K_{\SL{4}}$ can easily be computed to be
\begin{equation}
 K_{\SL{4}} = \mbf{64} \oplus \mbf{36} \oplus \mbf{20} \oplus \mbf{10} \oplus \obf{10} \oplus \mbf{6} \oplus \mbf{4} \,,
\end{equation}
as representations of $\SL{4}$, see for example \cite{Coimbra:2014uxa}. One can check that as a result
\begin{equation}
 \mathrm{Im}\tau_{\SL{4}} = \mbf{20} \oplus \mbf{10} \oplus \obf{10} \oplus \mbf{6} \oplus \mbf{4} \,,
\end{equation}
and hence the intrinsic torsion lives in the representations
\begin{equation}
 W_{\SL{4}} = \mbf{6} \oplus \mbf{4} \oplus \obf{4} \oplus \mbf{1} \,. \label{eq:SL4IT}
\end{equation}

One can find explicit expressions for these, using the nilpotent exterior derivative $d$ and the $\SL{4}$ structure, defined by $\K$, $\hK$ and $\kappa$. The only tensor combinations using these and only one derivative are
\begin{equation}
 \begin{split}
  \left(d\K\right)^{ab} &= \frac12 \epsilon^{abcde} \partial_{cd} \K_e = \kappa^2 \tilde{T}^{ab} + \kappa^{-1} T_3^{[a} \hK^{b]} \,, \\
  \left(d\hK\right)_a &= \partial_{ba} \hK^b = \kappa^3 P_{2\,a} + \kappa K_a P_1 \,, \label{eq:7DDilTorsion}
 \end{split}
\end{equation}
with
\begin{equation}
 \begin{split}
  \tilde{T}^{ab} K_b &= 0 \,, \\
  T_3^a K_a &= 0 \,, \\
  P_{2\,a} \hK^a &= 0 \,,
 \end{split}
\end{equation}
as in \eqref{eq:DilTorsionConstraints} and \eqref{eq:7DDilTorsionConstraints}. These imply that the irreducible components of the intrinsic torsion transform as
\begin{equation}
 \begin{split}
  \tilde{T}^{ab} &\in \mbf{6} \,, \\
  T_3^a &\in \mbf{4} \,, \\
  P_{2\,a} &\in \obf{4} \,, \\
  P_1 &\in \mbf{1} \,,
 \end{split}
\end{equation}
of $\SO{3,3} \simeq \SL{4} \subset \SL{5}$. From the decomposition $\mbf{10} \longrightarrow \mbf{6} \oplus \mbf{4}$ and $\obf{5} \longrightarrow \obf{4} \oplus \mbf{1}$ as $\SL{5} \longrightarrow \SL{4}$ one can see that we have given the most general form of \eqref{eq:7DDilTorsion}. Furthermore, comparing with \eqref{eq:SL4IT} we see that these indeed are all the components of the intrinsic torsion.

One can also express the irreducible components of the intrinsic torsion directly in terms of $d\K$ and $d\hK$ as follows
\begin{equation}
 \begin{split}
  T_3^a &= 2 \kappa^{-4} \left( \frac12 \epsilon^{abcde} \partial_{cd} \K_e \right) \K_b \,, \\
  \tilde{T}^{ab} &= \frac12 \kappa^{-2} \epsilon^{cdefg} \partial_{ef} \K_g \left( \delta^{ab}_{cd} - 2 \kappa^{-5} \delta^{[a}_{[c} \hK^{b]} \K_{d]} \right) \,, \\
  P_1 &= \kappa^{-6} \hK^a \partial_{ba} \hK^{b} \,, \\
  P_{2\,a} &= \kappa^{-3} \partial_{ba} \hK^b - \kappa^{-8} K_a \hK^c \partial_{dc} \hK^d \,.
 \end{split}
\end{equation}

\subsubsection{Intrinsic torsion of $\SO{3}$ structure} \label{A:7DIT}
Let us begin by calculating what representations are present in the intrinsic torsion. By repeating the analysis of subsection \ref{A:7DDilIT} we find in terms of $\SU{2}_S \times \SU{2}_R$ representations
\begin{equation}
 \begin{split}
  K_{\SU{2}} &= \left(\mbf{1},\mbf{1}\right) \oplus \left(\mbf{3},\mbf{1}\right) \oplus \left(\mbf{5},\mbf{1}\right) \oplus \left(\mbf{3},\mbf{3}\right) \oplus \left(\mbf{2},\mbf{2}\right) \oplus \left(\mbf{4},\mbf{2}\right) \,, \\
  W &= \left( \mbf{4},\mbf{2} \right) \oplus \left(\mbf{2},\mbf{4}\right) \oplus 4 \cdot \left(\mbf{2},\mbf{2}\right) \oplus 2 \cdot \left(\mbf{3},\mbf{3}\right) \oplus 2 \cdot \left(\mbf{1},\mbf{3}\right) \oplus 2 \cdot \left(\mbf{3},\mbf{1}\right) \oplus 3 \cdot \left(\mbf{1},\mbf{1}\right) \,, \\
  \mathrm{Im} \tau_{\SU{2}} &= \left(\mbf{4},\mbf{2}\right) \oplus \left(\mbf{2},\mbf{4}\right) \oplus \left(\mbf{3},\mbf{3}\right) \oplus \left(\mbf{3},\mbf{1}\right) \oplus \left(\mbf{1},\mbf{1}\right) \,,
 \end{split}
\end{equation}
and hence the intrinsic torsion
\begin{equation}
 \begin{split}
  W_{\SU{2}} &= W / \mathrm{Im} \tau_{\SU{2}} \\
  &= 4 \cdot \left(\mbf{2},\mbf{2}\right) \oplus \left(\mbf{3},\mbf{3}\right) \oplus 2 \cdot \left(\mbf{1},\mbf{3}\right) \oplus \left(\mbf{3},\mbf{1}\right) \oplus 2 \cdot \left(\mbf{1},\mbf{1}\right) \,.
 \end{split}
\end{equation}

As discussed in section \ref{s:IntTorsion}, the intrinsic torsion is given by
\begin{equation}
 \begin{split}
  d\K^{ab} &= \kappa^2 T_1^{ab} + \kappa\, \J_u{}^{ab}\, T_2{}^u + \kappa \hK^{[b} \bullet T_3^{a]} \,, \\
  d\hat{K}_a &= \kappa K_a P_{1} + \kappa^3 P_{2\,a} \,, \\
  \gL_{\J_{[u}} \J_{v]}{}^{ab} &= \kappa^2 R_{1\,uv}{}^{ab} + \kappa\, R_{2\,uvw} \J^{w\,ab} + \kappa\, T_{2[u} \J_{v]}{}^{ab} - \kappa \hK^{[b} \left( \LG_{uv}{}^{a]}{}_c T_3^{c} \right) \,, \\
  \gL_{\J_u} \hK^a &= \kappa^4 S_{1\,u}{}^a + \kappa^3 \J_u{}^{ab} S_{2\,b} + \kappa \left( U_u - T_{2\,u} \right) \hK^a \,, \\
  \gL_{\J_u} \kappa^5 &= \kappa^6 U_u \,, \label{eq:7DJTorsionClasses}
  \end{split}
\end{equation}
We have already derived the right-hand side of $d\K$, $\gL_{\J_{[u}} \J_{v]}$ and $\gL_{\J_u} \kappa^5$ in section \ref{s:IntTorsion}. $d\hK$ has already been discussed in the previous section of this appendix, \ref{A:7DDilIT}. The final expansion, $\gL_{\J_u} \hK^a$ is easily explained. We expect the representations
\begin{equation}
 \left[ \left(\mbf{2},\mbf{2}\right) \oplus \left(\mbf{1},\mbf{1}\right) \right] \otimes \left(\mbf{1},\mbf{3}\right) = \left(\mbf{2},\mbf{2}\right) \oplus \left(\mbf{2},\mbf{4}\right) \oplus \left(\mbf{1},\mbf{3}\right) \,,
\end{equation}
to appear. However, we already derived the term transforming in the $\left(\mbf{1},\mbf{3}\right)$ in section \ref{s:IntTorsion}: it is given by $\left(U_u - T_{2\,u} \right) \hK^a$. The $\left(\mbf{2},\mbf{2}\right)$ and $\left(\mbf{2},\mbf{4}\right)$ are then given by $S_{2\,a}$ and $S_{1\,u}{}^a$, respectively which thus must satisfy
\begin{equation}
 \begin{split}
  S_{1\,u}{}^a \K_a &= S_{1\,u}{}^a \J^u{}_{ab} = 0 \,, \\
  S_{2\,a} \hK^a &= 0 \,.
 \end{split}
\end{equation}

\section{Six-dimensional non-chiral half-maximal SUGRA} \label{s:D=6}

\subsection{Conventions}
Here we give the details for the case of $D = 6$ with $E_{5(5)} = \SO{5,5}$. The generalised Lie derivative of a section of ${\cal R}_1$, i.e. a generalised vector $V^M$, with $M = 1, \ldots, 16$ labelling the $R_1$ (spinor) representation of $\SO{5,5}$, is given by
\begin{equation}
 \begin{split}
  \gL_{\Lambda} V^M &= \Lambda^N \partial_N V^M - V^N \partial_N \Lambda^M + \frac12 \left(\gamma_I\right)^{MN} \left(\gamma^I\right)_{PQ} V^P \partial_N \Lambda^Q \,.
 \end{split}
\end{equation}
Here $I = 1, \ldots, 10$ represents the $R_2$ (vector) representation of $\SO{5,5}$ and these indices are raised/lowered by the constant $\SO{5,5}$ invariant metric $\eta_{IJ}$. Thus, the $Y$-tensor can be read off as
\begin{equation}
 Y^{MN}_{PQ} = \frac12 \left(\gamma_I\right)^{MN} \left(\gamma^I\right)_{PQ} \,.
\end{equation}
Similarly, for a section of ${\cal R}_2$, $V^I$, the generalised Lie derivative is given by
\begin{equation}
 \gL_{\Lambda} V^I = \Lambda^M \partial_M V^I + \frac12 \left(\gamma_J \gamma^I \right)_M{}^N V^J \partial_N \Lambda^M\,.
\end{equation}

Again, we follow similar conventions to \cite{Wang:2015hca} for the wedge products and nilpotent derivative and let $A_{1,2} \in \Gamma\left({\cal R}_1\right)$, $B_{1,2} \in \Gamma\left({\cal R}_2\right)$ and $C_{1,2} \in \left({\cal R}_3\right)$. Our conventions for the wedge products are
\begin{equation}
 \begin{split}
  \left( A_1 \wedge A_2 \right)^I &= \frac12 \left( \gamma^I \right)_{MN} A_1{}^M A_2{}^N \,, \\
  \left( A \wedge B \right)_M &= \frac12 \left(\gamma^I\right)_{MN} A^N B_I \,, \\
  A \wedge C &= A^M C_M \,, \\
  \left( A \wedge_P C \right)^{IJ} &= A^M C_N \left(\gamma^{IJ}\right)_M{}^N \,, \\
  B_1 \wedge B_2 &= B_1^I B_2^J \eta_{IJ} \,, \\
  \left( B_1 \wedge_P B_2 \right)^{IJ} &= B_1^{[I} B_2^{J]} \,.
 \end{split}
\end{equation}
We also make use of the bullet product
\begin{equation}
 \begin{split}
  \bullet: R_2 \otimes R_3 &\longrightarrow R_1 \,, \qquad \left( B \bullet C \right)^M = \frac12 B^I \left(\gamma_I\right)^{MN} C_N \,.
 \end{split}
\end{equation}
The nilpotent derivative is given by
\begin{equation}
 \begin{split}
  \left( dB \right)^M &= \left(\gamma_I\right)^{MN} \partial_N B^I \,, \\
  \left( dC \right)_I &= \frac12 \left(\gamma_I\right)^{MN} \partial_M C_N \,.
 \end{split}
\end{equation}

\subsection{Dilaton and $\SO{4}$ structure}
A dilaton structure is defined by a scalar density $\kappa$ of weight $\frac14$ and tensor fields $\K^I$, $\hK^I$ of weight $\frac12$. These must satisfy the compatibility conditions \eqref{eq:KPurity}, \eqref{eq:KCompatibility}
\begin{equation}
 \begin{split}
  \K^I \K^J \eta_{IJ} &= 0 \,, \\
  \hK^I \hK^J \eta_{IJ} &= 0 \,, \\
  \K^I \hK^J \eta_{IJ} &= \kappa^4 \,.
 \end{split} \label{eq:SO55KCompatibility}
\end{equation}
Under the subgroup $\SO{4,4} \times \mathbb{R}^+ \subset \SO{5,5}$, the $\mbf{10}$ branches as
\begin{equation}
 \mbf{10} \longrightarrow \mbf{8}^c_0 \oplus \mbf{1}_2 \oplus \mbf{1}_{-2} \,.
\end{equation}
The conditions \eqref{eq:SO55KCompatibility} imply that at each point the $\K$ and $\hK$ belong to the representations
\begin{equation}
 \K \in \mbf{1}_{2} \,, \qquad \hK \in \mbf{1}_{-2} \,,
\end{equation}
of $\SO{4,4} \times \mathbb{R}^+ \subset \SO{5,5}$. It is easy to check that these, with $\kappa$ are stabilised by $\SO{4,4} \subset \SO{5,5} \times \mathbb{R}^+$ and thus define a $\SO{4,4}$ structure.

The $\SO{4}$ structure is defined by the four additional nowhere-vanishing generalised vector fields $J_u{}^M$, $u = 1, \ldots, 4$, with $M = 1, \ldots, 16$ denoting the $R_1$ representation of $\SO{5,5}$. The compatibility conditions \eqref{eq:JCompatibility} are
\begin{equation}
 \begin{split}
  \frac12 \left(\gamma^I\right)_{MN} J_u{}^N \K_I &= 0 \,, \\
  \frac12 \left(\gamma^I\right)_{MN} J_u{}^M J_v{}^N &= \delta_{uv} \K^I \,. \label{eq:SO55JCompatibility}
 \end{split}
\end{equation}
Let us show that the first condition implies that at each point $J_u \in \mbf{8}^v{}_{1}$ of $\SO{4,4} \times \mathbb{R}^+ \subset \SO{5,5}$. For this we decompose the $\mbf{16}$ and $\obf{16}$ under $\SO{4,4} \times \mathbb{R}^+$ to find
\begin{equation}
 \begin{split}
  \mbf{16} &\longrightarrow \mbf{8}^v_1 \oplus \mbf{8}^s_{-1} \,, \\
  \obf{16} &\longrightarrow \mbf{8}^v_{-1} \oplus \mbf{8}^s_1 \,.
 \end{split}
\end{equation}
The first condition of \eqref{eq:SO55JCompatibility} can be viewed as a map from $\mbf{16} \longrightarrow \obf{16}$, and since $\K^I \in \mbf{1}_2$ its kernel is given by $\mbf{8}^v_1$. It follows that $J_u \in \mbf{8}^v_1$.

Decomposing further under $\SO{4}_S \times \SO{4}_R \simeq \left(\SU{2} \times \SU{2}\right)_S \times \left(\SU{2} \times \SU{2}\right)_R \subset \SO{4,4}$, where the subscripts $S/R$ denote the structure and R-symmetry group, the $\mbf{8}^v$ branches as
\begin{equation}
 \mbf{8}^v \longrightarrow \left(\mbf{2},\mbf{2},\mbf{1},\mbf{1}\right) \oplus \left(\mbf{1},\mbf{1},\mbf{2},\mbf{2}\right) \,.
\end{equation}
These two representations appear in the second equation of \eqref{eq:SO55JCompatibility} with opposite signs on the right-hand side, and thus, we see that $\J_u{}^M \in \left(\mbf{1},\mbf{1},\mbf{2},\mbf{2}\right)$ and thus form a quadruple of the $\SO{4}_R$ R-symmetry group. This breaks the structure group $\SO{5,5} \times \mathbb{R}^+ \longrightarrow \left( \SU{2} \times \SU{2} \right)_S \simeq \SO{4}_S$ and thus defines a $\SO{4}_S$ structure.

\subsection{Intrinsic torsion}

\subsubsection{Intrinsic torsion of the dilaton structure} \label{A:6DDilT}
Let us first find the representations expected in the dilaton structure, before giving their explicit expressions. The intrinsic torsion is given by \eqref{eq:WInt}, i.e.
\begin{equation}
 W_{\SO{4,4}} = W/\mathrm{Im} \tau_{\SO{4,4}} \,,
\end{equation}
where $W = \mbf{144} \oplus \mbf{16}$ of $\SO{5,5}$. Decomposing this under $\SO{4,4}$ one finds the representations
\begin{equation}
 W = \mbf{56}^v \oplus \mbf{56}^s \oplus 3 \cdot \mbf{8}^v \oplus 3 \cdot \mbf{8}^s \,.
\end{equation}
On the other hand, $\tau: K_{\SO{4,4}} \longrightarrow W$ maps from the space of $\SO{4,4}$ connections, $K_{\SO{4,4}}$ to the space of torsions. Following for example \cite{Coimbra:2014uxa}, $K_{\SO{4,4}}$ can easily be shown to be
\begin{equation}
 K_{\SO{4,4}} = \mbf{160}^v \oplus \mbf{160}^s \oplus \mbf{56}^v \oplus \mbf{56}^s \oplus \mbf{8}^v \oplus \mbf{8}^s \,,
\end{equation}
as representations of $\SO{4,4}$. One can check that this implies
\begin{equation}
 \mathrm{Im}\tau_{\SO{4,4}} =  \mbf{56}^v \oplus \mbf{56}^s \oplus \mbf{8}^v \oplus \mbf{8}^s \,,
\end{equation}
and hence the intrinsic torsion lives in the representations
\begin{equation}
 W_{\SO{4,4}} = 2 \cdot \mbf{8}^v \oplus 2 \cdot \mbf{8}^s \,. \label{eq:SO44IT}
\end{equation}

One can find explicit expressions for these, using the nilpotent exterior derivative $d$ and the $\SO{4,4}$ structure, defined by $\K$, $\hK$ and $\kappa$. The only tensor combinations using these and only one derivative are
\begin{equation}
 \begin{split}
  d\K^M &= \left(\gamma_I\right)^{MN} \partial_N \K^I = \kappa^2 \tilde{T}^M + \frac12 \left(\gamma_I \right)^{MN} \hK^I T_{3\,N} \,, \\
  d\hK^M &= \left(\gamma_I\right)^{MN} \partial_N \hK^I = \kappa^2 \tilde{P}^M + \frac12 \left(\gamma_I\right)^{MN} \hK^I P_{3\,N} \,, \label{eq:6DDilTorsion}
 \end{split}
\end{equation}
where according to \eqref{eq:DilTorsionConstraints} and \eqref{eq:6DDilTorsionConstraints}
\begin{equation}
 \begin{split}
  \K^I \left(\gamma_I\right)_{MN} \tilde{T}^N &= 0 \,, \\
  \K^I \left(\gamma_I\right)_{MN} \tilde{P}^N &= 0 \,, \\
  \K^I \left(\gamma_I\right)^{MN} T_{3\,N} &= 0 \,, \\
  \K^I \left(\gamma_I\right)^{MN} P_{3\,N} &= 0 \,.
 \end{split}
\end{equation}
This implies that
\begin{equation}
 \begin{split}
  \tilde{T}^N &\in \mbf{8}^v \,, \\
  \tilde{P}^N &\in \mbf{8}^v \,, \\
  T_{3\,N} &\in \mbf{8}^s \,, \\
  P_{3\,N} &\in \mbf{8}^s \,.
 \end{split}
\end{equation}
To see that this is the most general form of \eqref{eq:6DDilTorsion}, note that under $\SO{4,4} \subset \SO{5,5}$ we have the branching
\begin{equation}
 \mbf{16} \longrightarrow \mbf{8}^v \oplus \mbf{8}^s \,,
\end{equation}
and that these are exactly the representations appearing in \eqref{eq:6DDilTorsion}. Furthermore, by comparison with \eqref{eq:SO44IT} we see that we have captured all the elements of the intrinsic torsion.

\subsubsection{Intrinsic torsion of the $\SO{4}$ structure}
We will now repeat the above analysis for the intrinsic torsion of the $\SO{4}$ structure. We begin by computing
\begin{equation}
 \begin{split}
  K_{\SO{4}} &= \qbf{1}{3}{2}{2} \oplus \qbf{2}{3}{1}{2} \oplus \qbf{3}{1}{2}{2} \oplus \qbf{3}{2}{2}{1} \oplus 2 \cdot \qbf{2}{2}{1}{1} \\
  & \quad \oplus \qbf{1}{2}{2}{1} \oplus \qbf{2}{1}{1}{2} \oplus \qbf{2}{4}{1}{1} \oplus \qbf{4}{2}{1}{1} \oplus \qbf{1}{4}{2}{1} \oplus \qbf{4}{1}{1}{2} \,, \\
  W &= \qbf{3}{1}{2}{2} \oplus \qbf{1}{3}{2}{2} \oplus \qbf{2}{2}{1}{3} \oplus \qbf{2}{2}{3}{1} \oplus \qbf{2}{1}{3}{2} \oplus \qbf{1}{2}{2}{3} \\
  & \quad \oplus \qbf{2}{3}{1}{2} \oplus \qbf{3}{2}{2}{1} \oplus 4 \cdot \qbf{1}{1}{2}{2} \oplus 4 \cdot \qbf{2}{2}{1}{1} \oplus 4 \cdot \qbf{2}{1}{1}{2} \\
  & \quad \oplus 4 \cdot \qbf{1}{2}{2}{1} \,, \\
  \mathrm{Im} \tau_{\SO{4}} &= \qbf{3}{1}{2}{2} \oplus \qbf{3}{2}{2}{1} \oplus \qbf{2}{3}{1}{2} \oplus \qbf{1}{3}{2}{2} \oplus 2 \cdot \qbf{2}{2}{1}{1} \\
  & \quad \oplus \qbf{1}{2}{2}{1} \oplus \qbf{2}{1}{1}{2} \,,
 \end{split}
\end{equation}
in terms of $\left( \SU{2} \times \SU{2} \right)_S \times \left( \SU{2} \times \SU{2} \right)_R$ representations. Thus, the intrinsic torsion is given by
\begin{equation}
 \begin{split}
  W_{\SO{4}} &= W / \mathrm{Im} \tau_{\SO{4}} \\
  &= \qbf{2}{1}{3}{2} \oplus \qbf{1}{2}{2}{3} \oplus 2 \cdot \qbf{2}{2}{1}{1} \oplus 2 \cdot \qbf{1}{2}{2}{1} \\
  & \quad \oplus 2 \cdot \qbf{2}{1}{1}{2} \oplus 4 \cdot \qbf{1}{1}{2}{2} \,.
 \end{split} \label{eq:6DWint}
\end{equation}

As discussed in section \ref{s:IntTorsion}, the intrinsic torsion of the $\SO{4}$ structure is given by
\begin{equation}
 \begin{split}
  d\K^M &= \kappa^2 T_1{}^M + \kappa^2 T_2{}^u \J_u{}^M + \frac12 \left(\gamma_I\right)^{MN} \hK^I T_{3\,N} \,, \\
  d\hK^M &= \kappa^2 P_1{}^M + \kappa P_2{}^u \J_u{}^M + \frac12 \left(\gamma_I\right)^{MN} \hK^I P_{3\,N} \,, \\
  \gL_{\J_{[u}} \J_{v]}{}^{M} &= \kappa^2 R_{1\,uv}{}^{M} + \kappa\, R_{2\,uvw} \J^{w\,M} + \kappa\, T_{2[u} \J_{v]}{}^{M} - \frac12 \hK^I \left(\gamma_I\right)^{MN} \left( \LG_{uv}{}_N{}^P T_{3\,P} \right) \,, \\
  \gL_{\J_u} \hK^I &= \kappa^3 S_{1\,u}{}^I + \kappa^2 \J_u{}^M \left(\gamma^I\right)_{MN} S_{2}{}^N + \kappa \left(U_u - T_{2\,u} \right) \hK^I \,, \\
  \gL_{\J_u} \kappa^4 &= \kappa^5 U_u \,. \label{eq:6DTorsion}
 \end{split}
\end{equation}
The right-hand sides of $d\K^M$, $\gL_{\J_{[u}} \J_{v]}{}^M$ and $\gL_{\J_u} \kappa^4$ have already been derived in sections \ref{s:IntTorsion}. While we have discussed $d\hK^M$ in section \ref{A:6DDilT}, we give here a slightly different expression for it. This is simply because we have further decomposed the $8^v$ of $\SO{4,4}$ under $\SO{4}_S \times \SO{4}_R$, using $\J_u{}^M$ exactly as we have done for $d\K^M$ in going from \eqref{eq:6DDilTorsion} to \eqref{eq:6DTorsion}.

We are left with explaining the right-hand side of $\gL_{\J_u} \hK$. First, note that
\begin{equation}
 \hK_I \gL_{\J_u} \hK^I = \frac12 \gL_{\J_u} \left( \hK_I \hK^I \right) = 0 \,.
\end{equation}
Thus, for any fixed value of $u = 1, \ldots, 4$ we are only expecting the representations
\begin{equation}
 \begin{split}
  \mbf{8}^c \oplus \mbf{1} &\in \mbf{10} \,,
 \end{split}
\end{equation}
in terms of $\SO{4,4} \subset \SO{5,5}$ representations. Taking now into account the various values of $u$ we find that we are expecting the following representations of $\SO{4}_S \times \SO{4}_R$
\begin{equation}
 \begin{split}
  \left[ \right.&\!\!\left.\qbf{1}{2}{1}{2} \oplus \qbf{2}{1}{2}{1} \oplus \qbf{1}{1}{1}{1} \right] \otimes \qbf{1}{1}{2}{2} \\ &= \qbf{1}{2}{2}{3} \oplus \qbf{2}{1}{3}{2} \oplus \qbf{1}{2}{2}{1} 
  \oplus \qbf{2}{1}{1}{2} \oplus \qbf{1}{1}{2}{2} \,.
 \end{split}
\end{equation}
These are exactly the representations appearing in \eqref{eq:6DTorsion}. We have already discussed in section \ref{s:IntTorsion} that the $\qbf{1}{1}{2}{2}$ is just given by the $U_u - T_{2\,u}$. The other representations are contained in $S_{1\,u}{}^I$ and $S_2{}^M$ since from \eqref{eq:TorsionClassRepConstraint} these satisfy
\begin{equation}
 \begin{split}
  S_{1\,u}{}^I \K_I &= S_{1\,u}{}^I \hK_I = S_{1\,u}{}^I \left(\gamma_I\right)^{MN} \hJ^u{}_N = 0 \,, \\
  S_2{}^M \left(\gamma_I\right)_{MN} \hK^I &= 0 \,.
 \end{split}
\end{equation}
These equations imply that
\begin{equation}
 \begin{split}
  S_{1\,u}{}^I &\in \mbf{\left(1,2,2,3\right)} \oplus \mbf{\left(2,1,3,2\right)} \,, \\
  S_2{}^M &\in \mbf{\left(1,2,2,1\right)} \oplus \mbf{\left(2,1,1,2\right)} \,,
 \end{split}
\end{equation}
as required.

Finally, one can check that the torsion classes appearing in \eqref{eq:6DTorsion} are exactly those listed in \eqref{eq:6DWint}.

\section{Five-dimensional half-maximal SUGRA} \label{s:D=5}
\subsection{Conventions}
This appendix contains the details for the case of $D = 5$ with the relevant exceptional group $\EG{6}$. The generalised Lie derivative of a section of ${\cal R}_1$, i.e. a generalised vector $V^M$, with $M = 1, \ldots, 27$ labelling the fundamental representation of $\EG{6}$, is given by
\begin{equation}
 \gL_{\Lambda} V^M = \Lambda^N \partial_N V^M - V^N \partial_N \Lambda^M + 10 d_{NLP} d^{MKP} V^L \partial_K \Lambda^N \,,
\end{equation}
where $d^{MNP}$ and $d_{MNP}$ are the symmetric cubic invariant of $\EG{6}$. They are normalised to satisfy
\begin{equation}
 d_{MPQ} d^{NPQ} = \delta^N_M \,.
\end{equation}
From the generalised Lie derivative, the $Y$-tensor can be read off to be
\begin{equation}
 Y^{MN}_{PQ} = 10 d^{MNK} d_{PQK} \,.
\end{equation}
Let us also give the generalised Lie derivative for a section of ${\cal R}_2$, $B_M$,
\begin{equation}
 \gL_{\Lambda} B_M = \Lambda^N \partial_N B^M + B_N \partial_M \Lambda^N - 10 d_{MLP} d^{NKP} B_N \partial_K \Lambda^L + B_M \partial_N \Lambda^N \,.
\end{equation}

We use the following conventions for wedge products, similar to \cite{Wang:2015hca}.  For $A_{1,2} \in \Gamma\left({\cal R}_1\right)$, $B \in \Gamma\left({\cal R}_2\right)$, $C \in \left({\cal R}_3\right)$,
\begin{equation}
 \begin{split}
  \left( A_1 \wedge A_2 \right)_M &= d_{MNP} A_1^N A_2^P \,, \\
  A \wedge B &= A^M B_M \,, \\
  \left( A \wedge_P B \right)^\alpha &= \left(t^{\alpha}\right)^M{}_N A^N B_M \,,
 \end{split}
\end{equation}
where $\alpha = 1, \ldots, 78$ labels the adjoint representation of $\EG{6}$ and $t^\alpha$ its generators. The adjoint indices are raised/lowered with the Cartan-Killing metric $\kappa_{\alpha\beta} = \tr \left( t_\alpha t_\beta \right)$. The bullet product is given by
\begin{equation}
 \begin{split}
  \bullet: R_1 \otimes R_3 &\longrightarrow R_1 \,, \qquad \left( A \bullet C \right)^M = C_\alpha \left(t^\alpha\right)^M{}_N A^N \,, \\
  \bullet: R_2 \otimes R_3 &\longrightarrow R_2 \,, \qquad \left( B \bullet C \right)_M = C_\alpha \left(t^\alpha\right)^N{}_M B_N \,, \\
  \bullet_P: R_1 \otimes R_2 &\longrightarrow R_3 \,, \qquad \left( A \bullet B \right)_\alpha = A^M B_N \left(t_\alpha\right)^N{}_M \,.
 \end{split}
\end{equation}
The ``exterior derivative'' can only act on a section of the ${\cal R}_2$ bundle and is given by
\begin{equation}
 dB^M = 10 d^{MNP} \partial_N B_P \,,
\end{equation}
as in \cite{Wang:2015hca}.

The following identity is often useful
\begin{equation}
 \left(t_\alpha\right)_N{}^M \left(t^\alpha\right)_L{}^K = \frac1{18} \delta_N^M \delta_L^K + \frac16 \delta_N^K \delta_L^M - \frac53 d^{MKR} d_{NLR} \,. \label{eq:E6PAdj}
\end{equation}

\subsection{Dilaton and $\SO{5}$ structure}
A dilaton structure is defined by tensor fields $\K_M$, $\hK^M$ and $\kappa$ satisfying
\begin{equation}
 \begin{split}
  d^{MNP} \K_M \K_N &= 0 \,, \\
  d_{MNP} \hK^M \hK^N &= 0 \,, \\
  \K_M \hK^M &= \kappa^3 \,, \label{eq:5DDilCompat}
 \end{split}
\end{equation}
where $\hK^M$, $\K_M$ and $\kappa$ have weights $\frac13$, $\frac23$ and $\frac13$ respectively. To understand these conditions, decompose $\EG{6} \longrightarrow \SO{5,5} \times \mathbb{R}^+$, under which
\begin{equation}
 \begin{split}
  \mbf{27} &\longrightarrow \mbf{10}_2 \oplus \mbf{16}_{-1} \oplus \mbf{1}_{-4} \,, \\
  \obf{27} &\longrightarrow \mbf{10}_{-2} \oplus \obf{16}_1 \oplus \mbf{1}_4 \,.
 \end{split}
\end{equation}
It is easy to see, for example by looking at the $\mathbb{R}^+$ charges, that the first and second conditions of \eqref{eq:5DDilCompat} imply that
\begin{equation}
 \K \in \mbf{1}_{4} \,, \qquad \hK \in \mbf{1}_{-4} \,.
\end{equation}

The $\SO{5} \subset \SO{5,5} \subset \EG{6} \times \mathbb{R}^+$ structure is defined by further introducing five generalised vector fields $J_u{}^M$, with $u = 1, \ldots, 5$, satisfying
\begin{equation}
 \begin{split}
  \J_u{}^M \left(t_\alpha\right)^N{}_M \K_N &= J_u{}^M \K_M = 0 \,, \\
  \J_u{}^M \J_v{}^N d_{MNP} &= \delta_{uv} \K_P \,. \label{eq:5DSO5Compat}
 \end{split}
\end{equation}
Decomposing again under $\SO{5,5} \times \mathbb{R}^+ \subset \EG{6}$ the first condition implies that
\begin{equation}
 \J_u \in \mbf{10}_2 \,.
\end{equation}
Decomposing further under $\SO{5}_S \times \SO{5}_R \subset \SO{5,5}$, where the $S/R$ subsets denote the structure and R-symmetry groups, we find
\begin{equation}
 \mbf{10} \longrightarrow \dbf{5}{1} \oplus \dbf{1}{5} \,.
\end{equation}
These two irreducibles would appear in the second equation of \eqref{eq:5DSO5Compat} with opposite signs. Thus, the second condition of \eqref{eq:5DSO5Compat} implies that
\begin{equation}
 \J_u \in \left(\mbf{1},\mbf{5}\right) \,,
\end{equation}
and hence the $\J_u$'s form a quintuple under the $\SO{5}_R$ symmetry.

It is useful to introduce
\begin{equation}
 \hJ_{u\,M} = d_{MNP} \J_u{}^N \hK^P \,,
\end{equation}
which satisfies
\begin{equation}
 \hJ_{u\,M} \J_v{}^M = \kappa^3 \delta_{uv} \,,
\end{equation}
and
\begin{equation}
 10 d^{MNP} \hJ_{u\,M} \hJ_{v\,N} = \kappa^3 \delta_{uv} \hK^P \,.
\end{equation}
One can express also $\J_{u}{}^M$ in terms of $\hJ_{u\,M}$ and $\K_M$ via
\begin{equation}
 \J_u{}^M = 10 \kappa^{-3} d^{MNK} \K_N \hJ_{u\,K} \,. \label{eq:E6JhJ}
\end{equation}
These follow from \eqref{eq:E6PAdj}.

\subsection{Intrinsic torsion}

\subsubsection{Intrinsic torsion of the dilaton structure} \label{A:5DDilT}
Let us first calculate the representations expected in the dilaton structure, before giving their explicit expressions. Recall from \eqref{eq:WInt} that the intrinsic torsion is given by
\begin{equation}
 W_{\SO{5,5}} = W/\mathrm{Im} \tau_{\SO{5,5}} \,,
\end{equation}
where $W = \mbf{351} \oplus \mbf{27}$ of $\EG{6}$. Decomposing these under $\SO{5,5}$ one finds the representations
\begin{equation}
 W = \mbf{144} \oplus \mbf{120} \oplus \mbf{45} \oplus 2 \cdot \mbf{16} \oplus \obf{16} \oplus 2 \cdot \mbf{10} \oplus \mbf{1} \,.
\end{equation}
On the other hand, $\tau: K_{\SO{5,5}} \longrightarrow W$ maps from the space of $\SO{5,5}$ connections, $K_{\SO{5,5}}$ to the space of torsions. $K_{\SO{5,5}}$ can easily be computed to be
\begin{equation}
 K_{\SO{5,5}} = \mbf{560} \oplus \mbf{320} \oplus \mbf{144} \oplus \mbf{120} \oplus \mbf{45} \oplus \mbf{16} \oplus \mbf{10} \,,
\end{equation}
as representations of $\SO{5,5}$, see for example \cite{Coimbra:2014uxa}. One can check that as a result
\begin{equation}
 \mathrm{Im}\tau_{\SO{5,5}} = \mbf{144} \oplus \mbf{120} \oplus \mbf{45} \oplus \mbf{16} \oplus \mbf{10} \,,
\end{equation}
and hence the intrinsic torsion lives in the representations
\begin{equation}
 W_{\SO{5,5}} = \mbf{16} \oplus \obf{16} \oplus \mbf{10} \oplus \mbf{1} \,. \label{eq:SO55IT}
\end{equation}

Let us now turn to the explicit expression of the dilaton structure. The only covariant combinations with one derivative that we can form are given by
\begin{equation}
 \begin{split}
  \left(d\K\right)^M &= d^{MNP} \partial_N \K_P = \kappa^2 \tilde{T}^M + \kappa \left(t_\alpha\right)^M{}_N T_{3}{}^\alpha \hK^N \,, \\
  \gL_{\hK} \K_M &= \kappa \K_M P_1 + \kappa \left(t_\alpha\right)^N{}_M P_2{}^\alpha \K_N \,, \\
  \gL_{\hK} \kappa^3 &= P_1 \kappa^4 \,, \label{eq:D5DilatonTorsion}
 \end{split}
\end{equation}
as well as
\begin{equation}
 \gL_{\hK} \hK^M = 5 d^{MNP} \partial_N \left(d_{PQR} \hK^Q \hK^R \right) = 0 \,,
\end{equation}
due to the compatibility condition \eqref{eq:5DDilCompat}.

Let us now derive the right-hand side of \eqref{eq:D5DilTorsionClasses}. For $d\K$ we can in principle have the $\SO{5,5}$ representations
\begin{equation}
 \mbf{10} \oplus \mbf{16} \oplus \mbf{1} \subset \mbf{27} \textrm{ of } \EG{6} \,.
\end{equation}
However, the singlet component has to vanish since
\begin{equation}
 \K_M d^{MNP} \partial_N \K_P = \frac12 d^{MNP} \partial_N \left( \K_M \K_P \right) = 0 \,.
\end{equation}

Similarly, consider
\begin{equation}
 d^{MNP} \K_N \gL_{\hK} \K_P = \frac12 \gL_{\hK} \left( d^{MNP} \K_N \K_P \right) = 0 \,.
\end{equation}
However, $d^{MNP} \K_N$ projects the $\obf{27}$ of $\EG{6}$ onto its $\mbf{10}$ representation under the branching to $\SO{5,5}$. Thus, the second equation of \eqref{eq:D5DilatonTorsion} only has components of the intrinsic torsion in the $\SO{5,5}$ singlet and $\obf{16}$ representations. On the other hand, the singlets appearing in the second and third line of \eqref{eq:D5DilatonTorsion} are equal since
\begin{equation}
 P_1 = \kappa^{-4} \hK^M \gL_{\hK} \K_M = \kappa^{-4} \gL_{\hK} \left(\hK^M \K_M\right) = \kappa^{-4} \gL_{\hK} \kappa^3 \,.
\end{equation}

Putting all this together, we have
\begin{equation}
 \begin{split}
  \left(t_\alpha\right)^M{}_N \tilde{T}^N \K_M &= \tilde{T}^M \K_M = 0 \,, \\
  T_3{}^\alpha \left(t_\alpha\right)^M{}_N \K_M &= T_3{}^\alpha \left(t_\alpha\right)^M{}_N d^{NPQ} \K_P = 0 \,, \\
  P_{2\,\alpha} \left(t^\alpha\right)^M{}_N \hK^N &= P_{2\,\alpha} \left(t^\alpha\right)^M{}_N d_{MPQ} \hK^P = 0 \,. \label{eq:D5DilatonTorsionConstraints}
 \end{split}
\end{equation}
To see that the above really corresponds to the conditions \eqref{eq:5DDilTorsionConstraints} and \eqref{eq:5DDilTorsionConstraints2}, we need to show that
\begin{equation}
 X^{MN} \equiv T_3{}^{\alpha} \left(t_\alpha\right)^{M}{}_P d^{NPQ} K_Q \in \mbf{351} \,, \qquad X'_{MN} \equiv P_{2\,\alpha} \left(t^{\alpha}\right)^P{}_M d_{NPQ} \hK^Q \in \obf{351} \,.
\end{equation}
This follows from
\begin{equation}
 \begin{split}
  X^{(MN)} &= \frac32 T_3{}^{\alpha} \left(t_\alpha\right)^{(M}{}_Q d^{NP)Q} \K_P = 0 \,, \\
  X'_{(MN)} &= \frac32 P_3{}^{\alpha} \left(t_\alpha\right)^Q{}_{(M} d_{NP)Q} \hK^P = 0 \,,
 \end{split}
\end{equation}
using first that $T_3{}^\alpha \left(t_\alpha\right)^M{}_N \K_M = P_3{}^\alpha \left(t_\alpha\right)^N{}_M \hK^M = 0$ and second that $d^{MNP}$ and $d_{MNP}$ are $\EG{6}$ invariants. This shows that $X^{MQ} \in \mbf{351}$ and $X'^{MQ} \in \obf{351}$.

Equation \eqref{eq:D5DilatonTorsionConstraints} implies that
\begin{equation}
 \begin{split}
  \tilde{T}^M &\in \mbf{10} \,, \\
  T_3{}^\alpha &\in \mbf{16} \,, \\
  P_1 &\in \mbf{1} \,, \\
  P_{2\,\alpha} &\in \obf{16} \,.
 \end{split}
\end{equation}
We see that the representations appearing in \eqref{eq:D5DilatonTorsion} are exactly those in \eqref{eq:SO55IT}.

\subsubsection{Intrinsic torsion of the $\SO{5}$ structure} \label{A:5DSO5T}
Let us now study the intrinsic torsion of the $\SO{5}$ structure. We again begin by finding the representations that we expect. In terms of $\SO{5}_S \times \SO{5}_R$ representations we have
\begin{equation}
 \begin{split}
  K_{\SO{5}} &= \dbf{5}{10} \oplus \dbf{4}{20} \oplus \dbf{4}{16} \oplus \dbf{4}{4} \oplus \dbf{1}{35} \oplus 2 \cdot \dbf{1}{10} \oplus \dbf{1}{5} \,, \\
  W &= \dbf{16}{4} \oplus \dbf{10}{5} \oplus 2 \cdot \dbf{10}{1} \oplus \dbf{5}{5} \oplus 2 \cdot \dbf{5}{1} \oplus \dbf{4}{16} \\
  & \quad \oplus 4 \cdot \dbf{4}{4} \oplus 2 \cdot \dbf{1}{10} \oplus 2 \cdot \dbf{1}{5} \oplus \dbf{1}{1} \,, \\
  \mathrm{Im} \tau_{\SO{5}} &= \dbf{4}{16} \oplus \dbf{4}{4} \oplus 2 \cdot \dbf{1}{10} \oplus \dbf{1}{5} \,,
 \end{split}
\end{equation}
and hence the intrinsic torsion
\begin{equation}
 \begin{split}
  W_{\SO{5}} &= \dbf{16}{4} \oplus \dbf{10}{5} \oplus 2 \cdot \dbf{10}{1} \oplus \dbf{5}{5} \oplus 2 \cdot \dbf{5}{1} \oplus 3 \cdot \dbf{4}{4} \\
  & \quad \oplus \dbf{1}{5} \oplus \dbf{1}{1} \,. \label{eq:D5Wint}
 \end{split}
\end{equation}

Explicitly, the intrinsic torsion of the $\SO{5}$ structure is given by
\begin{equation}
 \begin{split}
  d\K^M &= \kappa^2 T_1^M + \kappa T_{2}{}^u \J_u{}^M + \kappa \left(t_\alpha\right)^M{}_N T_{3}{}^\alpha \hK^N \,, \\
  \gL_{\hK} K_M &= \kappa K_M P_1 + \kappa \left(t_\alpha\right)^N{}_M P_2{}^\alpha K_N \,, \\
  \gL_{\hK} J_u{}^M &= \kappa^2 P_{3\,u}{}^M + \kappa P_{4\,[uv]} \J^{v\,M} + \frac12 \kappa P_1 \J_u{}^M \\
  & \quad - \kappa^{-3} \left(t_\alpha\right)^M{}_N \hK^N \J_u{}^P \left(t^\beta t^\alpha\right)^Q{}_P P_{2\,\beta} \K_Q \,, \\
  \gL_{\hK} \kappa^3 &= P_1 \kappa^4 \,, \\
  \gL_{\J_{[u}} \J_{v]}{}^M &= \kappa^2 R_{1\,uv}{}^{M} + \kappa\, R_{2\,uvw} \J^{w\,M} + \kappa\, T_{2[u} \J_{v]}{}^{M} - \frac12 \hK^N \left[ \LG_{uv}\,,\,\, T_3 \right]_N{}^M \,, \\
  \gL_{\J_u} \hK^M &= \kappa^2 S_{1\,u}{}^M + S_2{}^\alpha \left(t_\alpha\right)^M{}_N \J_u{}^N + \kappa \hK^M \left(U_u - T_{2\,u}\right) \,, \\
  \gL_{\J_u} \kappa^3 &= U_u \kappa^4 \,. \label{eq:5DTorsion}
 \end{split}
\end{equation}
The expressions for $d\K$, $\gL_{\J_{[u}} \J_{v]}$ and $\gL_{\J_u} \kappa^3$ have been derived in section \ref{s:IntTorsion}, and those for $\gL_{\hK} \K$ and $\gL_{\hK} \kappa^3$ in section \ref{A:5DDilT}. Here we will derive the expressions for $\gL_{\hK} \J_u$ and $\gL_{\J_u} \hK$.

Let us begin with $\gL_{\hK} \J_u{}^M$. Using \eqref{eq:E6JhJ}, we find that
\begin{equation}
 \hK_M \gL_{\hK} \J_u{}^M = 10 \kappa^{-3} \hK_M d^{MNP} \gL_{\hK} \left( \hJ_{u\,N} \K_P \right) = 0 \,.
\end{equation}
Thus, for each $u = 1, \ldots, 5$ we can only have the representations
\begin{equation}
 \mbf{16} \oplus \mbf{10}
\end{equation}
of $\SO{5,5}$ appearing. Taking into account the quintuplet coming from $u$, we thus expect at most the representations
\begin{equation}
 \begin{split}
  \dbf{4}{4} & \otimes \dbf{1}{5} \oplus \dbf{1}{5} \otimes \dbf{1}{5} \oplus \dbf{5}{5} \\&= \dbf{5}{5} \oplus \dbf{4}{16} \oplus \dbf{4}{4} \oplus \dbf{1}{14} \oplus \dbf{1}{10} \oplus \dbf{1}{1} \,,
 \end{split}
\end{equation}
of $\SO{5}_S \times \SO{5}_R$ to appear.

Next, we look at the representations $\dbf{1}{14} \oplus \dbf{1}{10} \oplus \dbf{1}{1}$ which are given by
\begin{equation}
 Q_{uv} = \kappa^{-4} \hJ_{v\,M} \gL_{\hK} \J_u{}^M \,.
\end{equation}
However, recall that
\begin{equation}
 \gL_{\hK} \hK^M = 0 \,,
\end{equation}
and thus
\begin{equation}
 \hJ_{v\,M} \gL_{\hK} \J_u{}^M = \J_v{}^M \gL_{\hK} \hJ_{u\,M} \,.
\end{equation}
As a result, the symmetric of $Q_{uv}$ is given by
\begin{equation}
 Q_{(uv)} = \frac12 \kappa^{-4} \gL_{\hK} \left( \J_{u}{}^M \hJ_{v\,M} \right) = \frac12 \delta_{uv} \kappa^{-4} \gL_{\hK} \kappa^3 = \frac12 \delta_{uv} P_1 \,,
\end{equation}
i.e. the $\dbf{1}{14}$ vanishes and the $\dbf{1}{1}$ is given by $P_1$. Only the antisymmetric part gives a new component of the intrinsic torsion and we call it $P_{4\,uv} = Q_{[uv]}$.

Finally, we look at the representations $\dbf{4}{16} \oplus \dbf{4}{4}$ which are given by
\begin{equation}
 \begin{split}
  Q_{\alpha\,u} &= \kappa^{-4} \left(t_\alpha\right)^N{}_M \K_N \gL_{\hK} \J_u{}^M \\
  &= - \kappa^{-4} \J_u{}^M \left(t_\alpha\right)^N{}_M \gL_{\hK} \K_N \\
  &= - \kappa^{-3} \J_u{}^M \left(t_\alpha\right)^N{}_M \left(t_\beta\right)^P{}_N P_2{}^\beta \K_P \,.
 \end{split}
\end{equation}
We see that the $\dbf{4}{16}$ vanishes while the $\dbf{4}{4}$ is given by $P_2{}^\alpha$. This leads to
\begin{equation}
 \begin{split}
  \gL_{\hK} J_u{}^M &= \kappa^2 P_{3\,u}{}^M + \kappa P_{4\,[uv]} \J^{v\,M} + \frac12 \kappa P_1 \J_u{}^M \\
  & \quad - \kappa^{-3} \left(t_\alpha\right)^M{}_N \hK^N \J_u{}^P \left(t^\beta t^\alpha\right)^Q{}_P P_{2\,\beta} \K_Q \,,
 \end{split}
\end{equation}
as given in equation \eqref{eq:5DTorsion}. Here
\begin{equation}
 P_{3\,u}{}^M \K_M = P_{3\,u}{}^M \left(t_\alpha\right)^N{}_M \K_N = P_{3\,u}{}^M \hJ_{v\,M} = 0 \,,
\end{equation}
and hence $P_{3\,u}{}^M \in \dbf{5}{5}$ of $\SO{5}_S \times \SO{5}_R$. Note that one could also write the last term as
\begin{equation}
 \left(t_\alpha\right)^M{}_N \hK^N \J_u{}^P \left(t^\beta t^\alpha\right)^Q{}_P P_{2\,\beta} \K_Q = - \frac53 \left(t_\alpha\right)^L{}_P \K_L P_2{}^\alpha d^{MNP} \hJ_{u\,N} \,.
\end{equation}

Now, let us derive the right-hand side for $\gL_{\J_u} \hK^M$. To see why it contains only three torsion classes consider
\begin{equation}
 d_{MNP} \hK^N \gL_{\J_u} \hK^P = \frac12 \gL_{\J_u} \left( d_{MNP} \hK^N \hK^P \right) = 0 \,.
\end{equation}
This implies that for each fixed $u = 1, \ldots, 5$ $\gL_{\J_u} \hK^M$ only takes values in the representations $\mbf{1} \oplus \mbf{16}$ of $\SO{5,5}$. Now, including the varying $u$ we see that we can only have the following representations under $\SO{5}_S \times \SO{5}_R$
\begin{equation}
 \left[ \mbf{\left(4,4\right)} \oplus \mbf{\left(1,1\right)} \right] \otimes \mbf{\left(1,5\right)} = \mbf{\left(4,4\right)} \oplus \mbf{\left(4,16\right)} \oplus \mbf{\left(1,5\right)} \,,
\end{equation}
which correspond to the intrinsic torsion given in $\gL_{\J_u} \hK$ in equation \eqref{eq:5DTorsion}. Recall that, as shown in section \ref{s:IntTorsion}, the component in the $\dbf{1}{5}$ is given by $U_u - T_{2\,u}$. The other components of the intrinsic torsion satisfy
\begin{equation}
 \begin{split}
  S_{1\,u}{}^M \left(t_\alpha\right)^M{}_N \hJ^u{}_N &= S_{1\,u}{}^M \K_M = 0 \,, \\
  S_2{}^\alpha \left(t_\alpha\right)^M{}_N \hK^N &= 0 \,,
 \end{split}
\end{equation}
and thus belong to
\begin{equation}
 \begin{split}
  S_{1\,u}{}^M &\in \dbf{16}{4} \,, \\
  S_{2}{}^\alpha &\in \dbf{4}{4} \,.
 \end{split}
\end{equation}
We see that all together we obtain precisely the intrinsic torsion as in \eqref{eq:D5Wint}.

\section{Decomposition of the intrinsic torsion} \label{A:TorsionClassesDecomp}
Let us derive equation \eqref{eq:JTorsionClasses}. We begin with the decomposition of $\tc_{\K} = \kappa^2 \tilde{T} + \kappa^{6-D} \hK \bullet T_3$. Under $\SO{d-1,d-1} \longrightarrow \SO{d-1}_S \otimes \SO{d-1}_R$ we find that
\begin{equation}
 V_{d-1,d-1} \longrightarrow V_S \oplus V_{R} \,,
\end{equation}
where $V_{S/R}$ denotes the vector representation of $\SO{d-1}_{S/R}$. Furthermore the $\J_u$ form a basis for $V_R$ at each point and thus we can write $\tilde{T} = T_1 + J_u T_2{}^u$ as in the first equation of \eqref{eq:JTorsionClasses} with $T_1 \in V_S$ and $T_2{}^u \in V_R$.

For $\tc_{\J\,uv}$ we can instead have the following representations appearing
\begin{equation}
 \Lambda^2 V_R \otimes R_1 = \Lambda^2 V_R \otimes V_R \oplus \Lambda^2 V_R \otimes V_S \oplus \Lambda^2 V_R \otimes \phi_{d-1,d-1} \,,
\end{equation}
i.e.
\begin{equation}
 \gL_{\J_{[u}} \J_{v]} = \kappa^2 R_{1\,uv} + \kappa \tilde{R}_{uv,w} \J^w + \kappa^{6-D} \hK \bullet \tilde{R}_{uv} \,,
\end{equation}
where
\begin{equation}
 \begin{split}
  R_{1\,uv} &\in \Lambda^2 V_R \otimes V_S \,, \\
  \tilde{R}_{uv,w} &\in \Lambda^2 V_R \otimes V_R \,, \\
  \tilde{R}_{uv} &\in \Lambda^2 V_R \otimes \phi_{d-1,d-1} \,,
 \end{split}
\end{equation}
and thus they are given by
\begin{equation}
 \begin{split}
  \tilde{R}_{uv,w} &= \kappa^{1-D} \hJ_{w} \wedge \gL_{\J_{[u}} \J_{v]} \,, \\
  \tilde{R}_{uv} &= - 2 \kappa^{-4} \gL_{\J_{[u}} \J_{v]} \wedge \K \,, \\
  R_{1\,uv} &= \kappa^{-2} \gL_{\J_{[u}} \J_{v]} - \kappa^{-D} \J^w \left( \hJ_{w} \wedge \gL_{\J_{[u}} \J_{v]} \right) + 2 \kappa^{-D} \hK \bullet \left( \gL_{\J_{[u}} \J_{v]} \wedge \K \right) \,.
 \end{split}
\end{equation}
Both $\tilde{R}_{uv,w} \in \Lambda^2 V_R \otimes V_R$ and $\tilde{R}_{uv} \in \Lambda^2 V_R \otimes \phi_{d-1,d-1}$ are in principle reducible. Thus, we can write
\begin{equation}
 \tilde{R}_{uv,w} = R_{2\,uvw} + \hat{R}_{uv,w} \,,
\end{equation}
with $R_{2\,uvw} = \tilde{R}_{[uv,w]} \in \Lambda^3 V_R$ and $\hat{R}_{[uv,w]} = 0$. Thus,
\begin{equation}
 \hat{R}_{uv,w} = \frac23 \left( \tilde{R}_{u(v,w)} - \tilde{R}_{v(u,w)} \right) \,.
\end{equation}
We can calculate
\begin{equation}
 \begin{split}
  2 \kappa^{D-1} \tilde{R}_{u(v,w)} &= \left( \gL_{\J_u} \J_{(v} \right) \wedge \hJ_{w)} - \hJ_{(w} \wedge \gL_{\J_{v)}} \J_u \\
  &= \left( \gL_{\J_u} \J_{(v} \right) \wedge \J_{w)} \wedge \hK - \hK \wedge \J_{(w} \gL_{\J_{v)}} \J_u \\
  &= \hK \wedge \left[ \frac12 \gL_{\J_u} \left( \J_v \wedge \J_w \right) - \gL_{\J_{(v}} \left( \J_{w)} \J_u \right) + \left( \gL_{\J_{(v}} \J_{w)} \right) \wedge \J_u \right] \\
  &= \hK \wedge \left[ \frac12 \delta_{vw} \gL_{\J_u} \K - \delta_{u(w} \gL_{\J_{v)}} \K + \frac1{d-1} \delta_{vw} \left( \gL_{\J^x} \J_x \right)   \wedge \J_u \right] \\
  &= \delta_{vw} \frac{2}{d-1} \hJ_u \wedge \gL_{\J^x} \J_x - \frac{2}{d-1} \delta_{u(v} \hJ_{w)} \wedge \gL_{\J^x} \J_x \,,
 \end{split}
\end{equation}
which means that
\begin{equation}
 \begin{split}
  \hat{R}_{uv,w} &= \frac{2}{d-1} \kappa^{1-D} \delta_{w[v} \hJ_{u]} \wedge \gL_{J^x} J_x \\
  &= \kappa^{1-D} \delta_{w[v} \hJ_{u]} \wedge d\K \\
  &= T_{2[u} \delta_{v]w} \,,
 \end{split}
\end{equation}
and
\begin{equation}
 \tilde{R}_{uv,w} = R_{uvw} + T_{2[u} \delta_{v]w} \,.
\end{equation}

Finally, a long but straightforward computation shows that
\begin{equation}
 \tilde{R}_{uv} = \LG_{uv} \cdot T_3 \,.
\end{equation}
Thus, we find that
\begin{equation}
 \gL_{\J_{[u}} \J_{v]} = \kappa^2 R_{1\,uv} + \kappa\, R_{2\,uvw} \J^w + \kappa\, T_{2[u} \J_{v]} - \kappa^{6-D} \hK \bullet \left( \LG_{uv} \cdot T_3 \right) \,.
\end{equation}

\section{$\SO{d-1}_R$ invariance of the scalar potential} \label{a:PotentialRSymmetry}
Here we will show that the universal part of the scalar potential \eqref{eq:PotUniversal}, excluding the $T_1^2$ and $T_2^2$ terms, is fixed by requiring invariance under local $\SO{d-1}_R$ transformations. We begin with  the Ansatz
\begin{equation}
 \begin{split}
  V_0 &= \alpha_1 \kappa^{4-D} R_{1\,uv} \wedge R_1^{uv} \wedge \hK + \alpha_2 R_{2\,uvw} R_2^{uvw} + \alpha_3 \kappa^{-2} \gL_{\J_u} \left( U^u \kappa \right) + \alpha_4 U_u U^u + \alpha_5 U_u T_2^u \,. \label{eq:PotentialAnsatz}
 \end{split}
\end{equation}
We ignore $T_1^2$ and $T_2^2$ terms, because $T_1$ and $T_2$ are invariant under local $\SO{d-1}_R$ transformations.

Under the $\SO{d-1}_R$ symmetry, the $\J_u$'s transform as
\begin{equation}
 \delta_\lambda \J_u = \lambda_u{}^v \J_v \,.
\end{equation}
Using \eqref{eq:JTorsionClasses} one finds the anomalous transformations
\begin{equation}
 \begin{split}
  \Delta_\lambda R_{2\,uvw} &= 3 \kappa^{-1} \J_{[u}{}^M \partial_{|M|} \lambda_{vw]} \,, \\
  \Delta_\lambda R_{1\,uv}{}^M &= \kappa^{-2} \frac{1}{d-1} Y^{MN}_{PQ} \J_w^Q \J{}^{w\,P} \partial_N \lambda_{uv} \,, \\
  \Delta_\lambda U_u &= \kappa^{-1} \J^{v\,M} \partial_M \lambda_{uv} \,.
 \end{split} \label{eq:AnomalousTorsion}
\end{equation}
Here we will ignore the torsion classes $S_{1\,u}$ and $S_2$ as these will vanish when we have an honestly half-maximal theory.

Using \eqref{eq:AnomalousTorsion}, one finds that the terms in \eqref{eq:PotentialAnsatz} transform as
\begin{equation}
 \begin{split}
  \Delta_\lambda \left(R_{2\,uvw} R_2^{uvw} \right) &= 6 \kappa^{-D} \left( \hJ^u{}_N \gL_{\J^v} \J^{w\,N} \right) \J_u{}^M \partial_M \lambda_{vw} \,, \\
  \Delta_\lambda \left( R_{1\,uv} \wedge R_1^{uv} \wedge \hK \right) &= 2 \kappa^{D-4} R_2^{uv\,M} \partial_M \lambda_{uv} \\
  &= 2 \kappa^{D-6} \gL_{\J_u} \J_v{}^M \partial_M \lambda^{uv} - 2 \kappa^{-4} \hJ^w{}_N \gL_{\J^u} \J^{v\,N} \J_w{}^M \partial_M \lambda_{uv} \\
  & \quad - 2 \kappa^{D-5} T_{2\,u} \J_v{}^M \partial_M \lambda^{uv} \,, \\
  \Delta_\lambda U_u U^u &= 2 \kappa^{-1} U_u \J_v{}^M \partial_M \lambda^{uv} \,, \\
  \Delta_\lambda U_u T_2^u &= \kappa^{-1} T_{2\,u} \J_v{}^M \partial_M \lambda^{uv} \,, \\
  \Delta_\lambda \left( \gL_{\J_u} \left( \kappa U^u \right) \right) &= \J_u{}^M U_v \partial_M \lambda^{uv} + \kappa^{-1} \J^{u\,M} \partial_M \J^{v\,N} \partial_N \lambda_{uv} \\
  &= \J_u{}^M U_v \partial_M \lambda^{uv} + \frac12 \kappa^{-1} \gL_{\J_u} \J_v{}^M \partial_M \lambda^{uv} \,,
 \end{split}
\end{equation}
where going to the last line we used the section condition \eqref{eq:SectionCondition}. Thus, one finds that the scalar potential \eqref{eq:PotentialAnsatz} transforms as
\begin{equation}
 \begin{split}
  \Delta V_0 &= 2 \left( 3 \alpha_2 - \alpha_1 \right) \kappa^{-D} \left(\J^u{}_N \gL_{\J^v} \J^{w\,N} \right) \J_u{}^M \partial_M \lambda_{vw} \\
  &\quad + 2 \left( \alpha_1 + \frac{\alpha_3}{4} \right) \kappa^{-1} \gL_{\J_u} \J_v{}^M \partial_M \lambda^{uv} - 2 \left( \alpha_1 + \frac{\alpha_5}{2} \right) \kappa^{-1} T_{2\,u} \J_v{}^M \partial_M \lambda^{uv} \\
  & \quad + \left( 2 \alpha_4 - \alpha_3 \right) \kappa^{-1} U_u \J_v{}^M \partial_M \lambda^{uv} \,.
 \end{split}
\end{equation}
Thus the scalar potential is invariant when
\begin{equation}
 \begin{split}
  \alpha_2 &= \frac{\alpha_1}{3} \,, \\
  \alpha_3 &= - 4 \alpha_1 \,, \\
  \alpha_4 = a_5 &= - 2 \alpha_1 \,.
 \end{split}
\end{equation}

\section{Intrinsic torsion of $\ON{d} \subset \ON{d,d} \times \mathbb{R}^+$ structure} \label{A:DFTTorsion}

We begin by showing that the symmetric part of the first equation of \eqref{eq:DFTIntTorsionPre} vanishes. Using the compatibility condition \eqref{eq:DFTCompat}, it follows that
\begin{equation}
 \gL_{\J_{(u}} \J_{v)}{}^M = \frac12 \eta^{MN} \partial_N \left( \J_u{}^P \J_v{}^Q \eta_{PQ} \right) = 0 \,. \label{eq:ONInt1}
\end{equation}

Thus we see that
\begin{equation}
 \gL_{\J_u} \J_v{}^M = \gL_{\J_{[u}} \J_{v]}{}^M = R_{uv}{}^M + R_{uv,w} \J^{w,M} \,,
\end{equation}
where $R_{uv}{}^M \J_{w,M} = 0$ and
\begin{equation}
 \J_{w,M} \gL_{\J_{[u}} \J_{v]}{}^M = R_{uv,w} \,.
\end{equation}
We will now show that $R_{uv,w} = R_{uvw} = R_{[uvw]}$ is totally antisymmetric. We first write
\begin{equation}
 R_{uv,w} = R_{uvw} + \hat{R}_{uv,w} \,,
\end{equation}
where $R_{uvw}$ is totally antisymmetric and
\begin{equation}
 \hat{R}_{uv,w} = \frac23 \left( R_{u(v,w)} - R_{v(u,w)} \right) \,.
\end{equation}
However,
\begin{equation}
 \begin{split}
  R_{u(v,w)} &= \frac12 \left( \J_{(w|M|} \gL_{\J_{|u|}} \J_{v)}{}^M - \J_{(w|M} \gL_{\J_{v)}} \J_{u}{}^M \right) \\
  &= \frac14 \gL_{\J_u} \left( \J_{w,M} \J_v{}^M \right) - \frac12 \gL_{\J_{(v}} \left( \J_{w)}{}^M \J_{u,M} \right) + \frac12 \J_u{}^M \gL_{\J_{(v}} \J_{w)}{}^M \\
  &= 0 \,,
 \end{split}
\end{equation}
using \eqref{eq:DFTCompat} and \eqref{eq:ONInt1}. As a result we see that $\hat{R}_{uv,w} = 0$ and $R_{uv,w} = R_{uvw} = R_{[uvw]}$ is totally antisymmetric.

Thus, we can write
\begin{equation}
 \gL_{\J_u} \J_{v}{}^M = R_{uv}{}^M + R_{uvw} \J^{w,M} \,,
\end{equation}
with $R_{uv}{}^M \J_{w,M} = 0$. Similarly we could consider
\begin{equation}
 \gL_{\J_u} e^{-2d} = U_u e^{-2d} \,,
\end{equation}
where $U_u$ is a component of the intrinsic torsion.

\newpage

\bibliographystyle{JHEP}
\bibliography{NewBib}

\end{document}